\documentclass[%
 mph,%
 amsmath,amssymb,
 reprint,%
superscriptaddress
]{revtex4-1}

\usepackage{color}
\usepackage{bbm}
\usepackage{tikz}

\usepackage{graphicx}
\usepackage{dcolumn}
\usepackage{bm}
\usepackage{color}

\usepackage[
pagebackref=false,
colorlinks=true,
linkcolor=blue,
urlcolor=blue,
filecolor=black,
citecolor=red,
pdfstartview=FitV,
pdftitle={},
pdfauthor={},
pdfsubject={},
pdfkeywords={},
pdfpagemode=None,
bookmarksopen=true
]{hyperref}

\begin{document}

\newtheorem{definition}{Definition}[section]
\newcommand{\be}{\begin{equation}}
\newcommand{\ee}{\end{equation}}
\newcommand{\bea}{\begin{eqnarray}}
\newcommand{\eea}{\end{eqnarray}}
\newcommand{\LE}{\left[}
\newcommand{\R}{\right]}
\newcommand{\nn}{\nonumber}
\newcommand{\Tr}{\text{Tr}}
\newcommand{\N}{\mathcal{N}}
\newcommand{\G}{\Gamma}
\newcommand{\vf}{\varphi}
\newcommand{\LL}{\mathcal{L}}
\newcommand{\Op}{\mathcal{O}}
\newcommand{\HH}{\mathcal{H}}
\newcommand{\arctanh}{\text{arctanh}}
\newcommand{\up}{\uparrow}
\newcommand{\down}{\downarrow}
\newcommand{\ket}[1]{\left| #1 \right>}
\newcommand{\bra}[1]{\left< #1 \right|}
\newcommand{\ketbra}[1]{\left|#1\right>\left<#1\right|}
\newcommand{\rd}{\partial}
\newcommand{\de}{\partial}
\newcommand{\ba}{\begin{eqnarray}}
\newcommand{\ea}{\end{eqnarray}}
\newcommand{\db}{\bar{\partial}}
\newcommand{\we}{\wedge}
\newcommand{\ca}{\mathcal}
\newcommand{\lr}{\leftrightarrow}
\newcommand{\f}{\frac}
\newcommand{\s}{\sqrt}
\newcommand{\vp}{\varphi}
\newcommand{\hvp}{\hat{\varphi}}
\newcommand{\tvp}{\tilde{\varphi}}
\newcommand{\tp}{\tilde{\phi}}
\newcommand{\ti}{\tilde}
\newcommand{\pr}{\propto}
\newcommand{\mb}{\mathbf}
\newcommand{\ddd}{\cdot\cdot\cdot}
\newcommand{\no}{\nonumber \\}
\newcommand{\la}{\langle}
\newcommand{\lb}{\rangle}
\newcommand{\ep}{\epsilon}
\newcommand{\eqdef}{=\mathrel{\mathop:}}
 \def\we{\wedge}
 \def\lr{\leftrightarrow}
 \def\f {\frac}
 \def\ti{\tilde}
 \def\ap{\alpha}
 \def\pr{\propto}
 \def\mb{\mathbf}
 \def\ddd{\cdot\cdot\cdot}
 \def\no{\nonumber \\}
 \def\la{\langle}
 \def\lb{\rangle}
 \def\ep{\epsilon}
\newcommand{\mcl}{\mathcal}
 \def\g{\gamma}
\def\Tr{\text{tr}}

\title{Spatial deformation of many-body quantum chaotic systems and quantum information scrambling}

\author{Kanato Goto}
\affiliation{%
  Department of Physics, Princeton University, Princeton, New Jersey, 08544, USA}
\affiliation{
  Center for Gravitational Physics and Quantum Information,
  Yukawa Institute for Theoretical Physics,
  Kyoto University, Kyoto 606-8501, Japan}
\affiliation{
  RIKEN Interdisciplinary Theoretical and Mathematical Sciences (iTHEMS),
  Wako, Saitama 351-0198, Japan}

\author{Taozhi Guo}
\affiliation{%
  Department of Physics, Princeton University, Princeton, New Jersey, 08544, USA}

\author{Tomoki Nosaka}
\affiliation{%
  Kavli Institute for Theoretical Sciences, University of Chinese Academy of Sciences,
  Beijing 100190, China}

\author{Masahiro Nozaki}%
\affiliation{%
  Kavli Institute for Theoretical Sciences, University of Chinese Academy of Sciences,
  Beijing 100190, China}
\affiliation{
  RIKEN Interdisciplinary Theoretical and Mathematical Sciences (iTHEMS),
  Wako, Saitama 351-0198, Japan}

\author{Shinsei Ryu}
\affiliation{%
Department of Physics, Princeton University, Princeton, New Jersey, 08544, USA}

\author{Kotaro Tamaoka}
\affiliation{%
Department of Physics, College of Humanities and Sciences, Nihon University, Sakura-josui, Tokyo 156-8550, Japan}

\date{\today}

\begin{abstract}
  We study the effect of spatial inhomogeneity on quantum information scrambling,
  a process of spreading and locally hiding quantum information in quantum many-body systems.
  As a paradigmatic example, we consider the quantum chaotic Ising spin chain and
  its inhomogeneous counterpart that is obtained by modulating the Hamiltonian density.
  Specifically, we consider the so-called M\"obius and sine-square deformations
  that were previously studied in the context of (1+1)-dimensional conformal field theories ($1+1$ d CFTs).  
  In the spatial region where the 
  modulated energy density is small, these deformations
  prevent the spreading of quantum information while in the region where the modulated energy density is large
  quantum information scrambling is accelerated. 
  This suggests that we can control the scrambling and butterfly effect by spatially modulating the Hamiltonian density. 
 We also found that the time dependence of energy density exhibits the signature 
 of black-hole-like excitation found in the $1+1$ d CFTs even 
 in the chaotic spin chain.
\end{abstract}

\maketitle



\section{Introduction and summary}

Quantum information scrambling -- a dynamical phenomenon where quantum information spreads and becomes locally
hidden in complex quantum many-body systems -- has become a center of broad interest in theoretical physics. 
Recent years also have seen the rapid development of experimental techniques to
measure scrambling in laboratories
(e.g., \cite{2019Natur.567...61L,2020PhRvL.124x0505J,2021PhRvX..11b1010B,2016PhRvA..94d0302S,PhysRevA.94.062329,2016arXiv160701801Y,2017PhRvA..95a2120Y,2018PhRvA..97d2105Y,2017PhRvE..95f2127C,2017arXiv171003363Y,2017NatPh..13..781G,2016arXiv161205249W,PhysRevX.7.031011,2017arXiv170506714M}).
Quantum information scrambling underlies thermalization and information retrieval from a black hole.
In the former, the information of the initial state spreads over the entire
system and can no longer be deduced by local measurements of few-body operators. 
Thus, irreversibility emerges even in unitary dynamics
(e.g., \cite{PhysRevA.43.2046,PhysRevE.50.888,2008Natur.452..854R,2011RvMP...83..863P,2005JSMTE..04..010C}).
It has been argued
\cite{2007JHEP...09..120H,2017arXiv171003363Y,2020arXiv200700895N}
that a black hole scrambles quantum information dumped into the black hole, and
then emits the Hawking radiations.
It has been discussed how to reconstruct the quantum state from the emitted Hawking radiations.
Information retrieval from the typical state is one of the central challenges in non-equilibrium physics.


Various measures diagnosing quantum information scrambling have been introduced, including 
out-of-time-ordered correlators (OTOCs), operator mutual information (OMI), (subsystem) spectral form factor, etc.
OTOC is one measure of scrambling and characterizes the growth of the Heisenberg operator.
The behavior of OTOC has been extensively studied across a variety of systems (e.g., \cite{2014JHEP...03..067S,2016JHEP...08..106M,PhysRevLett.115.131603,2016JHEP...10..009S,PhysRevD.96.065005,PhysRevX.7.031047,2017PhRvX...7c1016N,2018PhRvX...8b1014N,2018PhRvX...8c1057K,2018PhRvX...8c1058R,PhysRevB.96.020406,2017NJPh...19f3001B,2018PhRvL.121a6801H,PhysRevB.97.144304,2020NatPh..16..199X,Khetrapal:2022dzy}).

In this paper, we propose and study a system where the scrambling capability of dynamics
is controlled by the inhomogeneous deformation of the Hamiltonian. 
The deformations of our interest are the so-called M\"obius and sine-square deformation (SSD).
In these deformations, the energy scale is spatially modified by an envelope function.
The SSD was originally introduced as a simple way to remove the boundary effect in a finite-size system
\cite{PhysRevB.83.060414,PhysRevB.84.165132,2012JPhA...45k5003K,2009PThPh.122..953G,2011PhRvA..83e2118G,2011PhRvB..84k5116S,2011JPhA...44y2001K,PhysRevB.86.041108,PhysRevB.87.115128}.
Subsequently, much progress has been made on the M\"obius and Sine-Square (SS) deformations in
two-dimensional conformal field theories ($2$d CFTs) \cite{2015JPhA...48E5402I,2016IJMPA..3150170I,2016arXiv160309543O,PhysRevB.93.235119,2017arXiv170906238T,2018PTEP.2018f1B01T,2018JSP...172..353G,PhysRevLett.122.020201,Gluza_2022}.
Furthermore, various non-equilibrium processes have been studied by using the M\"obius/SS deformations
\cite{PhysRevB.97.184309,2019JPhA...52X5401M,Goto:2021sqx} including Floquet dynamics
\cite{PhysRevLett.118.260602,2018arXiv180500031W,2020PhRvX..10c1036F,2020PhRvB.102t5125H,2021PhRvR...3b3044W,2020arXiv201109491F,2021arXiv210910923W,Lapierre_2020,Lapierre_2020_1,Lapierre_2021}.
Ref.\ \cite{Goto:2021sqx} studied a quantum quench by the M\"obius/SSD Hamiltonian from the thermal initial state.
In particular, it was found that the quench process creates a local excitation,
dubbed black-hole like excitation, that carries the entropy of the total system. 
The streaming-free quasi-particles give a nice explanation for the findings on this entanglement dynamics.
In the case of the SSD Hamiltonian, the quasi-particles propagate at the non-uniform traveling speed to the point
where the envelope function vanishes, accumulate around this point, and serve as a source of quantum entanglement.
This suggests that the information about the initial state localizes at this point.
The inhomogeneous deformation may offer the key to understanding the information retrieval from the typical state. 
On the other hand, in the rest of the system, the temperature becomes lower lecause the number of
quasi-particles 
decreases there.
This opens up the possibility of quenches where the sub-regions are cooled down under the evolution by the inhomogeneous Hamiltonian \cite{2016arXiv161104591Z,2018PhRvL.120u0604A,2019PhRvB..99j4308M,Goto:2021sqx,2022arXiv221100040W}.

These previous findings are largely universal in the sense that they depend only on conformal symmetry.
They are insensitive to the details of CFTs, such as the operator content
and the distinction between rational and irrational (holographic) CFTs.
In the regular spatially uniform dynamics, it is known that these theories exhibit
different quantum chaotic behaviors (different quantum information scrambling capabilities)
\cite{2015JHEP...09..110A,2014PhRvD..89f6015A,2019JSMTE..09.3107N,2020JHEP...01..031K,2021JHEP...03..146K,2022JHEP...06..100G}.
In a separate work, we investigate the effects of the spatial deformations on quantum information spreading of different CFTs \cite{2023arXiv230208009G}.
In this work, we investigate the effects of spatial inhomogeneity in a wider context
in systems that do not have conformal symmetry.
Specifically, we consider the quantum Ising chain in the presence of both
transverse and longitudinal magnetic fields \cite{2011PhRvL.106e0405B,2016JHEP...02..004H,2020arXiv201214609M,2020PhRvB.101q4313C}
and its M\"obius/SS deformations. 
This is a prototypical example of chaotic many-body quantum systems
that may be easier to realize in quantum simulators
and allows us to study quantum information scrambling behaviors (without conformal symmetry). 
We note that the SSD deformation has been utilized in non-conformal systems in numerical simulations to remove the boundary effects 
\cite{Nishimoto_2013,Ito_2018}.

Another issue we plan to address is if the formation of a black-hole-like excitation
found in Ref.\ \cite{Goto:2021sqx} 
can also be realized in the spin chain model. The analysis in Ref.\ \cite{Goto:2021sqx}
relied on the quasi-particle picture and conformal symmetry.
It is therefore of our interest to see if a black-hole-like excitation forms
even when the quasi-particle picture does not apply, i.e., in the chaotic regime.

In the present paper, we study the time-evolution of information-theoretic quantities
that diagnose the quantum information scrambling 
during the time-evolution induced by the M\"obius/SS deformed Hamiltonian.
The spin system considered consists of $L$ sites with the periodic boundary condition.
We found that on the sites where the Hamiltonian densities are amplified by the envelope function, quantum information scrambling is accelerated, while on the sites where the envelope function makes these densities smaller, this process becomes slower. 
Furthermore, during the SSD evolution, the time dependence of the energy density exhibits the signature of the emergence of black-hole-like excitations around the bond where the envelope function vanishes as in $2$d CFT \cite{Goto:2021sqx,2023arXiv230208009G}.
During the M\"obius evolution, the energy density does not exhibit the periodic oscillation in time, unlike in $2$d CFT.
This might be because the system considered in this paper does not have conformal symmetry.
\if[0]
*********
In this paper, we consider the M\"obius/SS deformation of a one-dimensional Ising chain with magnetic fields in the chaotic regime \cite{2011PhRvL.106e0405B}. 
The goal of this paper is to qualitatively study how the M\"obius/SS deformation temporally and spatially modifies scrambling.
We study the non-local correlations evolving with time according to the M\"obius/SS deformed chaotic Hamiltonian.
We also study how Heisenberg operators grow under the evolution induced by this Hamiltonian.

We study the scrambling properties of unitary time evolution operators by information-theoretic quantities such as operator mutual information (OMI), out-of-time ordered correlator (OTOC), and so on.
For OMI, we define a dual state by mapping the time evolution operators to a state in the doubled Hilbert space.
From the perspective of the state, we start from a state consisting of Bell pairs \cite{PhysRev.47.777} and evolve the system with the M\"obius/SS deformed Hamiltonian.
We study the time evolution of non-local correlations associated with subsystems of the dual state by using the OMI and two-point function.
In addition to these quantities, we study the effect of inhomogeneous deformation on the spectrum of the reduced density matrix by using the subsystem spectral form factor (SSF).
We also study the growth of Heisenberg operators under time evolution by the M\"obius/SS deformed Hamiltonian by using the OTOC and return amplitude. 
It may be worth mentioning in this paper the effect of the deformation on the free theory.
We report the time-dependence of bipartite operator mutual information (BOMI) under the evolution by inhomogeneously-deformed Hamiltonian of the $2$d free fermionic theory in Appendix \ref{Section:BOMI_FF}.  
\fi

The paper is organized as follows. 
In Section \ref{Section:preliminary}, we will describe the definition of the spin model considered in this paper
and the M\"obius/SS deformation of this model, and then report the chaoticity of this inhomogeneously deformed spin model.
In Section \ref{Section:ED-BHlike-exitaion}, we will report the time dependence of the energy density during the evolution induced by the M\"obius/SS deformed Hamiltonian.
There are some indications that the black-hole-like excitations emerge around the bond where the envelope unction vanishes.
In Section \ref{Section:BOMIandtTP}, we will present the numerical analyses of the time-dependence of the bipartite operator mutual information (BOMI) and compare it with the results of the two-dimensional holographic conformal field theory ($2$d holographic CFT), the CFT having the gravity dual.
In Section \ref{Section:SFF}, we also discuss the subsystem spectral form factor (SSF) of the SSD and  M\"obius/SS deformed Hamiltonian.
In Section \ref{Section:speed}, we will present numerical analyses of the time dependence of the return amplitude and OTOCs under this evolution.
In Section \ref{Section:Discussion_and_FD}, we will discuss the results of this paper, and comment on a few future directions.

\section{Preliminaries \label{Section:preliminary}}

\begin{figure}[t]
  \begin{center}
    \includegraphics[width=6cm]{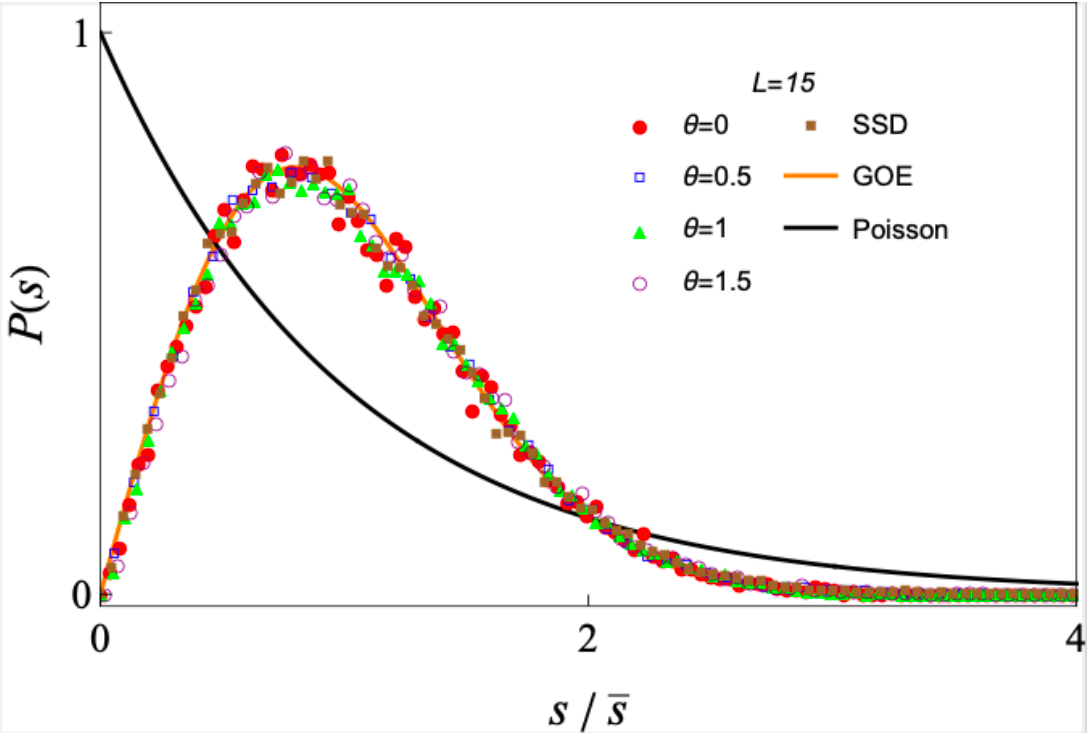}
    \caption{
      Distribution of the normalized nearest-neighbor level spacings
      \eqref{normalizedspacings} of the M\"obius/SS deformed and un-deformed
      chaotic spin chains in \eqref{220413Mobius} and \eqref {220413SSD}. 
      Here, $(h_x,h_z)=(-1.05,0.5)$. We have identified the global structure
      of $\bar\rho_\alpha(E)$ by fitting the energy level density of each sector with an eighth-order polynomial of $E$.
      Black/Orange curves in the plot are the Wigner-Dyson distribution for
      Gaussian orthogonal ensemble (GOE) $P_{\text{GOE}}(s)=\frac{\pi s}{2}e^{-\frac{\pi s^2}{4}}$
      and the Poisson distribution $P_{\text{Poisson}}(s)=e^{-s}$.
    }
    \label{fig_NNSD1}
  \end{center}
\end{figure}


The model of our interest in this paper is the one-dimensional Ising spin system
with both transverse and longitudinal magnetic fields,
\be
H_0= \sum_{a=1}^L\left(\sigma_{z,a}\sigma_{z,a+1}+h_x\sigma_{x,a}+h_z\sigma_{z,a}\right).
\label{Hclosed}
\ee
Here,
$\sigma_{\alpha, a}$ ($\alpha=x,y,z$) are Pauli operators at the $a$-th site,
and we have employed the periodic boundary condition $\sigma_{z,L+1}=\sigma_{z,1}$.
For $(h_x,h_z)=(-1.05,0.5)$, this Ising spin chain is in the chaotic regime \cite{2011PhRvL.106e0405B}.

As in \cite{PhysRevB.84.165132}, we define the M\"obius deformed Hamiltonian of the chaotic chain as
\begin{align}
&
H_{\text{M\"obius}}=\sum_{a=1}^{L}\left[\Bigl(1-\tanh 2\theta \cos{\left(\frac{2\pi a}{L}\right)}\Bigr)\sigma_{z,a}\sigma_{z,a+1}\right.
\nonumber \\
&\quad \left. +\Bigl(1-\tanh 2\theta \cos{\left(\frac{\pi(2a-1)}{L}\right)}\Bigr)
(h_x\sigma_{x,a}+h_z\sigma_{z,a})\right],
\label{220413Mobius}
\end{align}
which depends on the inhomogeneity parameter $\theta$.
For $L$ even and $\theta>0$, the envelope function takes its minimum $1-\tanh 2\theta$ on the bond connecting the first site and the $L$-th site (which we call $b_0$), and takes the maximum $1+\tanh 2\theta$ on the bound connecting the $\frac{L}{2}$-th site and the $(\frac{L}{2}+1)$-th site (which we call $b_1$).
For $\theta=0$ the M\"obius deformed Hamiltonian reduces to the undeformed
Hamiltonian $H_0$ \eqref{Hclosed},
while in the SSD limit $\theta \rightarrow \infty$, the M\"obius deformed Hamiltonian becomes the following SSD Hamiltonian
\begin{align}
  &H_{\text{SSD}}=\sum_{a=1}^{L}
    \Big[2\sin^2\Bigl(\frac{\pi a}{L}\Bigr)\sigma_{z,a}\sigma_{z,a+1}
    \nonumber \\
&\quad 
    +2\sin^2\Bigl(\frac{\pi(2a-1)}{2L}\Bigr)(h_x\sigma_{x,a}+h_z\sigma_{z,a})\Big].
\label{220413SSD}
\end{align}
Note that the envelope function of the SSD Hamiltonian vanishes on the bond $b_0$

In Fig.\ \ref{fig_NNSD1}, we plot the distribution of the nearest-neighbor level spacings
(see Appendix \ref{Section:levelstatics} for details)
of the chaotic Ising chain.
The M\"obius/SS deformation does not affect the quantum chaoticity of the
system at least globally. In the following sections, we will try to probe
chaotic behaviors locally by studying operator entanglement,
subsystem spectral form factor, and OTOCs.

\begin{figure*}[t]
  \begin{center}
    \includegraphics[width=0.3\textwidth]{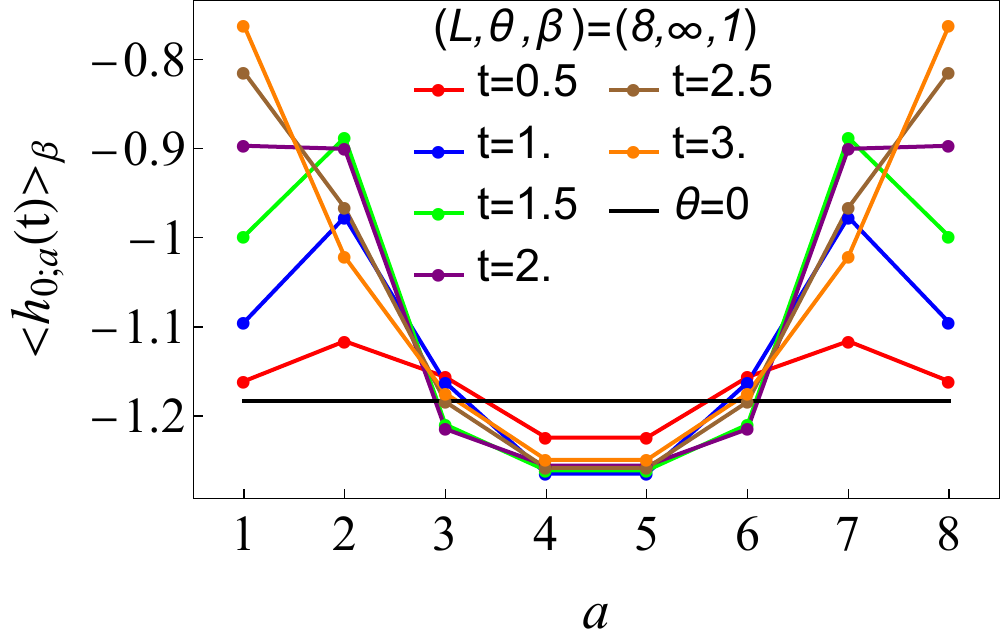}
        \includegraphics[width = 0.38\textwidth]{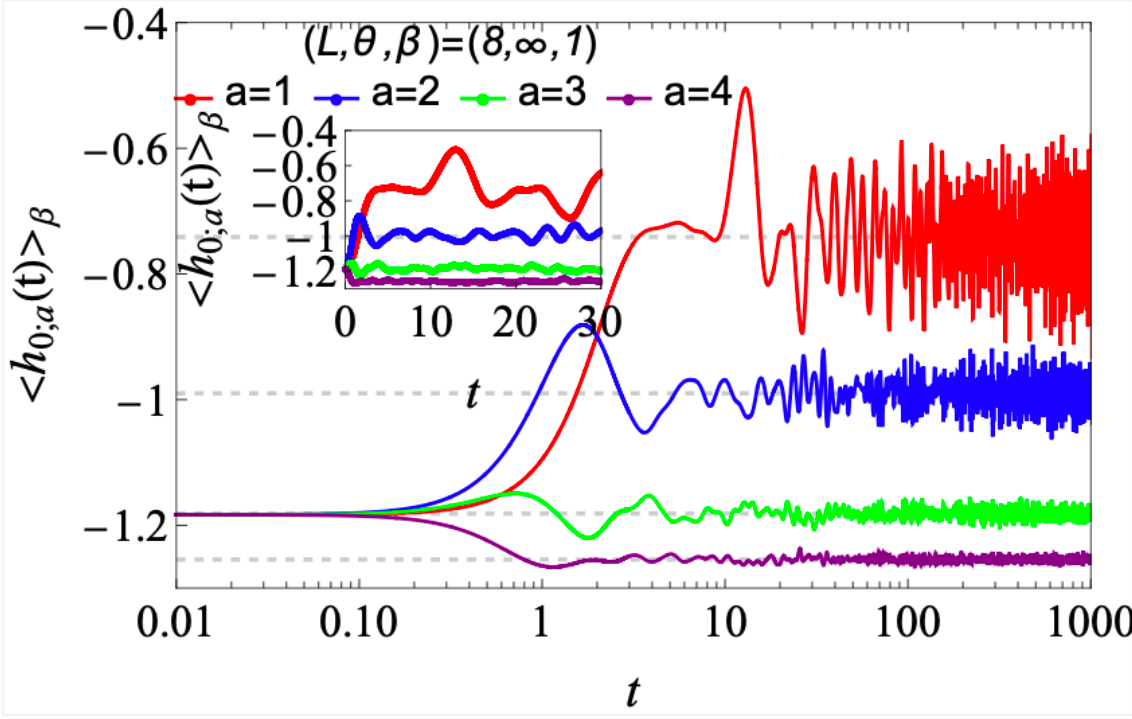}
            \includegraphics[width = 0.3\textwidth]{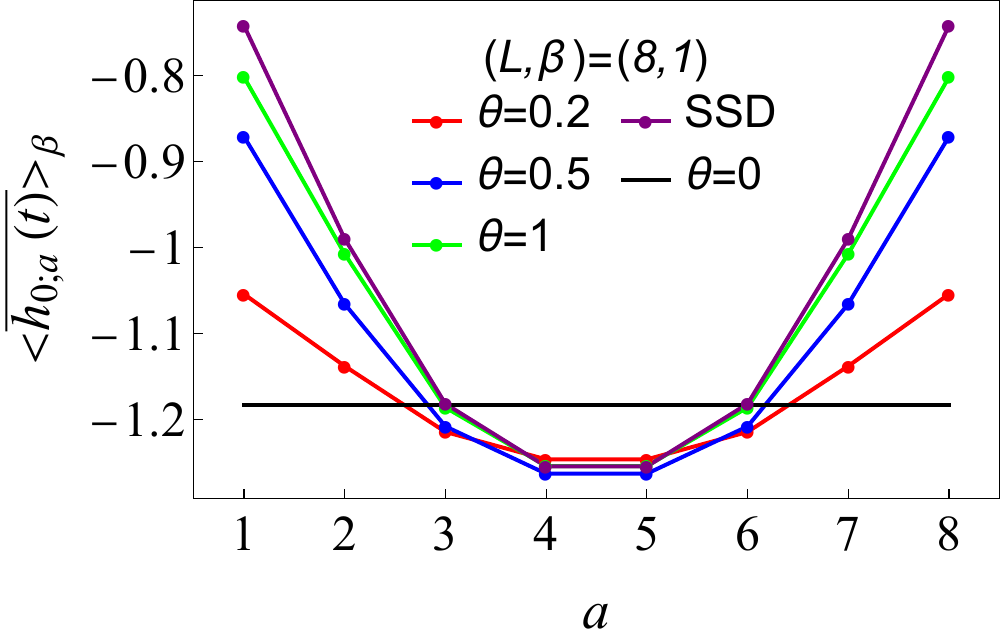}
    \caption{
      Left panel: time-dependence of $\langle h_{0,a}(t)\rangle_\beta$ in the evolution induced  by $H_{\mathrm{SSD}}$ \eqref{220413SSD} in the chaotic regime $(h_x,h_z)=(-1.05,0.5)$, as a function of $a$.
      The black line is the time-dependence of $\langle h_{0,a}(t)\rangle_\beta$ in the evolution induced by the undeformed Hamiltonian in the chaotic regime.
      Center panel: time-dependence of the energy density $\left\langle h_{0,a}(t)\right\rangle_\beta$ under SSD evolution and position/$\theta$-dependence of the averaged energy density during the time-evolution by (\ref{220413Mobius}) and (\ref{220413SSD}).
      For visibility, we have connected the discrete data points in $30<t<1000$ with $\Delta \log t=1.62\times 10^{-3}$ with lines.
      Right panel: the energy density averaged over late times ($100<t<100000$ with $\Delta \log t=1.62\times 10^{-3}$).
      The average values are indicated with dashed horizontal lines in the center panel for SSD and in Fig.~\ref{theta_and_position_dependence_of_h0a} for $\theta=0.2,0.5,1$.
    }
    \label{fig:ted}
  \end{center}
\end{figure*}

\section{Energy density and black-hole like excitation \label{Section:ED-BHlike-exitaion}}

Let us begin with the time-dependence 
of the energy density during the evolution 
induced by $H_{\text{M\"obius}}$ and $H_{\text{SSD}}$.
The energy density operator $h_{0,a}$ at $a$-th site 
of the un-deformed spin chain is 
\begin{align}
h_{0,a}=\frac{1}{2}
(\sigma_{z,a-1}\sigma_{z,a}
+\sigma_{z,a}\sigma_{z,a+1})
+h_x\sigma_{x,a}+h_z\sigma_{z,a},
\label{energydensityunderSSDquench}
\end{align}
so that $H_0=\sum_{a=1}^Lh_{0,a}$.
Let us start from a thermal state 
with temperature $\beta^{-1}$, 
and then evolve it with the M\"obius/SS-deformed 
Hamiltonian.
In Fig.\ \ref{fig:ted},
we plot the expectation value
$
\langle h_{0,a}(t)\rangle_\beta=
\text{tr}\,[e^{iH_{\text{Inho}}t}
h_{0,a}e^{-iH_{\text{Inho}}t}e^{-\beta H_0}]/
\text{tr}\, [e^{-\beta H_0}]
$
for various $t$ as a function of $a$.
Here, $H_{\text{Inho}}$ is either $H_{\text{M\"obius}}$ or $H_{\text{SSD}}$.
(In Fig.\ \ref{fig:ted}, we only show the time evolution under
the SSD Hamiltonian.
See Appendix \ref{app_energydensity} for the M\"obius evolution.).
As expected, the inhomogeneous evolution accumulates
the energy density around the bond $b_0$ while it depletes the energy density around the bond $b_1$.
At the early times, $\langle h_{0,a}(t)\rangle_{\beta}$ increases for the sites close to the bond $b_0$ and decreases for the sites close to the bond $b_1$, hence the energy is accumulated around $b_0$.
During the subsequent time evolution,
$\langle h_{0,a}(t)\rangle_{\beta}$
oscillates around a certain value
that depends on $a$ and $\theta$.
This value is larger for $a$ being closer to $b_0$.
In the right panel of Fig.\ \ref{fig:ted},
we plot the energy density averaged over late times
for various $\theta$ as a function of $a$.
For all $\theta$ considered in this paper, the energy density
around $b_0$ is larger than those around $b_1$.
We can see from Figs.\ \ref{fig:ted}, 
the spatial energy distribution becomes sharper as $\theta$ is increased.
Overall, the behavior under the SSD Hamiltonian
is consistent with what was observed in CFT
\cite{Goto:2021sqx}:
the SSD Hamiltonian pushes excitations and entropy toward
the weakest bond $b_0$. In the case of CFT,
this process results in a pair of black-hole-like excitations
that merge eventually at the origin $b_0$
in $t\to \infty$.

On the other hand, in the case of CFT under the M\"obius time evolution,
a periodic oscillation with periodicity
that depends on the deformation parameter and the total length of the chain was observed in \cite{2023arXiv230208009G}.
The periodic oscillation originates from the integrability of CFT even under
the M\"obius deformation.
Such periodic oscillation was not observed for the chaotic spin chain,
although erratic oscillations were seen at later times as shown in Fig.~\ref{fig:ted}. 
Such difference may be a result of the lack of integrability or the lattice nature of the spin chain model. The limited system size might also be a factor.  

\section{Operator entanglement \label{Section:BOMIandtTP}}


We now turn to BOMI and discuss
quantum information scrambling. 
Consider a time evolution operator with a time-independent Hamiltonian,
$
U(t)=e^{-iHt}
$,
acting on the Hilbert space 
${\cal H}_1$ of the spin system.
By introducing another copy of the Hilbert space ${\cal H}_2$
and using the channel-state map \cite{nielsen_chuang_2010}, we define the dual state of this time evolution operator as 
\be \label{TFD_eigen}
\ket{\Psi (t)} =\f{1}{2^{\f{L}{2}}}(e^{-iHt}\otimes 1)
\sum_{i=1}^{\mathrm{dim}\, \mathcal{H}_1}
\ket{E_i}_1\otimes \ket{E_i^*}_2,
\ee
where
$\mathrm{dim}\, \mathcal{H}_1=2^L$, 
$\ket{E_i}_1$ are the eigenstates
of $H$ in the original Hilbert space ${\cal H}_1$
and $\ket{E_i}_2$ represent the corresponding states
in the other Hilbert space ${\cal H}_2$.
The symbol $\ket{\cdot^*}$ stands for the CPT conjugate of a state $\ket{\cdot}$.
Note that the Hamiltonian $H$ acts only on $\mathcal{H}_1$.
In the following, let ${\cal O}^{(I)}$ denote the operator acting only on ${\cal H}_I$, i.e., ${\cal O}^{(1)}={\cal O}\otimes 1$ and ${\cal O}^{(2)}=1\otimes {\cal O}$.
By expanding the right-hand side of \eqref{TFD_eigen} with the spin basis $\ket{s}_{I,a}$ (eigenstate of $\sigma_{z,a}^{(I)}$ with eigenvalue $s$), we can also write $\ket{\Psi(t)}$ with the EPR pairs $\ket{\text{EPR},a}=\f{1}{\sqrt{2}}\sum_{s=1,-1}\ket{s}_{1,a}\otimes\ket{s}_{2,a}$ as
$
\ket{\Psi(t)}=e^{-iH^{(1)}t} \bigotimes_{a=1}^L \ket{\text{EPR},a}.
$
As indicated in (\ref{TFD_eigen})
the information of $U(t)$ is encoded in the entanglement structure of the dual state $\ket{\Psi(t)}$.


We now divide the doubled spin chain associated with ${\cal H}_1\otimes {\cal H}_2$ into $\mathcal{V}$ and $\overline{\mathcal{V}}$, the complement to $\mathcal{V}$, and then define an operator entanglement entropy (OEE) associated with the reduced density matrix, $\rho_{\mathcal{V}}=\Tr_{\overline{\mathcal{V}}}\ket{\Psi(t)}\bra{\Psi(t)}$ for the subsystem $\mathcal{V}$ as 
$
S_{\mathcal{V}} = -\Tr_{\mathcal{V}}\left(\rho_{\mathcal{V}}\log{\rho_{\mathcal{V}}}\right).
$
BOMI of the disjoint intervals, $A$ and $B$, is defined by the linear combination of OEEs, 
\be
I_{A,B}=S_A+S_B -S_{A\cup B},
\ee
where $A \cup B$ denotes the union of $A$ and $B$.
Take $A$ and $B$ to be the subsets of the spin chain associated with $\mathcal{H}_1$ and $\mathcal{H}_2$, respectively.
Let $l_{\mathcal{V}}$ and $\mathcal{V}_c$ denote the size and center of the subsystem $\mathcal{V}$.
In this case, the values of $S_A$ and $S_B$ are those for a maximally entangled state,
$
S_A= l_A\log{2},\quad S_B=l_B\log{2}.
$
Thus, the time-dependence of $I_{A,B}$ is governed only by $S_{A\cup B}$.
%

\begin{figure*}[tbp]
    \begin{tabular}{cc}
      \begin{minipage}[t]{0.5\hsize}
        \centering
        \includegraphics[keepaspectratio, scale=0.6]{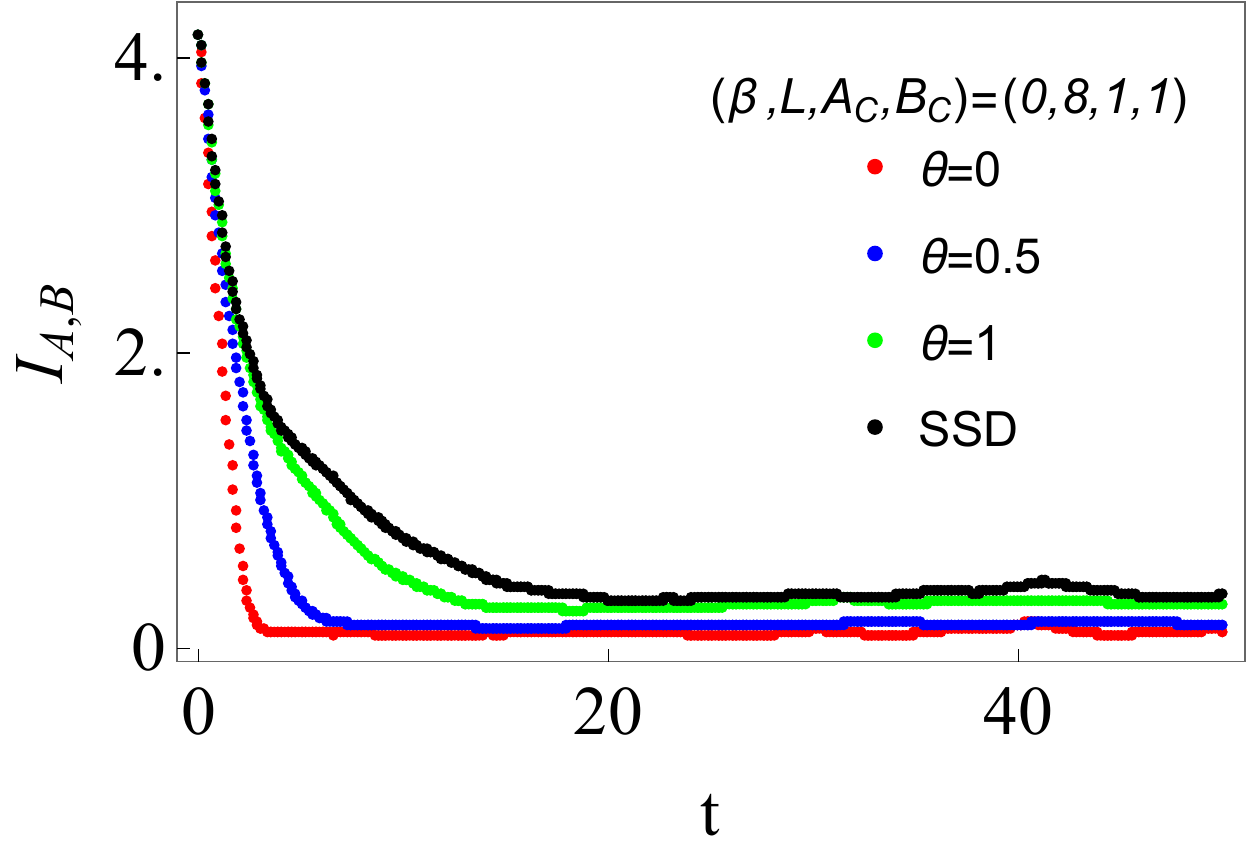}
    
       [a] $\theta$-dependence of $I_{A,B}$.
    
      \end{minipage} &
      \begin{minipage}[t]{0.5\hsize}
        \centering
        \includegraphics[keepaspectratio, scale=0.6]{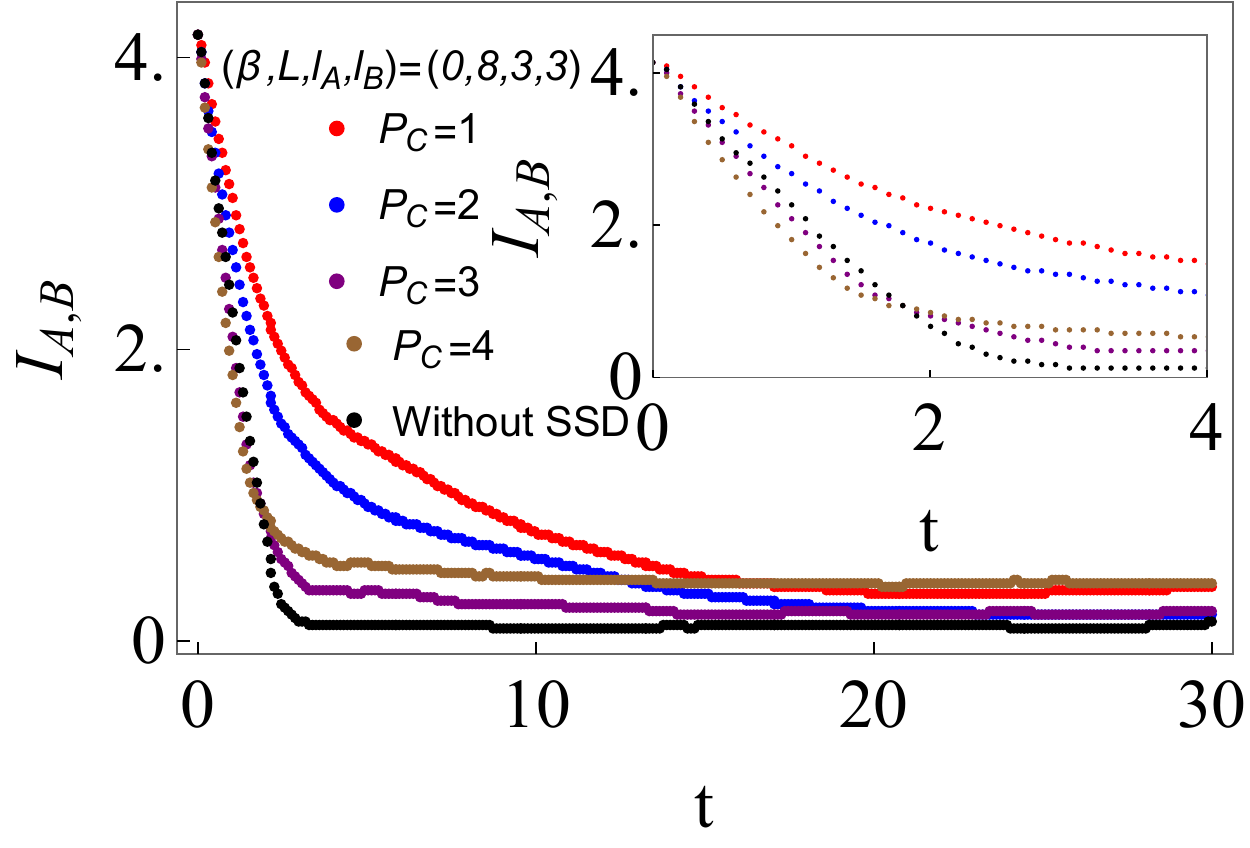}

        [b] Position-dependence of $I_{A,B}$.
      \end{minipage} \\
    \end{tabular}
    \caption{The $\theta$- and position-dependence of the time evolution of $I_{A,B}$, as functions of $t$. Here, $(L,l_A,l_B)=(8,3,3)$. The panel [a] shows the $t$-dependence of $I_{A,B}$ for $\theta=0,0.5,1,$ and $\infty$, and $P_c=1$.
    The panel [b] shows $t$-dependence of BOMI for $P_c=1,2,3,$ and $4$ and $\theta=2.5$. 
The black point is the time dependence of $I_{A,B}$ for $\theta=0$. In the inset of [b], we focus on the time-dependence of $I_{A,B}$ in the early-time region, $0 \le t \le 4$. 
} 
    \label{theta_and_position_dependence}
  \end{figure*}

Throughout this section, we set $L=8$.
We set
the centers and sizes of $A$ and $B$ as $A_c=B_c\eqdef P_c$ and $l_A=l_B=3$, 
and discuss the $\theta$- and $P_c$-dependence of BOMI $I_{A,B}$. 
In Fig.\ \ref{theta_and_position_dependence},
we plot BOMI obtained from numerical exact diagonalization
for various values of $\theta$ and $P_c$.
We found that for $\theta\ge 2.5$,
the time evolution of $I_{A,B}$ is almost independent of $\theta$.
Hence,
in Fig.\ \ref{theta_and_position_dependence}[b]
we show $I_{A,B}$ only for $\theta= 2.5$.

For all the choices of $\theta$ we studied, $I_{A,B}$ monotonically decreases with $t$
and then approximately saturates to a certain value that weakly depends on $\theta$.
For $\theta=0$, the early-time decay is approximately linear in $t$.
For large values of $\theta$,
the decay of $I_{A,B}$ is slower and
deviates from the linear decay. 
In the SSD limit, $\theta \rightarrow \infty$, the early-time decay of $I_{A,B}$
is the slowest among all $\theta$ considered.
So is the late-time saturation value,
although the variation of the saturation value as a function of $\theta$ is
rather small. 
The non-zero saturation values
for $\theta>0$, albeit being small,
is indicative of
the destruction/weakening of scrambling 
by the M\"obius/SSD deformation. 
Further investigation is needed to determine if the late time saturation values remain finite in the thermodynamic limit.
The late time values of 
the two-point function
(Appendix \ref{App:twopointfunction})
are similarly 
indicative of 
the inhibition of
the factorization of the density matrix
$\rho_{1^{(1)}\cup 1^{(2)}}
\approx \rho_{1^{(1)}} \otimes \rho_{1^{(2)}}$ 
at late times, 
i.e., scrambling.
Here, $1^{(1/2)}$ are the local Hilbert space
at site 1 for the first/second copy of the Hilbert space.
Once again, whether this factorization persists in the thermodynamic limit or not remains to be investigated.

%
%


As for the position-dependence of BOMI,
for all $P_c$ considered, 
$I_{A,B}$ monotonically decreases with $t$, and saturates to a value that is practically constant 
as in Fig.\ \ref{theta_and_position_dependence}[b].
When $P_c$ gets further away from $b_0$ and closer to $b_1$, the decay rate of $I_{A,B}$ in the early time-regime ($0\le t < t_m \approx 1.8$) becomes larger. 
We can see from the numerical plot of $I_{A,B}$ in
Fig.\ \ref{theta_and_position_dependence}[b]
that in the early-time region,
the M\"obius/SS deformation may make scrambling in the subregion near $b_0$
slower, while it may make scrambling in the subspace near $b_1$ faster.





Let us now make a comparison with BOMI
in the two-dimensional holographic conformal field theories ($2$d holographic CFT) with the gravity dual. 
The $2$d holographic CFT is known by its 
strong scrambling dynamics \cite{2014JHEP...03..067S,2016JHEP...08..106M}.
In \cite{2023arXiv230208009G}, some of the authors in this paper have studied the time dependence of the BOMI under the evolution by $2$d M\"obius/SS deformed holographic CFT. 
In quantum field theory,
it is natural to regularize 
the dual state \eqref{TFD_eigen}
by introducing, for example,
an exponential suppression $e^{-\beta H_0}$ for high energy states,
and consider
\be \label{TFD_in_2dCFT}
\ket{\Psi(t)}=\f{1}{\Tr e^{-2\beta H}} \sum_{i}
e^{-i H^{(1)}_{\text{M\"obius/SSD}}t}
e^{-\beta H^{(1)}_0}\ket{E_i}_1 \ket{E_i^*}_2
\ee
where $H_{\text{M\"obius/SSD}}^{(1)}$ and $H_0^{(1)}$
act on $\mathcal{H}_1$.
The symbol $\beta$ is a regularization parameter.
Let $x$ be the coordinate on the spatial circle,
$0\le x \le L$.
In the SSD limit, the envelope function vanishes at $x=0$, while the function at $x=\f{L}{2}$ is two.
Define $A$ and $B$ as the subsystems of $\mathcal{H}_{2}$ and $\mathcal{H}_1$, respectively. 
For simplicity, let us take the size and center of subsystems to be $l_A=l_B$ and $A_c=B_c=P_c=0$.

One of the main findings in Ref.\ \cite{2023arXiv230208009G}
is that, for a given choice of subsystems $A$ and $B$,
there is a threshold value of $\theta$ that separates
the following two types of behaviors of $I_{A,B}$.
For smaller $\theta$,
$I_{A,B}$ decreases monotonically with time,
and then it approximately vanishes.
The larger $\theta$ is,
the slower the early-time decay of $I_{A,B}$ is.
For large $\theta$ and large times,
$I_{A,B}$ grows with time and oscillates periodically in time
with the period $T=L\cosh{2\theta}$.
In the SSD limit, $I_{A,B}$ returns to the initial value of $I_{A,B}$ for the large times.
The recovery of $I_{A,B}$ in the large time regime is described by the quasi-particle picture where under the SSD time evolution, the quasi-particles of $\mathcal{H}_1$ move to $x=0$, and then accumulate around $x=0$.
For the large times, $S_B$ is much larger than $S_{A\cup B}$, and $S_B$ is approximated by thermal entropy.
Consequently, $I_{A,B}$ returns to the initial value.
Ref.\ \cite{2023arXiv230208009G} also develops the line tension picture for operator entanglement of inhomogeneously-deformed time evolution operators.

While the early time behaviors of $I_{A,B}$
are similar,
for large $t$, the time-dependence of $I_{A,B}$ in $2$d holographic CFT is different from that in the M\"obius/SSD deformed spin chain.
One of the reasons that may cause this difference of $I_{A,B}$ is that the state considered is a thermofield double state with $\beta=0$.
For $\beta=0$, the reduced density matrix for $B$ is the identity operator on $B$,
and $S_B$ is the OEE for a maximally-entangled state.
Thus, 
$S_B$ does not increase any more, and it is constant unlike that in $2$d holographic CFT.
As for $I_{A,B}$,
the contribution to the early-time decay of $I_{A,B}$ in the CFT case 
is different from that in the spin system considered.
In the $2$d M\"obius/SSD holographic CFT, the early-time dependence of $I_{A,B}$ is determined by the positive contribution from 
$S_B$ and the negative contribution from 
$-S_{A\cup B}$.
In the M\"obius/SSD spin chain, only $-S_{A\cup B}$ depends on time and negatively contributes to the early time decay of $I_{A,B}$.
This difference in the time dependence of $S_B$ between $2$d holographic CFT and the spin system induces that of the late-time behavior of $I_{A,B}$.
For example,
in the $2$d holographic CFT
in the SSD limit, 
for large $t$, the positive contribution from the time dependence of $S_A +S_B$ overcomes the negative one from that of  $-S_{A\cup B}$.
As a consequence, the late-time $I_{A,B}$ grows with time, and then saturates to the initial value.
In contrast, in the SSD spin system, the positive contribution from $S_A +S_B$ is maximal and independent of time. 
The negative contribution from the time dependence of $-S_{A\cup B}$ 
grows in time and then approximately saturates to the maximal value, $S_A+S_B$, for the large times.
Consequently, $I_{A,B}$ is approximately zero at late times.

We can similarly compare our results with the 
so-called line-tension picture (membrane picture),
which is an effective description 
of the entanglement production in the $2$d holographic CFT with $\beta\rightarrow 0$
\cite{2018arXiv180300089J,PhysRevX.8.031058,PhysRevX.8.021013,Mezei:2018jco}.
In line-tension picture, the entanglement entropy $S_{A\cup B}$ of the unitary operator  is computed by the curve ${\cal C}$ connecting the edges of $A$ and $B$, which minimizes
$
S_{A\cup B}=\min _{\mathcal{C}} \int_{\mathcal{C}} d t \mathcal{T}(v), 
$
where 
$\mathcal{T}(v)$ is line-tension associated to two disconnected pieces of $\mathcal{C}$ with velocity $v = {dx}/{dt}$. $\mathcal{T}(v)$ in the $2$d holographic CFT is given by $\mathcal{T}(v)= \log q $ for $v<1$ and $v\log q$ for $v>1$ with $q \sim e^{\frac{c \pi}{3\beta}}$. While the line-tension picture was originally designed for 
infinite homogeneous space, it can be generalized to a compact space \cite{Goto:2021gve}, with a projection measurement \cite{Goto:2022fec}, and further to an inhomogeneous space generated by 
the SS deformation \cite{Goto:2023wai}. 
The SSD Hamiltonian deforms the spacetime in the line-tension picture inhomogeneously, and entanglement around the fixed point gets denser. In the case of $l_A=l_B$, $A_c=B_c=P_c=0$, the time-evolved subregion $A$ includes the fixed point with a large amount of entanglement and the curve that gives the trivial mutual information is no longer minimal. At late times, it is efficient to take the disconnected configuration that is not homologous to each subregion $A$ and $B$. Namely, to avoid the vicinity of the fixed point with a large amount of entanglement, the minimal curve of $S_{A\cup B}$ consists of two non-contractible pieces that wrap around a cylinder, and reduces a large amount of entanglement, which $S_A$ and $S_B$ count.
This reproduces the behavior of the late-time
$I_{A,B}$ in the holographic CFTs, which grows with time, and then saturates at the initial value.
However, this is different from the time-dependence of $I_{A,B}$ in the SSD spin chain as we explained,
presumably by the similar reason 
why the results for the chaotic spin chain and the holographic CFT disagree.

Let us now consider the state with the finite $\beta$.
Now, we discuss which parameter region we should consider exploring the difference of the behavior, irrelevant to the Hilbert space size, of $I_{A,B}$ in $2$d CFT and non-CFT.
In the spin system considered, the dimension of the local Hilbert space on a single site is two. 
The upper bound of entanglement entropy for the subsystem $B$ of the size $l_B$ is
$
S_B \le l_B\log{2}.
$
In the parameter region where $L \gg l_B \gg \beta \gg 1$, 
we assume that the entanglement entropy for $\mathcal{H}_1$ may be approximated by
$
S_{\mathcal{H}_1} \approx C_{\text{Th}}\times \beta \times L, 
$
where $C_{\text{Th}}$ is supposed to be independent of $\beta$ and $L$.
In the SSD limit of $2$d CFT, for all quasi-particles to accumulate near $b_0$, $S_{\mathcal{H}_1}$ 
should be smaller than $l_B\log{2}$:
$
S_{\mathcal{H}_1} \le l_B\log{2}.
$
In the parameter region where $ C_{\text{Th}}\times \beta \times L \ll l_B\log{2}$,
we can elucidate the difference of the dynamics between $2$d inhomogeneous CFT and the inhomogeneous non-CFT by investigating the time dependence of $I_{A,B}$.
In other words, we can study if the revival of $I_{A,B}$ is induced by the dynamics of non-CFT with M\"obius/SS deformation.

The $\theta$ and position dependence of BOMI of free fermion presented in Appendix \ref{Section:BOMI_FF} also shows similar results as the chaotic spin chain except for the late-time behavior. During the M\"obius evolution, the late-time behavior of BOMI exhibits periodic oscillation. The period of this oscillation might be due to the integrability of free fermion and might depend on $\theta$. During the SSD evolution, the late-time value of BOMI is non-zero and it is larger than that in the chaotic region.

\section{Subsystem spectral form factor}\label{Section:SFF}

We now turn to the subsystem spectral form factor. 
Suppose that we start from the systems
a product state $\ket{\psi^{p}}$
and evolve them with the Hamiltonian $H$:
$
\ket{\Psi^{p}(t)}=e^{-it H} \ket{\psi^{p}}.
$
The subsystem spectral form factor (SSF) for subsystem $A$ is defined as 
\cite{Chen_2018}
\begin{equation}
	g(\tau) =
  \Big\langle \sum_{i,j} e^{-\tau(\lambda_i - \lambda_j)}
    \Big\rangle,
\end{equation}
where
$\lambda_i $ is the eigenvalue of the reduced density matrix $\rho^p_A =
\Tr_{\overline{A}} \ket{\Psi^{p}(t)} \bra{\Psi^{p}(t)}$ and
$\left\langle \cdot \right\rangle$ denotes the average over the density
matrices of the different initial states, $\rho^p =  \ket{\Psi^{p}(t)} \bra{\Psi^{p}(t)}$.

\begin{figure}[t]
\centering
\includegraphics[width=0.48\columnwidth]{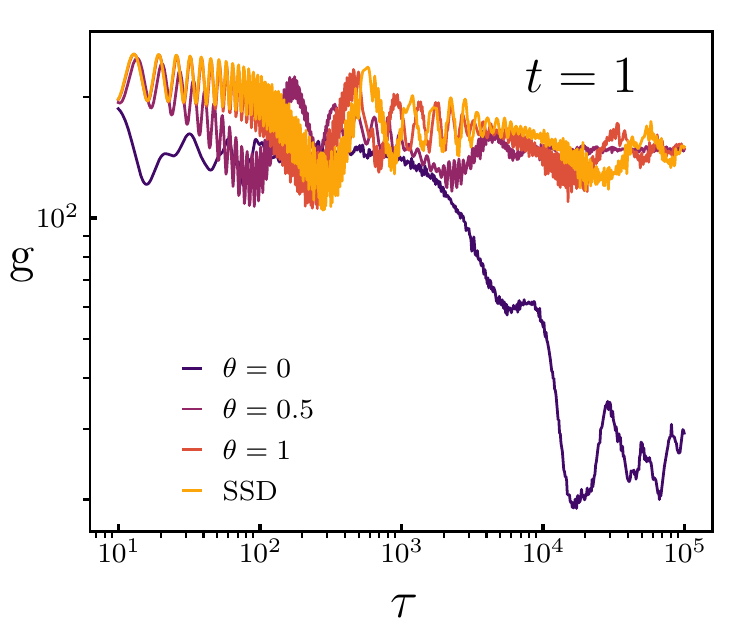}
\includegraphics[width=0.48\columnwidth]{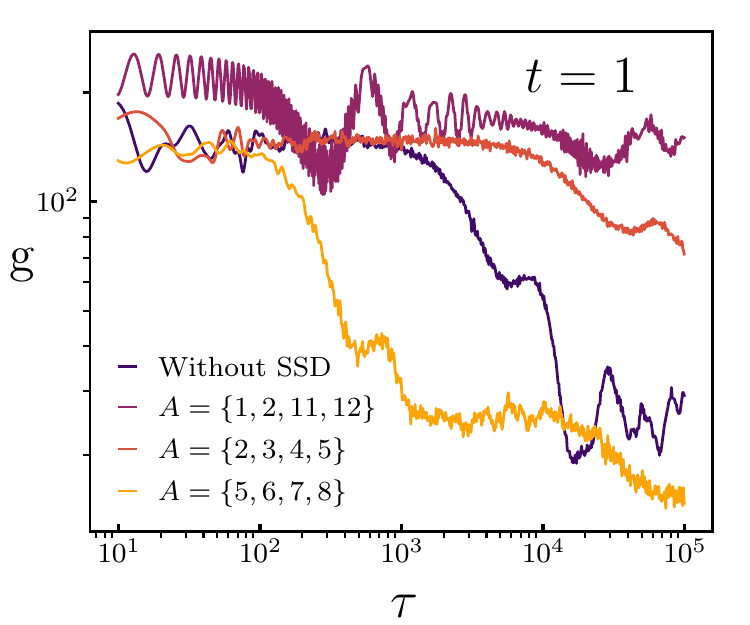}

\includegraphics[width=0.48\columnwidth]{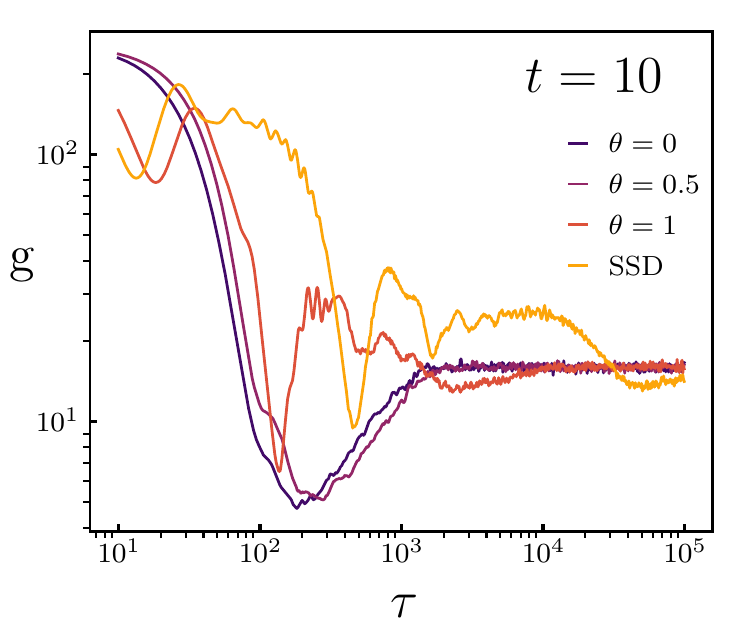}
\includegraphics[width=0.48\columnwidth]{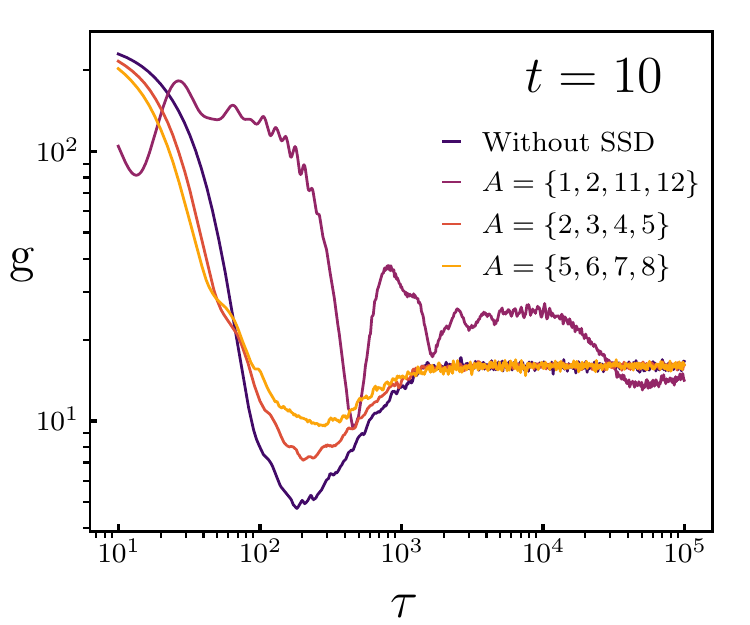}

\includegraphics[width=0.48\columnwidth]{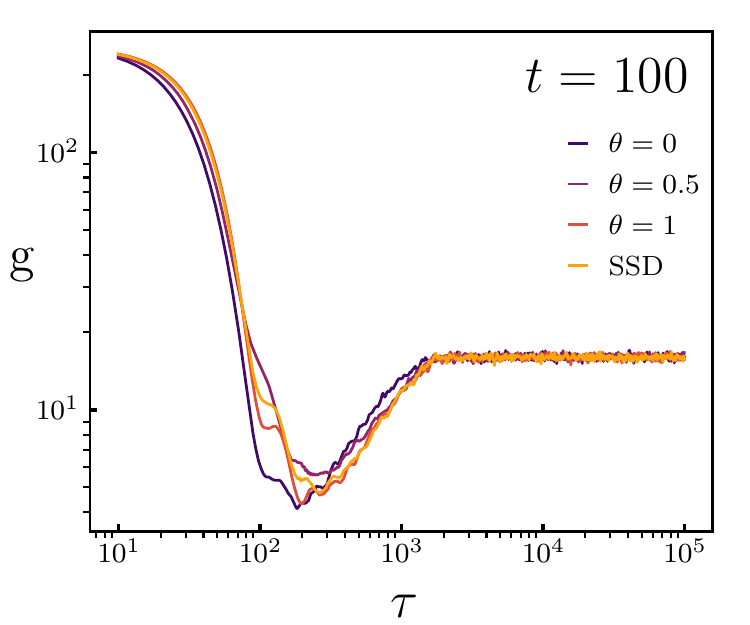}
\includegraphics[width=0.48\columnwidth]{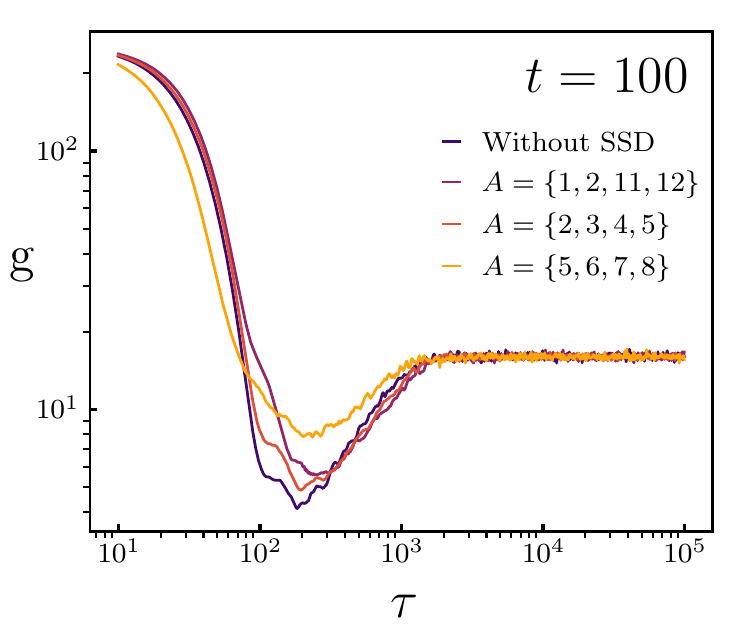}
\caption{ 
(Left) SSF of the M\"obius deformed chaotic spin chain of length $L = 12$  at different time evolution point different $\theta$. The subsystem site choice is $A = \{1,2,11,12\}$.
(Right)
SSF of the SSD deformed chaotic spin chain of length $L = 12$  at different time evolution points with different subsystem sites labeled in the plots.
}
\label{Fig:SFF_spin_the_dep}
\end{figure}

The chaotic behavior which may be represented by the dip-ramp-plateau feature of the SSF is observed in Fig.\ \ref{Fig:SFF_spin_the_dep} at late times. For a chaotic system, if the SSF does not show such a feature, then the subsystem might be in the pre-thermal regime or developing chaos. Therefore, we could use the time when the SSF exhibits the chaotic feature to infer how fast the information is scrambled. 

Let us focus on the numerical analysis of the SSF.
We plot the SSF under the evolution induced by $H_{\text{M\"obius}}$ and $H_{\text{SSD}}$ as a function of $\tau$ in Figs.\ \ref{Fig:SFF_spin_the_dep}.
First, we present the $\theta$-dependence of the SSF associated with the subsystem consisting of $1$, $2$, $11$, and $12$-th sites.
A subsystem with larger $\theta$ shows the dip-ramp-plateau feature earlier and thermalizes faster. 
The results indicate that the more the envelope function suppresses the Hamiltonian, the slower information scrambles and thermalizes. 
Such a feature is also observed in BOMI 
(Fig.\ \ref{theta_and_position_dependence}). 
As for the subsystem position dependence, 
the right panel of Fig.\ \ref{Fig:SFF_spin_the_dep} shows that for the large $t$, 
each subsystem thermalizes. 
Particularly, the subsystem consisting of $5$ to $8$-th sites thermalizes fastest as the dip-ramp-plateau feature is observed while the other subsystem is developing chaos or at a pre-thermal regime. 
It also thermalizes faster than the undeformed system. 
The subsystem consisting of $1$, $2$, $11$, and $12$-th sites thermalizes slowest as the  dip-ramp-plateau feature takes a longer time to develop. 
Such results confirm the BOMI results that the envelope function boosts the thermalization near $b_1$ and suppresses the thermalization near $b_0$. 
The scrambling behavior near $b_0$ observed with BOMI and SSF suggests that the black-hole like excitations shown in Section \ref{Section:ED-BHlike-exitaion} emerge as the information accumulates near $b_0$ due to slow scrambling speed.

\section{The effect of inhomogeneous deformations on the growth of operators \label{Section:speed}}
In this section,
we discuss two quantities
that can detect operator growth 
during the evolution induced by the M\"obius and SS deformed Hamiltonians, 
the time dependence of the return amplitude and OTOCs.
\subsection{Return amplitude \label{sec:returnamp}}

Let us define the return amplitude as the inner product of
$\sigma^{(1)}_{\alpha,a}(t) \ket{\text{TFD}}$ and $\sigma^{(1)}_{\alpha,a}(t=0) \ket{\text{TFD}}$,
\be
A_{\alpha,a}
= \left \langle \sigma_{\alpha,a}(t=0)\sigma_{\alpha,a}(t) \right \rangle_{\text{thermal}}.
\ee
Here, $\ket{\text{TFD}}$ is defined by (\ref{TFD_in_2dCFT}) for $t=0$.
Due to the growth of the Heisenberg operators,
the system may evolve from the initial state to an orthogonal state to it,
so that $A_{\alpha,a}$ decreases and approximately vanishes at $t\approx t_{\text{cp}}$.
As in \cite{Margolus:1997ih}, $t_{\text{cp}}$ can be thought of as
the time by which a single step of quantum computation is completed
\footnote{In \cite{Margolus:1997ih}, the initial and time-evolved states are
  defined as $\ket{\Psi}$ and $\ket{\Psi(t)}=U(t)\ket{\Psi}$, respectively. The
  return amplitude is defined as the inner product of them, $\left \langle \Psi|
    \Psi(t)\right \rangle$. This definition is different from ours. If we define
  the initial and time-evolved states as
  $\ket{\Psi}=\sigma^{(1)}_{\alpha,a}\ket{\text{TFD}}$ and
  $\ket{\Psi(t)}=U(t)\sigma^{(1)}_{\alpha,a}\ket{\text{TFD}}$, the
  time-dependence of the inner product is governed by a global quench, not the
  growth of the Heisenberg operators. Therefore, we introduce a generalized
  return amplitude that may capture the growth of the Heisenberg operators.}.
The computational time $t_{\text{cp}}$ can diagnose 
how fast the Heisenberg operator grows with time under evolution.
Under the evolution by the M\"obius and SS deformed Hamiltonians,
the computational speed may depend on the location of the Pauli operators and $\theta$. 

\begin{figure*}[t]
  \includegraphics[width = 0.3\textwidth]{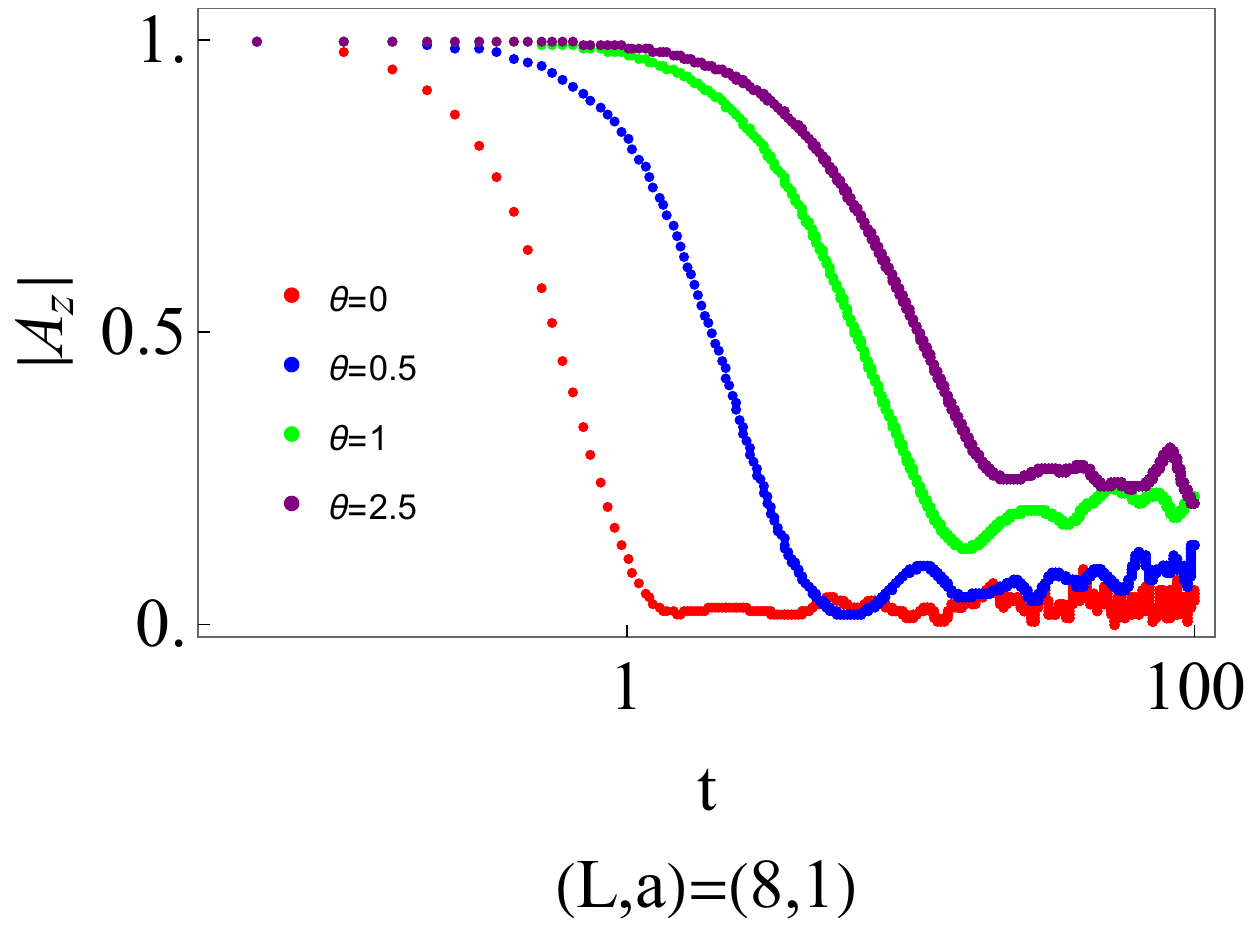}
  \includegraphics[width = 0.3\textwidth]{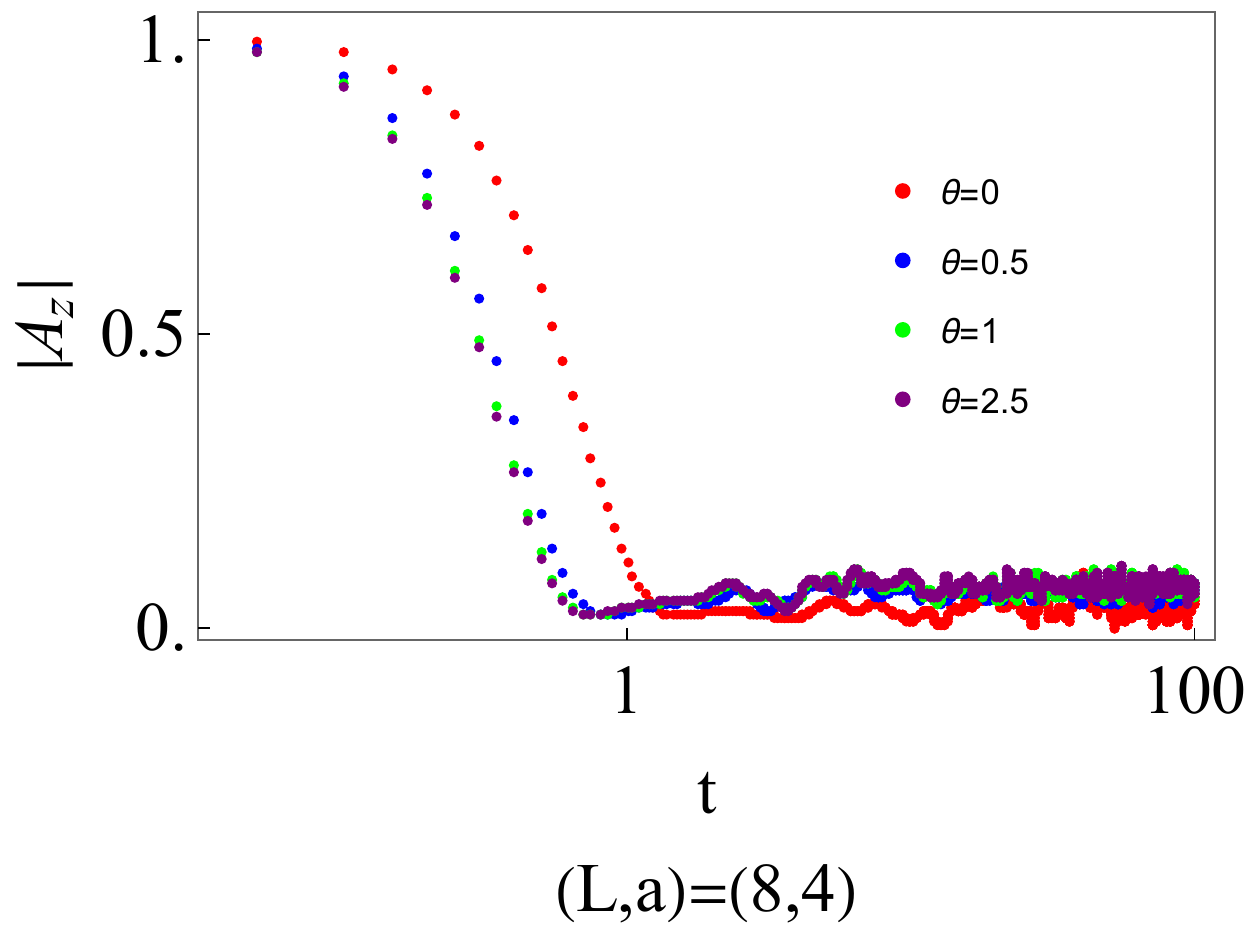}    
  \includegraphics[width = 0.3\textwidth]{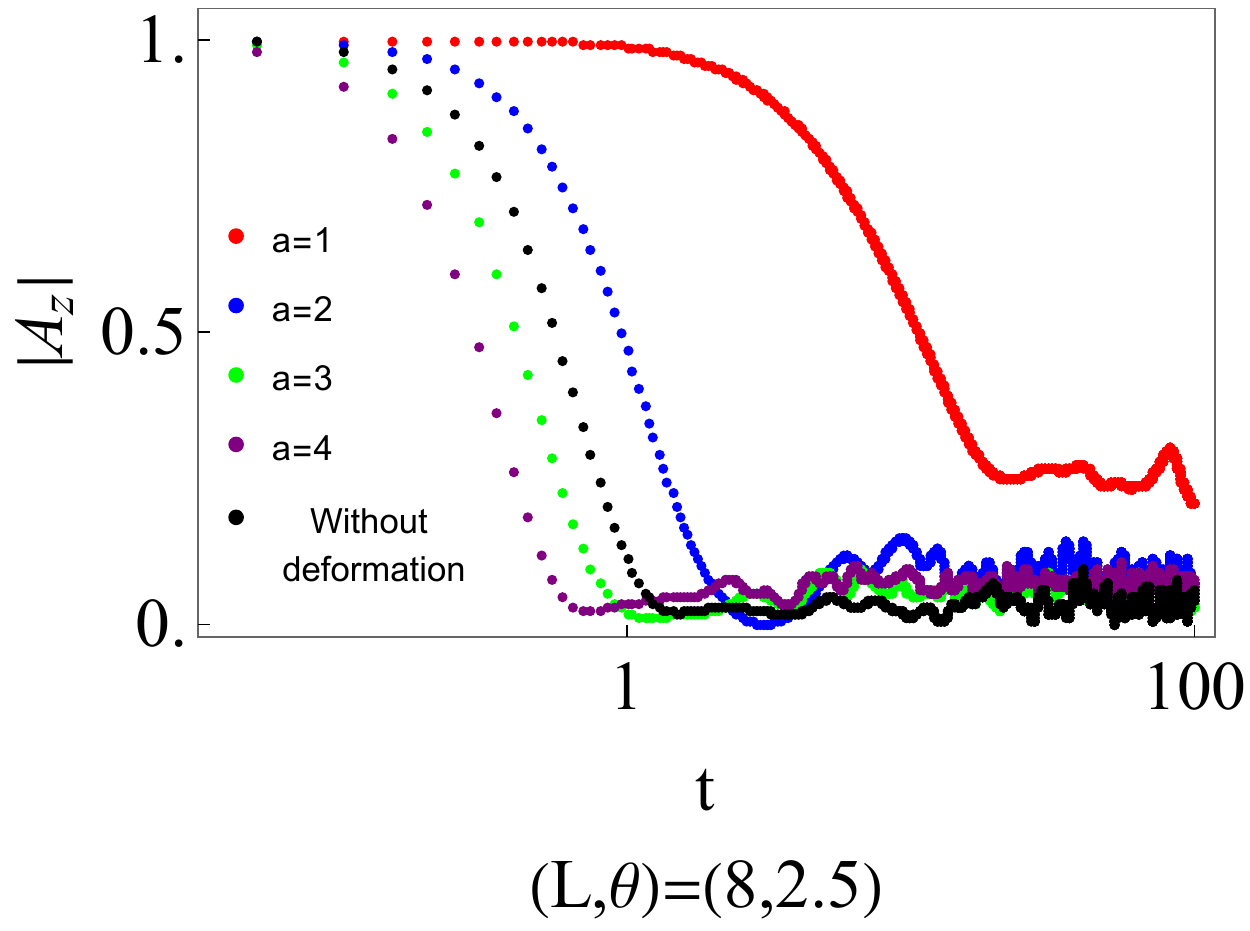}
  \caption{
    The $\theta$- and position-dependence of $|A_{z,a}|$ as function of $t$.
    In [a](top left) and [b](top right) we show how the time-dependence
    of $\left|A_{z,a=1}\right|$ and $\left|A_{z,a=4}\right|$ depends on $\theta$.
    Here, $\theta=0,0.5,1,2.5$. In [c](bottom),
    we show the time-dependence of $\left|A_{\alpha}\right|$ depends
    on the location of the Pauli operator. Here, $a=1,2,3,4$.}
  \label{theta_dependence_of_computational_speed}
\end{figure*}

    
    
    
    
        
    
    

In Fig.\ \ref{theta_dependence_of_computational_speed},
we plot $\left|A_{z,a}\right|$ as a function of $t$. 
The time-dependence of $\left|A_{\alpha=x,y,a}\right|$ is reported in Appendix \ref{App:ReturnAmplitude}.
The return amplitude, $\left|A_{\alpha=z,a=1,4}\right|$, monotonically decays with $t$,
and fluctuates around a finite value that depends on $a$ and $\theta$.
For $a=1$, this late-time value becomes larger when $\theta$ becomes larger.
If $\theta$ becomes larger, the time for the decay of $\left|A_{z,a=1}\right|$ to end becomes larger,
while the time for the decay of $\left|A_{z,a=4}\right|$ to end becomes smaller.
For $\theta=2.5$, the time for the decay of $\left|A_{z,a}\right|$ to end becomes larger
when $a$ gets closer to $a=1$.
As expected, the Heisenberg operator near $b_0$ grows slower
under the evolution induced by the inhomogeneously-deformed Hamiltonian
than under the evolution by the un-deformed one,
while the operator near $b_1$ grows more rapidly under the evolution
by the inhomogeneously-deformed one than by the un-deformed one.
Furthermore, the late-time behavior of $\left|A_{z,a=1}\right|$ for $\theta$
is interpreted as the prevention of operator growth by inhomogeneous deformation.

\subsection{OTOCs}

\begin{figure*}[tbh!]
  \centering 
  \includegraphics[width = 0.32\textwidth]{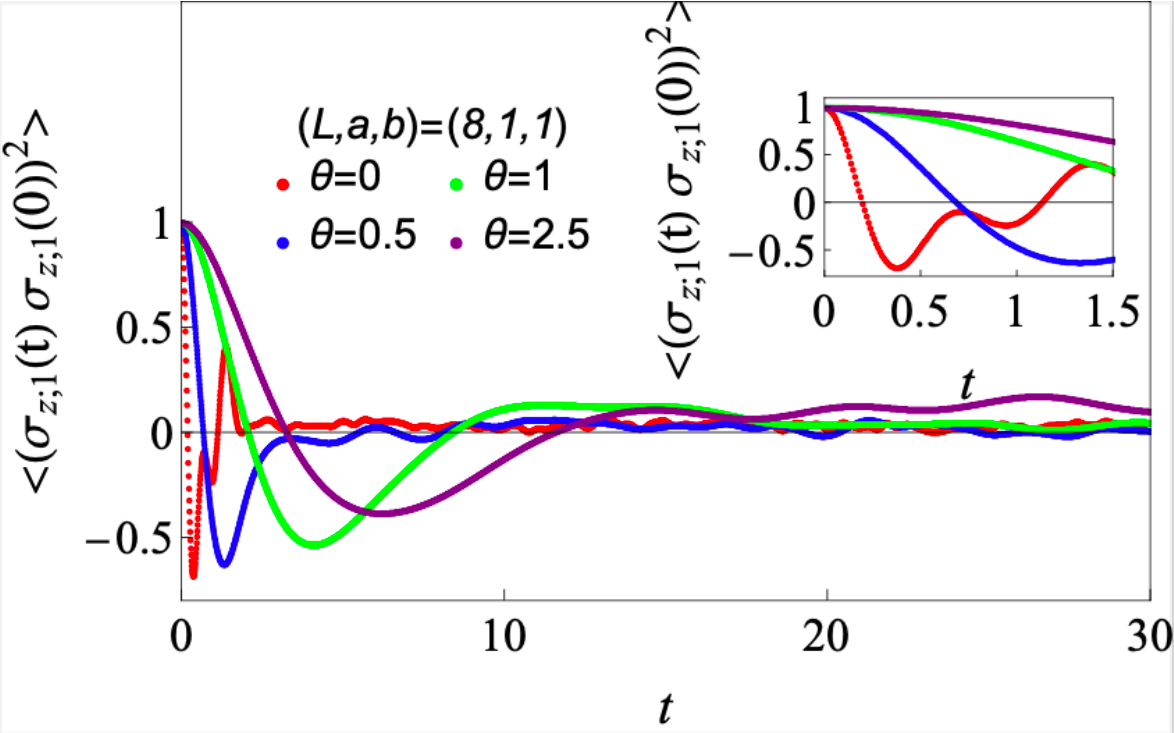}
  \includegraphics[width = 0.32\textwidth]{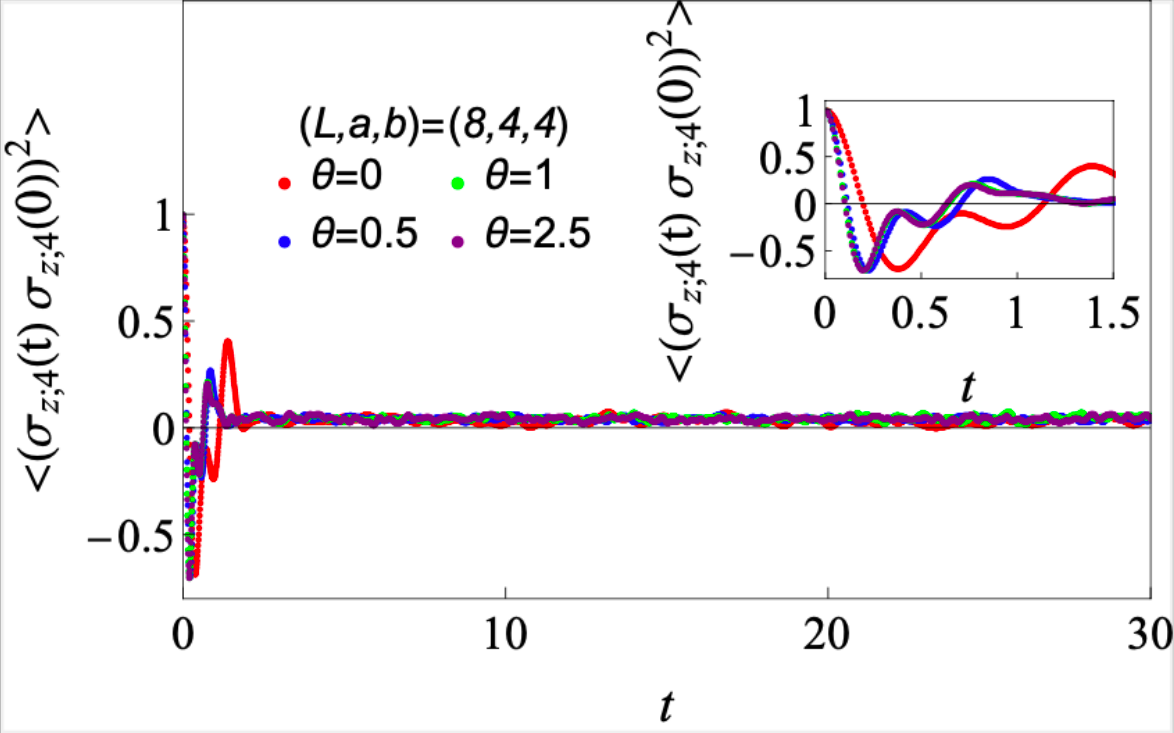}
  \includegraphics[width = 0.32\textwidth]{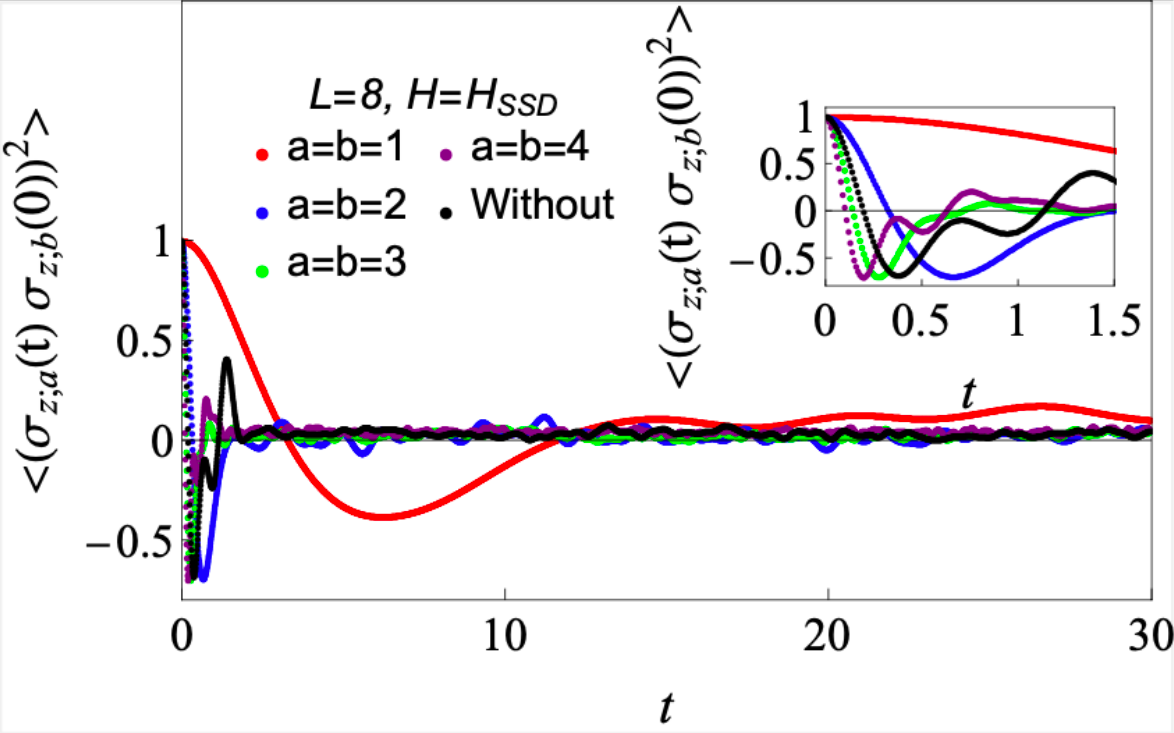}
  \caption{
    The $\theta$- and position-dependence of the OTOCs
    $\left\langle (\sigma_{z,a}(t)\sigma_{z,a}(0))^2\right\rangle
    =2^{-L}\text{tr}((\sigma_{z,a}(t)\sigma_{z,a}(0))^2)$
    during the evolution by the M\"obius/SS deformed Hamiltonian.
    In the top and middle panels, we show how
    $\left\langle (\sigma_{z,1}(t)\sigma_{z,1}(0))^2\right\rangle$ and
    $\left\langle (\sigma_{z,4}(t)\sigma_{z,4}(0))^2\right\rangle$ depend on $\theta$. 
    The bottom panel shows how $\left\langle
      (\sigma_{\mu,a}(t)\sigma_{\mu,a}(0))^2\right\rangle$
    depends on $a$ under the SSD time evolution.
    In the insets, we concentrate on the time evolution of the OTOC in $0\le t
    \le 1.5$.
  }
  \label{theta_and_position_dependence_OTOC}
\end{figure*}

Finally, let us turn to another measure of scrambling, OTOCs,
\begin{align}
  F(t)=\frac{\text{tr}(\sigma_{\alpha,a}(t)\sigma_{\alpha,b}\sigma_{\alpha,a}(t)\sigma_{\alpha,b})}{\text{tr} e^{-2\beta H_{0}}}
\label{OTOC_munusame}.
\end{align}
For simplicity, we set $b=a$ and $\beta=0$. 
In Fig.\ \ref{theta_and_position_dependence_OTOC},
we plot $\left\langle (\sigma_{z,a}(t)\sigma_{z,a}(0))^2\right\rangle$
under the M\"obius and SSD time evolution as a function of $t$.   
For all $\theta$ and $a$ considered, $\left\langle (\sigma_{z,1}(t)\sigma_{z,1}(0))^2\right\rangle$
decays with $t$, 
approaching a certain negative value that depends on $a$ and $\theta$.
During the subsequent evolution, it increases with $t$ and then fluctuates around zero.
For $a=1$, when $\theta$ becomes larger,
$\left\langle (\sigma_{z,1}(t)\sigma_{z,1}(0))^2\right\rangle$
decays with $t$ slower. 
For $a=4$, when $\theta$ becomes larger,
$\left\langle (\sigma_{z,4}(t)\sigma_{z,4}(0))^2\right\rangle$ decays with $t$ faster.
Under the M\"obius and SSD time evolution, when $a$ gets further away from $a=1$ and closer to $a=4$,
$\left\langle (\sigma_{z,4}(t)\sigma_{z,4}(0))^2\right\rangle$ for the early times 
 decays with $t$ faster. 
The interpretation for the early-time behavior in $t$ of $|A_{\alpha,a}|$ works for that of OTOCs.

\section{Discussions and Future directions \label{Section:Discussion_and_FD}}

Partly motivated by recent studies in (1+1)d CFT, we studied the inhomogeneous deformations of 
the chaotic quantum Ising spin chain.
As in the case of CFT, we observed the creation of black hole-like excitations
as signaled by the local accumulation of energy density. 
Away from the SSD limit, we however do not observe the coherent oscillation
with period $L \cosh 2\theta$, which is expected 
since the oscillation is tied to the integral spectrum of CFT.
This may indicate that to create black-hole like excitations
in chaotic spin chains (without conformal symmetry),
we do not need to tune $\theta$ at the SSD point.
Similarly, we can selectively cool/heat different 
parts of chaotic spin chains without tuning $\theta$ to the SSD point. 
We also discussed various indicators of quantum information 
scrambling. BOMI does not fully agree with the line tension picture and holography.
It is a future problem to make a more detailed comparison
by using large systems or quantum simulators.

Finally, we comment on a few future directions.

--{\it  Computational speed and butterfly velocity in the inhomogeneous time
  evolution}:
In this paper, we studied the qualitative behaviors of the return amplitudes and
OTOCs. Their behavior suggests that the butterfly velocity may depend on the
inhomogeneous parameter and the location of the operators.
It would be interesting to study the computational speed and butterfly velocity
by generalizing the velocity in the spin chain defined in
\cite{2020NatPh..16..199X}
to the one depending on the location.
In \cite{2018PhRvB..98n4304K,2019ScPP....7...45H},
the Lyapunov exponent that depends on the velocity is discussed.
It would be also interesting to study the Lyapunov exponent that depends on the velocity under the evolution by the M\"obius/SS deformed Hamiltonian.

-- {\it Recovery in spin system}:
As in $2$d holographic CFT, it would be interesting to construct the spin system
where the value of $I_{A,B}$ returns back to the initial one. 
This is expected to lead to the proposal of experimental systems in which
nonlocal correlations can be controlled by spatially modulating the Hamiltonian density.
 Furthermore, it would be also interesting to provide the spin and experimentally-realizable systems that have genuine tripartite entanglement introduced by the authors in \cite{2023arXiv230208009G}. Let us divide the total system into $A_{i=1,2,3}$ and the complement to them. If $I_{A_1,A_{i=2,3}}\approx 0$, $I_{A_2,A_3}\approx 0$, but the value of $I_{A_i,A_{j\neq i}\cup A_{k\neq i,j}}$ is large, then the system has the genuine tripartite entanglement. 

    
\section*{Acknowledgements}

We thank useful discussions with Masaki Tezuka and Huajia Wang.
K.G.~is supported by JSPS KAKENHI Grant-in-Aid for Early-Career Scientists (21K13930) and Research Fellowships of Japan Society for the Promotion of Science for Young Scientists (22J00663).
M.N.~is supported by funds from the University of Chinese Academy of Sciences (UCAS), funds from the Kavli
Institute for Theoretical Sciences (KITS).
K.T.~is supported by JSPS KAKENHI Grant No.~21K13920 and MEXT KAKENHI Grant No.~22H05265.
S.R.~is supported by the National Science Foundation under 
Award No.\ DMR-2001181, and by a Simons Investigator Grant from
the Simons Foundation (Award No.~566116).
This work is supported by
the Gordon and Betty Moore Foundation through Grant
GBMF8685 toward the Princeton theory program. 
Part of the results was computed by using the high-performance computing facilities provided by Yukawa Institute for Theoretical Physics (Sushiki server). Part of the calculations presented in this article was performed on computational resources managed and supported by Princeton Research Computing, a consortium of groups including the Princeton Institute for Computational Science and Engineering (PICSciE) and the Office of Information Technology's High-Performance Computing Center and Visualization Laboratory at Princeton University.

\bibliographystyle{ieeetr}
\bibliography{reference}

\begin{thebibliography}{100}

\bibitem{2019Natur.567...61L}
K.~A. {Landsman}, C.~{Figgatt}, T.~{Schuster}, N.~M. {Linke}, B.~{Yoshida},
  N.~Y. {Yao}, and C.~{Monroe}, ``{Verified quantum information scrambling},''
  {\em \nat}, vol.~567, pp.~61--65, Mar. 2019.

\bibitem{2020PhRvL.124x0505J}
M.~K. {Joshi}, A.~{Elben}, B.~{Vermersch}, T.~{Brydges}, C.~{Maier},
  P.~{Zoller}, R.~{Blatt}, and C.~F. {Roos}, ``{Quantum Information Scrambling
  in a Trapped-Ion Quantum Simulator with Tunable Range Interactions},'' {\em
  \prl}, vol.~124, p.~240505, June 2020.

\bibitem{2021PhRvX..11b1010B}
M.~S. {Blok}, V.~V. {Ramasesh}, T.~{Schuster}, K.~{O'Brien}, J.~M.
  {Kreikebaum}, D.~{Dahlen}, A.~{Morvan}, B.~{Yoshida}, N.~Y. {Yao}, and
  I.~{Siddiqi}, ``{Quantum Information Scrambling on a Superconducting Qutrit
  Processor},'' {\em Physical Review X}, vol.~11, p.~021010, Apr. 2021.

\bibitem{2016PhRvA..94d0302S}
B.~{Swingle}, G.~{Bentsen}, M.~{Schleier-Smith}, and P.~{Hayden}, ``{Measuring
  the scrambling of quantum information},'' {\em \pra}, vol.~94, p.~040302,
  Oct. 2016.

\bibitem{PhysRevA.94.062329}
G.~Zhu, M.~Hafezi, and T.~Grover, ``Measurement of many-body chaos using a
  quantum clock,'' {\em Phys. Rev. A}, vol.~94, p.~062329, Dec 2016.

\bibitem{2016arXiv160701801Y}
N.~Y. {Yao}, F.~{Grusdt}, B.~{Swingle}, M.~D. {Lukin}, D.~M. {Stamper-Kurn},
  J.~E. {Moore}, and E.~A. {Demler}, ``{Interferometric Approach to Probing
  Fast Scrambling},'' {\em arXiv e-prints}, p.~arXiv:1607.01801, July 2016.

\bibitem{2017PhRvA..95a2120Y}
N.~{Yunger Halpern}, ``{Jarzynski-like equality for the out-of-time-ordered
  correlator},'' {\em \pra}, vol.~95, p.~012120, Jan. 2017.

\bibitem{2018PhRvA..97d2105Y}
N.~{Yunger Halpern}, B.~{Swingle}, and J.~{Dressel}, ``{Quasiprobability behind
  the out-of-time-ordered correlator},'' {\em \pra}, vol.~97, p.~042105, Apr.
  2018.

\bibitem{2017PhRvE..95f2127C}
M.~{Campisi} and J.~{Goold}, ``{Thermodynamics of quantum information
  scrambling},'' {\em \pre}, vol.~95, p.~062127, June 2017.

\bibitem{2017arXiv171003363Y}
B.~{Yoshida} and A.~{Kitaev}, ``{Efficient decoding for the Hayden-Preskill
  protocol},'' {\em arXiv e-prints}, p.~arXiv:1710.03363, Oct. 2017.

\bibitem{2017NatPh..13..781G}
M.~{G{\"a}rttner}, J.~G. {Bohnet}, A.~{Safavi-Naini}, M.~L. {Wall}, J.~J.
  {Bollinger}, and A.~M. {Rey}, ``{Measuring out-of-time-order correlations and
  multiple quantum spectra in a trapped-ion quantum magnet},'' {\em Nature
  Physics}, vol.~13, pp.~781--786, Aug. 2017.

\bibitem{2016arXiv161205249W}
K.~X. {Wei}, C.~{Ramanathan}, and P.~{Cappellaro}, ``{Exploring Localization in
  Nuclear Spin Chains},'' {\em arXiv e-prints}, p.~arXiv:1612.05249, Dec. 2016.

\bibitem{PhysRevX.7.031011}
J.~Li, R.~Fan, H.~Wang, B.~Ye, B.~Zeng, H.~Zhai, X.~Peng, and J.~Du,
  ``Measuring out-of-time-order correlators on a nuclear magnetic resonance
  quantum simulator,'' {\em Phys. Rev. X}, vol.~7, p.~031011, Jul 2017.

\bibitem{2017arXiv170506714M}
E.~J. {Meier}, J.~{Ang'ong'a}, F.~A. {An}, and B.~{Gadway}, ``{Exploring
  quantum signatures of chaos on a Floquet synthetic lattice},'' {\em arXiv
  e-prints}, p.~arXiv:1705.06714, May 2017.

\bibitem{PhysRevA.43.2046}
J.~M. Deutsch, ``Quantum statistical mechanics in a closed system,'' {\em Phys.
  Rev. A}, vol.~43, pp.~2046--2049, Feb 1991.

\bibitem{PhysRevE.50.888}
M.~Srednicki, ``Chaos and quantum thermalization,'' {\em Phys. Rev. E},
  vol.~50, pp.~888--901, Aug 1994.

\bibitem{2008Natur.452..854R}
M.~{Rigol}, V.~{Dunjko}, and M.~{Olshanii}, ``{Thermalization and its mechanism
  for generic isolated quantum systems},'' {\em \nat}, vol.~452, pp.~854--858,
  Apr. 2008.

\bibitem{2011RvMP...83..863P}
A.~{Polkovnikov}, K.~{Sengupta}, A.~{Silva}, and M.~{Vengalattore},
  ``{Colloquium: Nonequilibrium dynamics of closed interacting quantum
  systems},'' {\em Reviews of Modern Physics}, vol.~83, pp.~863--883, July
  2011.

\bibitem{2005JSMTE..04..010C}
P.~{Calabrese} and J.~{Cardy}, ``{Evolution of entanglement entropy in
  one-dimensional systems},'' {\em Journal of Statistical Mechanics: Theory and
  Experiment}, vol.~4, p.~04010, Apr. 2005.

\bibitem{2007JHEP...09..120H}
P.~{Hayden} and J.~{Preskill}, ``{Black holes as mirrors: quantum information
  in random subsystems},'' {\em Journal of High Energy Physics}, vol.~9,
  p.~120, Sept. 2007.

\bibitem{2020arXiv200700895N}
Y.~{Nakata}, E.~{Wakakuwa}, and M.~{Koashi}, ``{Black holes as clouded mirrors:
  the Hayden-Preskill protocol with symmetry},'' {\em arXiv e-prints},
  p.~arXiv:2007.00895, July 2020.

\bibitem{2014JHEP...03..067S}
S.~H. {Shenker} and D.~{Stanford}, ``{Black holes and the butterfly effect},''
  {\em Journal of High Energy Physics}, vol.~3, p.~67, Mar. 2014.

\bibitem{2016JHEP...08..106M}
J.~{Maldacena}, S.~H. {Shenker}, and D.~{Stanford}, ``{A bound on chaos},''
  {\em Journal of High Energy Physics}, vol.~8, p.~106, Aug. 2016.

\bibitem{PhysRevLett.115.131603}
D.~A. Roberts and D.~Stanford, ``Diagnosing chaos using four-point functions in
  two-dimensional conformal field theory,'' {\em Phys. Rev. Lett.}, vol.~115,
  p.~131603, Sep 2015.

\bibitem{2016JHEP...10..009S}
D.~{Stanford}, ``{Many-body chaos at weak coupling},'' {\em Journal of High
  Energy Physics}, vol.~2016, p.~9, Oct. 2016.

\bibitem{PhysRevD.96.065005}
D.~Chowdhury and B.~Swingle, ``Onset of many-body chaos in the $o(n)$ model,''
  {\em Phys. Rev. D}, vol.~96, p.~065005, Sep 2017.

\bibitem{PhysRevX.7.031047}
A.~A. Patel, D.~Chowdhury, S.~Sachdev, and B.~Swingle, ``Quantum butterfly
  effect in weakly interacting diffusive metals,'' {\em Phys. Rev. X}, vol.~7,
  p.~031047, Sep 2017.

\bibitem{2017PhRvX...7c1016N}
A.~{Nahum}, J.~{Ruhman}, S.~{Vijay}, and J.~{Haah}, ``{Quantum Entanglement
  Growth under Random Unitary Dynamics},'' {\em Physical Review X}, vol.~7,
  p.~031016, July 2017.

\bibitem{2018PhRvX...8b1014N}
A.~{Nahum}, S.~{Vijay}, and J.~{Haah}, ``{Operator Spreading in Random Unitary
  Circuits},'' {\em Physical Review X}, vol.~8, p.~021014, Apr. 2018.

\bibitem{2018PhRvX...8c1057K}
V.~{Khemani}, A.~{Vishwanath}, and D.~A. {Huse}, ``{Operator Spreading and the
  Emergence of Dissipative Hydrodynamics under Unitary Evolution with
  Conservation Laws},'' {\em Physical Review X}, vol.~8, p.~031057, July 2018.

\bibitem{2018PhRvX...8c1058R}
T.~{Rakovszky}, F.~{Pollmann}, and C.~W. {von Keyserlingk}, ``{Diffusive
  Hydrodynamics of Out-of-Time-Ordered Correlators with Charge Conservation},''
  {\em Physical Review X}, vol.~8, p.~031058, July 2018.

\bibitem{PhysRevB.96.020406}
D.~J. Luitz and Y.~Bar~Lev, ``Information propagation in isolated quantum
  systems,'' {\em Phys. Rev. B}, vol.~96, p.~020406, Jul 2017.

\bibitem{2017NJPh...19f3001B}
A.~{Bohrdt}, C.~B. {Mendl}, M.~{Endres}, and M.~{Knap}, ``{Scrambling and
  thermalization in a diffusive quantum many-body system},'' {\em New Journal
  of Physics}, vol.~19, p.~063001, June 2017.

\bibitem{2018PhRvL.121a6801H}
M.~{Heyl}, F.~{Pollmann}, and B.~{D{\'o}ra}, ``{Detecting Equilibrium and
  Dynamical Quantum Phase Transitions in Ising Chains via Out-of-Time-Ordered
  Correlators},'' {\em \prl}, vol.~121, p.~016801, July 2018.

\bibitem{PhysRevB.97.144304}
C.-J. Lin and O.~I. Motrunich, ``Out-of-time-ordered correlators in a quantum
  ising chain,'' {\em Phys. Rev. B}, vol.~97, p.~144304, Apr 2018.

\bibitem{2020NatPh..16..199X}
S.~{Xu} and B.~{Swingle}, ``{Accessing scrambling using matrix product
  operators},'' {\em Nature Physics}, vol.~16, pp.~199--204, Jan. 2020.

\bibitem{Khetrapal:2022dzy}
S.~Khetrapal, ``{Chaos and operator growth in 2d CFT},'' {\em JHEP}, vol.~03,
  p.~176, 2023.

\bibitem{PhysRevB.83.060414}
T.~Hikihara and T.~Nishino, ``Connecting distant ends of one-dimensional
  critical systems by a sine-square deformation,'' {\em Phys. Rev. B}, vol.~83,
  p.~060414, Feb 2011.

\bibitem{PhysRevB.84.165132}
I.~Maruyama, H.~Katsura, and T.~Hikihara, ``Sine-square deformation of free
  fermion systems in one and higher dimensions,'' {\em Phys. Rev. B}, vol.~84,
  p.~165132, Oct 2011.

\bibitem{2012JPhA...45k5003K}
H.~{Katsura}, ``{Sine-square deformation of solvable spin chains and conformal
  field theories},'' {\em Journal of Physics A Mathematical General}, vol.~45,
  p.~115003, Mar. 2012.

\bibitem{2009PThPh.122..953G}
A.~{Gendiar}, R.~{Krcmar}, and T.~{Nishino}, ``{Spherical Deformation for
  One-Dimensional Quantum Systems},'' {\em Progress of Theoretical Physics},
  vol.~122, pp.~953--967, Oct. 2009.

\bibitem{2011PhRvA..83e2118G}
A.~{Gendiar}, M.~{Dani{\v{s}}ka}, Y.~{Lee}, and T.~{Nishino}, ``{Suppression of
  finite-size effects in one-dimensional correlated systems},'' {\em \pra},
  vol.~83, p.~052118, May 2011.

\bibitem{2011PhRvB..84k5116S}
N.~{Shibata} and C.~{Hotta}, ``{Boundary effects in the density-matrix
  renormalization group calculation},'' {\em \prb}, vol.~84, p.~115116, Sept.
  2011.

\bibitem{2011JPhA...44y2001K}
H.~{Katsura}, ``{Exact ground state of the sine-square deformed XY spin
  chain},'' {\em Journal of Physics A Mathematical General}, vol.~44,
  p.~252001, June 2011.

\bibitem{PhysRevB.86.041108}
C.~Hotta and N.~Shibata, ``Grand canonical finite-size numerical approaches: A
  route to measuring bulk properties in an applied field,'' {\em Phys. Rev. B},
  vol.~86, p.~041108, Jul 2012.

\bibitem{PhysRevB.87.115128}
C.~Hotta, S.~Nishimoto, and N.~Shibata, ``Grand canonical finite size numerical
  approaches in one and two dimensions: Real space energy renormalization and
  edge state generation,'' {\em Phys. Rev. B}, vol.~87, p.~115128, Mar 2013.

\bibitem{2015JPhA...48E5402I}
N.~{Ishibashi} and T.~{Tada}, ``{Infinite circumference limit of conformal
  field theory},'' {\em Journal of Physics A Mathematical General}, vol.~48,
  p.~315402, Aug. 2015.

\bibitem{2016IJMPA..3150170I}
N.~{Ishibashi} and T.~{Tada}, ``{Dipolar quantization and the infinite
  circumference limit of two-dimensional conformal field theories},'' {\em
  International Journal of Modern Physics A}, vol.~31, p.~1650170, Nov. 2016.

\bibitem{2016arXiv160309543O}
K.~{Okunishi}, ``{Sine-square deformation and Mobius quantization of
  two-dimensional conformal field theory},'' {\em arXiv e-prints},
  p.~arXiv:1603.09543, Mar. 2016.

\bibitem{PhysRevB.93.235119}
X.~Wen, S.~Ryu, and A.~W.~W. Ludwig, ``Evolution operators in conformal field
  theories and conformal mappings: Entanglement hamiltonian, the sine-square
  deformation, and others,'' {\em Phys. Rev. B}, vol.~93, p.~235119, Jun 2016.

\bibitem{2017arXiv170906238T}
S.~{Tamura} and H.~{Katsura}, ``{Zero-energy states in conformal field theory
  with sine-square deformation},'' {\em arXiv e-prints}, p.~arXiv:1709.06238,
  Sept. 2017.

\bibitem{2018PTEP.2018f1B01T}
T.~{Tada}, ``{Conformal quantum mechanics and sine-square deformation},'' {\em
  Progress of Theoretical and Experimental Physics}, vol.~2018, p.~061B01, June
  2018.

\bibitem{2018JSP...172..353G}
K.~{Gawedzki}, E.~{Langmann}, and P.~{Moosavi}, ``{Finite-Time Universality in
  Nonequilibrium CFT},'' {\em Journal of Statistical Physics}, vol.~172,
  pp.~353--378, July 2018.

\bibitem{PhysRevLett.122.020201}
E.~Langmann and P.~Moosavi, ``Diffusive heat waves in random conformal field
  theory,'' {\em Phys. Rev. Lett.}, vol.~122, p.~020201, Jan 2019.

\bibitem{Gluza_2022}
M.~Gluza, P.~Moosavi, and S.~Sotiriadis, ``Breaking of
  huygens{\textendash}fresnel principle in inhomogeneous
  tomonaga{\textendash}luttinger liquids,'' {\em Journal of Physics A:
  Mathematical and Theoretical}, vol.~55, p.~054002, jan 2022.

\bibitem{PhysRevB.97.184309}
X.~Wen and J.-Q. Wu, ``Quantum dynamics in sine-square deformed conformal field
  theory: Quench from uniform to nonuniform conformal field theory,'' {\em
  Phys. Rev. B}, vol.~97, p.~184309, May 2018.

\bibitem{2019JPhA...52X5401M}
I.~{MacCormack}, A.~{Liu}, M.~{Nozaki}, and S.~{Ryu}, ``{Holographic duals of
  inhomogeneous systems: the rainbow chain and the sine-square deformation
  model},'' {\em Journal of Physics A Mathematical General}, vol.~52,
  p.~505401, Dec. 2019.

\bibitem{Goto:2021sqx}
K.~Goto, M.~Nozaki, K.~Tamaoka, M.~T. Tan, and S.~Ryu, ``{Non-Equilibrating a
  Black Hole with Inhomogeneous Quantum Quench},'' 12 2021.

\bibitem{PhysRevLett.118.260602}
W.~Berdanier, M.~Kolodrubetz, R.~Vasseur, and J.~E. Moore, ``Floquet dynamics
  of boundary-driven systems at criticality,'' {\em Phys. Rev. Lett.},
  vol.~118, p.~260602, Jun 2017.

\bibitem{2018arXiv180500031W}
X.~{Wen} and J.-Q. {Wu}, ``{Floquet conformal field theory},'' {\em arXiv
  e-prints}, p.~arXiv:1805.00031, Apr. 2018.

\bibitem{2020PhRvX..10c1036F}
R.~{Fan}, Y.~{Gu}, A.~{Vishwanath}, and X.~{Wen}, ``{Emergent Spatial Structure
  and Entanglement Localization in Floquet Conformal Field Theory},'' {\em
  Physical Review X}, vol.~10, p.~031036, July 2020.

\bibitem{2020PhRvB.102t5125H}
B.~{Han} and X.~{Wen}, ``{Classification of S L$_{2}$ deformed Floquet
  conformal field theories},'' {\em \prb}, vol.~102, p.~205125, Nov. 2020.

\bibitem{2021PhRvR...3b3044W}
X.~{Wen}, R.~{Fan}, A.~{Vishwanath}, and Y.~{Gu}, ``{Periodically,
  quasiperiodically, and randomly driven conformal field theories},'' {\em
  Physical Review Research}, vol.~3, p.~023044, Apr. 2021.

\bibitem{2020arXiv201109491F}
R.~{Fan}, Y.~{Gu}, A.~{Vishwanath}, and X.~{Wen}, ``{Floquet conformal field
  theories with generally deformed Hamiltonians},'' {\em arXiv e-prints},
  p.~arXiv:2011.09491, Nov. 2020.

\bibitem{2021arXiv210910923W}
X.~{Wen}, Y.~{Gu}, A.~{Vishwanath}, and R.~{Fan}, ``{Periodically,
  Quasi-periodically, and Randomly Driven Conformal Field Theories (II):
  Furstenberg's Theorem and Exceptions to Heating Phases},'' {\em arXiv
  e-prints}, p.~arXiv:2109.10923, Sept. 2021.

\bibitem{Lapierre_2020}
B.~Lapierre, K.~Choo, C.~Tauber, A.~Tiwari, T.~Neupert, and R.~Chitra,
  ``Emergent black hole dynamics in critical floquet systems,'' {\em Physical
  Review Research}, vol.~2, Apr 2020.

\bibitem{Lapierre_2020_1}
B.~Lapierre, K.~Choo, A.~Tiwari, C.~Tauber, T.~Neupert, and R.~Chitra, ``Fine
  structure of heating in a quasiperiodically driven critical quantum system,''
  {\em Physical Review Research}, vol.~2, Sep 2020.

\bibitem{Lapierre_2021}
B.~Lapierre and P.~Moosavi, ``Geometric approach to inhomogeneous floquet
  systems,'' {\em Physical Review B}, vol.~103, jun 2021.

\bibitem{2016arXiv161104591Z}
M.~P. {Zaletel}, A.~M. {Kaufman}, D.~M. {Stamper-Kurn}, and N.~Y. {Yao},
  ``{Preparation of Low Entropy Correlated Many-body States via Conformal
  Cooling Quenches},'' {\em arXiv e-prints}, p.~arXiv:1611.04591, Nov. 2016.

\bibitem{2018PhRvL.120u0604A}
K.~{Agarwal}, R.~N. {Bhatt}, and S.~L. {Sondhi}, ``{Fast Preparation of
  Critical Ground States Using Superluminal Fronts},'' {\em \prl}, vol.~120,
  p.~210604, May 2018.

\bibitem{2019PhRvB..99j4308M}
P.~{Mitra}, M.~{Ippoliti}, R.~N. {Bhatt}, S.~L. {Sondhi}, and K.~{Agarwal},
  ``{Cooling arbitrary near-critical systems using hyperbolic quenches},'' {\em
  \prb}, vol.~99, p.~104308, Mar. 2019.

\bibitem{2022arXiv221100040W}
X.~{Wen}, R.~{Fan}, and A.~{Vishwanath}, ``{Floquet's Refrigerator: Conformal
  Cooling in Driven Quantum Critical Systems},'' {\em arXiv e-prints},
  p.~arXiv:2211.00040, Oct. 2022.

\bibitem{2015JHEP...09..110A}
C.~T. {Asplund}, A.~{Bernamonti}, F.~{Galli}, and T.~{Hartman}, ``{Entanglement
  scrambling in 2d conformal field theory},'' {\em Journal of High Energy
  Physics}, vol.~2015, p.~110, Sept. 2015.

\bibitem{2014PhRvD..89f6015A}
C.~T. {Asplund} and A.~{Bernamonti}, ``{Mutual information after a local quench
  in conformal field theory},'' {\em \prd}, vol.~89, p.~066015, Mar. 2014.

\bibitem{2019JSMTE..09.3107N}
L.~{Nie}, M.~{Nozaki}, S.~{Ryu}, and M.~{Tian Tan}, ``{Signature of quantum
  chaos in operator entanglement in 2d CFTs},'' {\em Journal of Statistical
  Mechanics: Theory and Experiment}, vol.~9, p.~093107, Sept. 2019.

\bibitem{2020JHEP...01..031K}
J.~{Kudler-Flam}, M.~{Nozaki}, S.~{Ryu}, and M.~T. {Tan}, ``{Quantum vs.
  classical information: operator negativity as a probe of scrambling},'' {\em
  Journal of High Energy Physics}, vol.~2020, p.~31, Jan. 2020.

\bibitem{2021JHEP...03..146K}
J.~{Kudler-Flam}, Y.~{Kusuki}, and S.~{Ryu}, ``{The quasi-particle picture and
  its breakdown after local quenches: mutual information, negativity, and
  reflected entropy},'' {\em Journal of High Energy Physics}, vol.~2021,
  p.~146, Mar. 2021.

\bibitem{2022JHEP...06..100G}
K.~{Goto}, A.~{Mollabashi}, M.~{Nozaki}, K.~{Tamaoka}, and M.~T. {Tan},
  ``{Information scrambling versus quantum revival through the lens of operator
  entanglement},'' {\em Journal of High Energy Physics}, vol.~2022, p.~100,
  June 2022.

\bibitem{2023arXiv230208009G}
K.~{Goto}, M.~{Nozaki}, S.~{Ryu}, K.~{Tamaoka}, and M.~{Tian Tan},
  ``{Scrambling and Recovery of Quantum Information in Inhomogeneous Quenches
  in Two-dimensional Conformal Field Theories},'' {\em arXiv e-prints},
  p.~arXiv:2302.08009, Feb. 2023.

\bibitem{2011PhRvL.106e0405B}
M.~C. {Ba{\~n}uls}, J.~I. {Cirac}, and M.~B. {Hastings}, ``{Strong and Weak
  Thermalization of Infinite Nonintegrable Quantum Systems},'' {\em Physical
  Review Letters}, vol.~106, p.~050405, Feb. 2011.

\bibitem{2016JHEP...02..004H}
P.~{Hosur}, X.-L. {Qi}, D.~A. {Roberts}, and B.~{Yoshida}, ``{Chaos in quantum
  channels},'' {\em Journal of High Energy Physics}, vol.~2016, p.~4, Feb.
  2016.

\bibitem{2020arXiv201214609M}
E.~{Mascot}, M.~{Nozaki}, and M.~{Tezuka}, ``{Local Operator Entanglement in
  Spin Chains},'' {\em arXiv e-prints}, p.~arXiv:2012.14609, Dec. 2020.

\bibitem{2020PhRvB.101q4313C}
B.~{Craps}, M.~{De Clerck}, D.~{Janssens}, V.~{Luyten}, and C.~{Rabideau},
  ``{Lyapunov growth in quantum spin chains},'' {\em \prb}, vol.~101,
  p.~174313, May 2020.

\bibitem{Nishimoto_2013}
S.~Nishimoto, N.~Shibata, and C.~Hotta, ``Controlling frustrated liquids and
  solids with an applied field in a kagome heisenberg antiferromagnet,'' {\em
  Nature Communications}, vol.~4, aug 2013.

\bibitem{Ito_2018}
T.~Ito, C.~Iino, and N.~Shibata, ``Thermodynamic properties of the s=1/2
  twisted triangular spin tube,'' {\em Physical Review B}, vol.~97, may 2018.

\bibitem{nielsen_chuang_2010}
M.~A. Nielsen and I.~L. Chuang, {\em Quantum Computation and Quantum
  Information: 10th Anniversary Edition}.
\newblock Cambridge University Press, 2010.

\bibitem{2018arXiv180300089J}
C.~{Jonay}, D.~A. {Huse}, and A.~{Nahum}, ``{Coarse-grained dynamics of
  operator and state entanglement},'' {\em ArXiv e-prints}, Feb. 2018.

\bibitem{PhysRevX.8.031058}
T.~Rakovszky, F.~Pollmann, and C.~W. von Keyserlingk, ``Diffusive hydrodynamics
  of out-of-time-ordered correlators with charge conservation,'' {\em Phys.
  Rev. X}, vol.~8, p.~031058, Sep 2018.

\bibitem{PhysRevX.8.021013}
C.~W. von Keyserlingk, T.~Rakovszky, F.~Pollmann, and S.~L. Sondhi, ``Operator
  hydrodynamics, otocs, and entanglement growth in systems without conservation
  laws,'' {\em Phys. Rev. X}, vol.~8, p.~021013, Apr 2018.

\bibitem{Mezei:2018jco}
M.~Mezei, ``{Membrane theory of entanglement dynamics from holography},'' {\em
  Phys. Rev. D}, vol.~98, no.~10, p.~106025, 2018.

\bibitem{Goto:2021gve}
K.~Goto, A.~Mollabashi, M.~Nozaki, K.~Tamaoka, and M.~T. Tan, ``{Information
  Scrambling Versus Quantum Revival Through the Lens of Operator
  Entanglement},'' 12 2021.

\bibitem{Goto:2022fec}
K.~Goto, M.~Nozaki, K.~Tamaoka, and M.~T. Tan, ``{Entanglement dynamics of the
  non-unitary holographic channel},'' {\em JHEP}, vol.~03, p.~101, 2023.

\bibitem{Goto:2023wai}
K.~Goto, M.~Nozaki, S.~Ryu, K.~Tamaoka, and M.~T. Tan, ``{Scrambling and
  Recovery of Quantum Information in Inhomogeneous Quenches in Two-dimensional
  Conformal Field Theories},'' 2 2023.

\bibitem{Chen_2018}
X.~Chen and A.~W.~W. Ludwig, ``Universal spectral correlations in the chaotic
  wave function and the development of quantum chaos,'' {\em Physical Review
  B}, vol.~98, aug 2018.

\bibitem{Margolus:1997ih}
N.~Margolus and L.~B. Levitin, ``{The Maximum speed of dynamical evolution},''
  {\em Physica D}, vol.~120, pp.~188--195, 1998.

\bibitem{Note1}
In \cite {Margolus:1997ih}, the initial and time-evolved states are defined as
  $\left | \Psi \right >$ and $\left | \Psi (t) \right >=U(t)\left | \Psi
  \right >$, respectively. The return amplitude is defined as the inner product
  of them, $\left \langle \Psi | \Psi (t)\right \rangle $. This definition is
  different from ours. If we define the initial and time-evolved states as
  $\left | \Psi \right >=\sigma ^{(1)}_{\alpha ,a}\left | \protect \text {TFD}
  \right >$ and $\left | \Psi (t) \right >=U(t)\sigma ^{(1)}_{\alpha ,a}\left |
  \protect \text {TFD} \right >$, the time-dependence of the inner product is
  governed by a global quench, not the growth of the Heisenberg operators.
  Therefore, we introduce a generalized return amplitude that may capture the
  growth of the Heisenberg operators.

\bibitem{2018PhRvB..98n4304K}
V.~{Khemani}, D.~A. {Huse}, and A.~{Nahum}, ``{Velocity-dependent Lyapunov
  exponents in many-body quantum, semiclassical, and classical chaos},'' {\em
  Physical Review B}, vol.~98, p.~144304, October 2018.

\bibitem{2019ScPP....7...45H}
X.~{Han} and S.~{Hartnoll}, ``{Quantum scrambling and state dependence of the
  butterfly velocity},'' {\em SciPost Physics}, vol.~7, p.~045, Oct. 2019.

\bibitem{Goto:2021anl}
K.~Goto, T.~Nosaka, and M.~Nozaki, ``{Chaos by Magic},'' 12 2021.

\bibitem{Peschel_2009}
I.~Peschel and V.~Eisler, ``Reduced density matrices and entanglement entropy
  in free lattice models,'' {\em Journal of Physics A: Mathematical and
  Theoretical}, vol.~42, p.~504003, dec 2009.

\bibitem{MacCormack_2021}
I.~MacCormack, M.~T. Tan, J.~Kudler-Flam, and S.~Ryu, ``Operator and
  entanglement growth in nonthermalizing systems: Many-body localization and
  the random singlet phase,'' {\em Physical Review B}, vol.~104, dec 2021.

\bibitem{Note2}
When $L$ is even, $P=-1$ sector also splits into $T=\pm 1$ sectors. See \cite
  {Goto:2021anl}.

\bibitem{PhysRevLett.52.1}
O.~Bohigas, M.~J. Giannoni, and C.~Schmit, ``Characterization of chaotic
  quantum spectra and universality of level fluctuation laws,'' {\em Phys. Rev.
  Lett.}, vol.~52, pp.~1--4, Jan 1984.

\bibitem{1984LNP...209....1B}
O.~{Bohigas} and M.-J. {Giannoni}, {\em {Chaotic motion and random matrix
  theories}}, vol.~209, pp.~1--99.
\newblock 1984.

\bibitem{Guhr:1997ve}
T.~Guhr, A.~Muller-Groeling, and H.~A. Weidenmuller, ``{Random matrix theories
  in quantum physics: Common concepts},'' {\em Phys. Rept.}, vol.~299,
  pp.~189--425, 1998.

\bibitem{1977RSPSA.356..375B}
M.~V. {Berry} and M.~{Tabor}, ``{Level Clustering in the Regular Spectrum},''
  {\em Proceedings of the Royal Society of London Series A}, vol.~356,
  pp.~375--394, Sept. 1977.

\end{thebibliography}

\appendix

\section{BOMI of deformed free Fermion}\label{Section:BOMI_FF}
\begin{figure}[t]
\centering
\includegraphics[width=0.48\columnwidth]{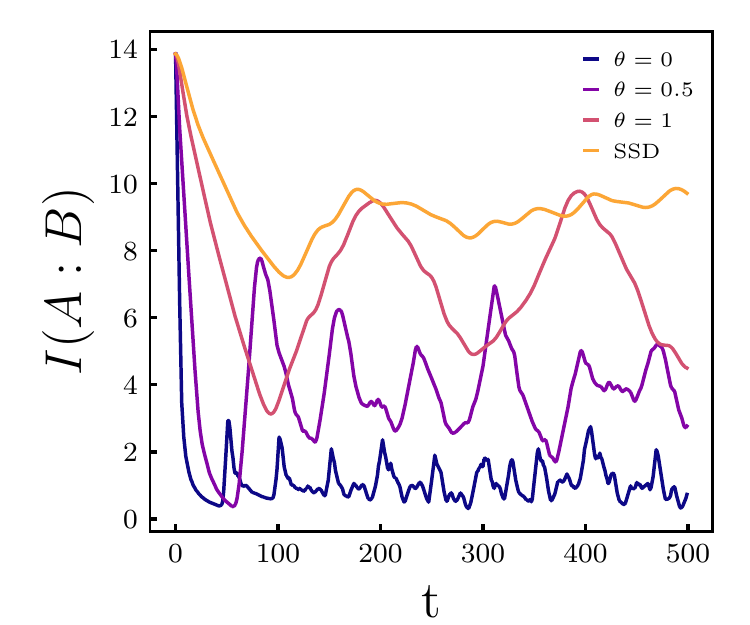}
\includegraphics[width=0.48\columnwidth]{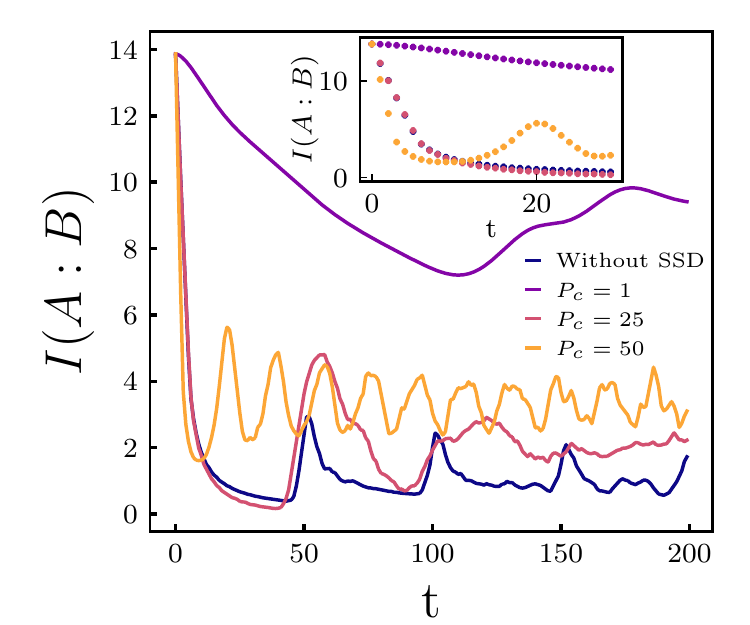}
\caption{(Left): BOMI  of Mobius deformed free Fermion $L = 100$ with symmetric subsystem $A = B= \{1...5\}\cup \{96...100\}$ for different $\theta$ defined in Eq \ref{eq:FF_mobi}. (Right): BOMI  of SSD deformed free Fermion with different subsystem choices $A = B = \{P_c-5,..., P_c+5\}$. 
Inset shows the early time result}
\label{Fig:FF_Bomi}
\end{figure}

In this section, we consider the bipartite operator entanglement of a free Fermion system
\begin{equation}
    H = \sum_{i,j} H_{i,j} c_i^{\dagger}c_j . 
\end{equation}
Generally, the entanglement entropy for a free Fermion system could be calculated using the correlation matrix method \cite{Peschel_2009}. Particularly for operator entanglement, 
following \cite{MacCormack_2021}, we consider the unitary operator 
\begin{equation}
    | U \rangle = e^{ - it (H_A + H_B)}| \text{TFD}\rangle ,
\end{equation}
with 
\begin{equation}
    |\text{TFD} \rangle = \frac{1}{Z} \prod_{k} (1 + e^{-\frac{\beta \epsilon_k}{2}}\psi_{Ak}^{\dagger}\psi_{Bk}^{\dagger}  ) |0 \rangle ,
\end{equation}
where $\psi_{Ak}^k$ is the Fermion creation operator in space A with momentum $k$. 
Define $\sin{\theta} = \frac{\frac{e^{\beta \epsilon_k}}{2}}{\sqrt{1+e^{\beta \epsilon_k
}}}$, 
the element of the
correlator matrix is given by
\begin{align}
   & C(x,x',I,J) =\\
   &\sum_{k} V_{xk}^{*}
    \begin{pmatrix}
\sin^2{\theta_k} & \sin{\theta_k}\cos{\theta_k}e^{it\epsilon_k} \\
\sin{\theta_k}\cos{\theta_k}e^{-it\epsilon_k} & \cos^2{\theta_k}  
\end{pmatrix}
V_{kx'}^{t},
\end{align}
where $x,x'$ labels position in space $I,J$. $V_{xk}$ is the matrix that diagonalizes the Hamiltonian in momentum space
\begin{equation}
    H = \sum_{x,y} \chi_x^{\dagger} H_{xy} \chi_y = \sum_{x,y} \chi_x^{\dagger} V_{xk} D_{kq} V_{qy}^{\dagger}\chi_y  .
\end{equation}
$D_{kq}$ is a diagnoal matrix. 
With such a method, we could compute the entanglement entropy of the subsystem as 
\begin{equation}
    S_A = -\sum_{i}\nu_i \ln(\nu_i) + (1 - \nu_i) \ln(1-\nu_i) ,
\end{equation}
where $\nu_i$ is the eigenvalue of the correlator matrix of the subsystem.
Then, the bipartite operator entanglement could be calculated as 
\begin{equation}
    I(A:B) = S_A + S_B - S_{A\cup B} .
\end{equation}
For the free Fermion system with Mobius deformation
\begin{equation}
    H_{\text{Mobius}} = -\sum_{x} \bigg(1 - \tanh(2\theta)\cos\Big(\frac{2\pi x}{L}\Big)\bigg) (c_i^{\dagger} c_{i+1} + h.c.) .
    \label{eq:FF_mobi}
\end{equation}
We plot BOMI as a function of $t$ in Fig.\ \ref{Fig:FF_Bomi}. 

First, we focus on the subsystem consisting of $1$ to 5-th and $96$ to 100-th sites.
The $\theta$-dependence plot on the left panel of Fig \ref{Fig:FF_Bomi} shows that as $\theta$ increases, the BOMI decays faster and saturates to a larger final value. Even though the system is integrable, it still indicates a similar feature as the chaotic spin chain as is shown in Fig.\ \ref{theta_and_position_dependence}. As $\theta$ increases, the thermalization of the subsystem is more suppressed and it took the subsystem longer to thermalize.

Then, we consider different subsystem choices and focus on the SSD limit of $\theta \rightarrow \infty$. 
The results in the right panel of Fig.\ \ref{Fig:FF_Bomi} show that at the early time, the subsystem near $P_c = 50$ thermalizes faster than the original system, and the subsystem near $P_c = 1$ thermalizes slower. 
Such results indicate that the envelope function that boosts and suppresses the Hamiltonian near $P_c = 50$ and near $P_c = 1$ also boosts and suppresses the thermalization process of the subsystem near $P_c = 50$ and $P_c = 1$. 
The phenomena shown are also similar to what is seen in the chaotic spin chain as in Fig.\ \ref{theta_and_position_dependence}.
Compared with the CFT calculation for the free fermion in \cite{2023arXiv230208009G}, the early time behavior of both $\theta$ and position dependence of BOMI show similar behavior as the CFT results as the thermalization happens faster for large $\theta$ and near $b_0$. 
The difference appears in the late time as the BOMI revivals back to its original value in CFT calculation for large system size which is not shown in the lattice fermion calculation.
The threshold value of $\theta_c$ is also not observed in the lattice fermion calculation. 
Such disagreement might be due to the small system size of the lattice fermion system. 

\section{Level statistics \label{Section:levelstatics}}
Let us consider the distribution of the nearest-neighbor level spacings
\begin{align}
s_{\alpha,i}=E^{\text{(unfolded)}}_{\alpha,i+1}-E^{\text{(unfolded)}}_{\alpha,i}.
\label{s}
\end{align}
Here, $\alpha$ is a label for the eigensectors of the full spectrum with respect to the discrete symmetry acting on the Hamiltonians in \eqref{220413Mobius} and \eqref{220413SSD}.
For $\theta>0$ with generic values of $(h_x,h_z)$, the discrete symmetry consists only of the reflection symmetry $P$
\begin{align}
P: |s_1\rangle \otimes |s_2\rangle \otimes \cdots \otimes |s_L\rangle
\rightarrow |s_L\rangle \otimes |s_{L-1}\rangle \otimes \cdots \otimes |s_1\rangle,
\end{align}
hence the two sectors are labelled by $\alpha=\text{``}P=\pm \text{''}$.
For $\theta=0$ with generic values of $(h_x,h_z)$, there is an additional symmetry $T$ (translation)
\begin{align}
T: |s_1\rangle \otimes |s_2\rangle \otimes \cdots \otimes |s_L\rangle
\rightarrow |s_L\rangle \otimes |s_1\rangle \otimes \cdots \otimes |s_{L-1}\rangle,
\end{align}
with which the $P=+1$ sector further splits into $L$ sectors \footnote{
When $L$ is even, $P=-1$ sector also splits into $T=\pm 1$ sectors. See \cite{Goto:2021anl}.
}
$\alpha=\text{``}P=+1, T=e^{\frac{2\pi i\ell}{L}}\text{''}$.
In \eqref{s}, we have also used, instead of the original eigenvalues $E_{\alpha,i}$, the unfolded energy levels $E^{\text{(unfolded)}}_{\alpha,i}$ which are obtained by the redefinition
\begin{align}
E_{\alpha,i}^{\text{(unfolded)}}=\int_{E_{\alpha,1}}^{E_{\alpha,i}}\bar{\rho}_\alpha(E')dE',
\end{align}
with the global shape of the energy level density in $\alpha$-sector $\rho_\alpha(E)$.
The separation into the eigensectors of symmetry and the unfolding is necessary for the level statics to be physically reasonable characterizations of the quantum chaotic property of the system \cite{PhysRevLett.52.1,1984LNP...209....1B,Guhr:1997ve,1977RSPSA.356..375B}.

In Fig.\ \ref{fig_NNSD1}, we have displayed the distribution of the normalized nearest-neighbor level spacings that are refined as
\begin{align}
\frac{s_{\alpha,i}}{\bar{s}},\quad \bar{s}=\text{mean}\{s_{\alpha,i}\}_{\alpha,i}
\label{normalizedspacings}
\end{align}
for the closed spin chain in \eqref{220413Mobius} and \eqref{220413SSD} in  the chaotic regime $(h_x,h_z)=(-1.05,0.5)$ with and without M\"obius/SSD deformation.
We observe that the M\"obius/SSD deformation does not affect the chaotic property characterized by the level statistics.

\section{Position-dependence the return amplitudes in the chaotic region \label{App:ReturnAmplitude}}
\begin{figure*}[tbh!]
    \centering
\includegraphics[width=0.32\textwidth]{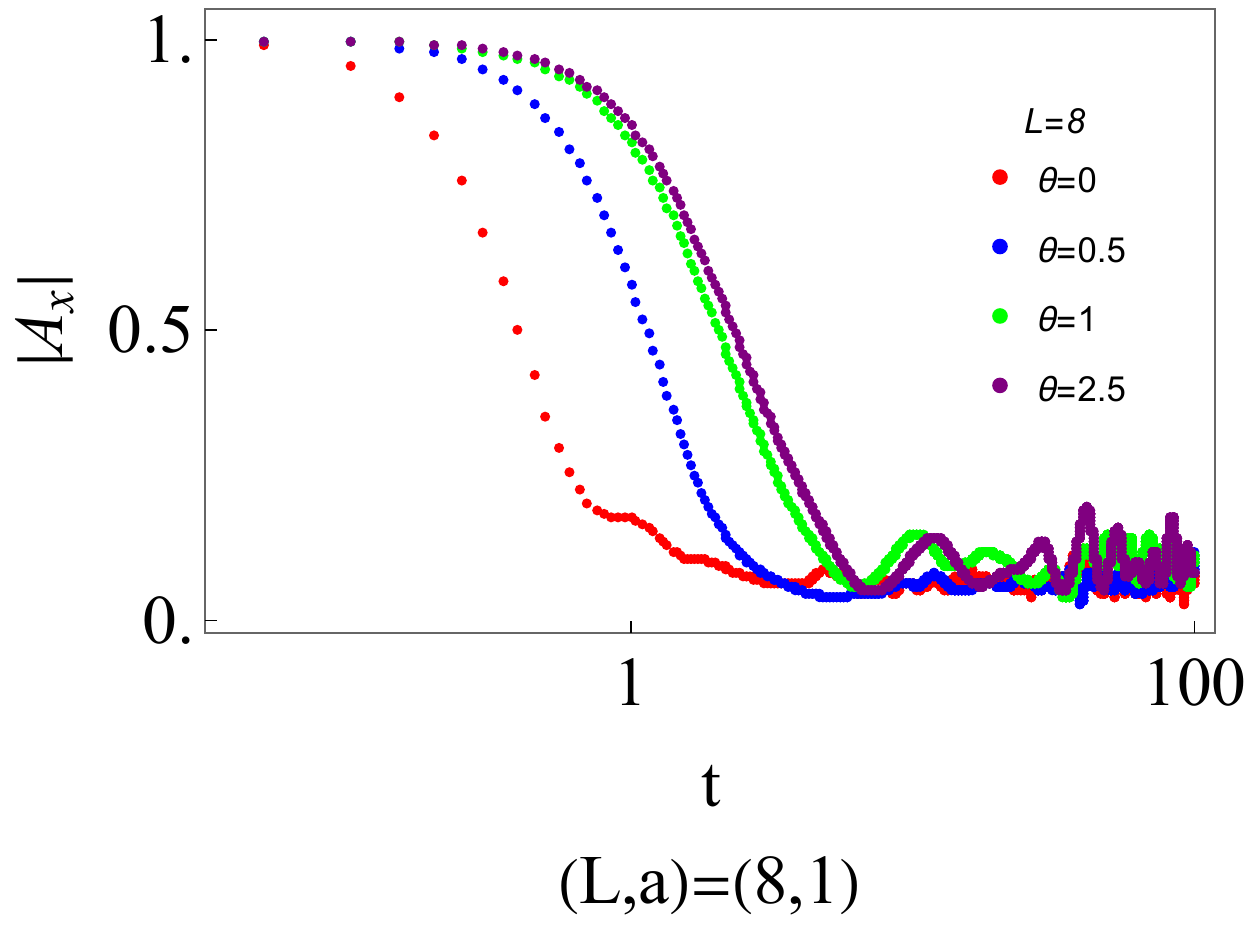}
\includegraphics[width=0.32\textwidth]{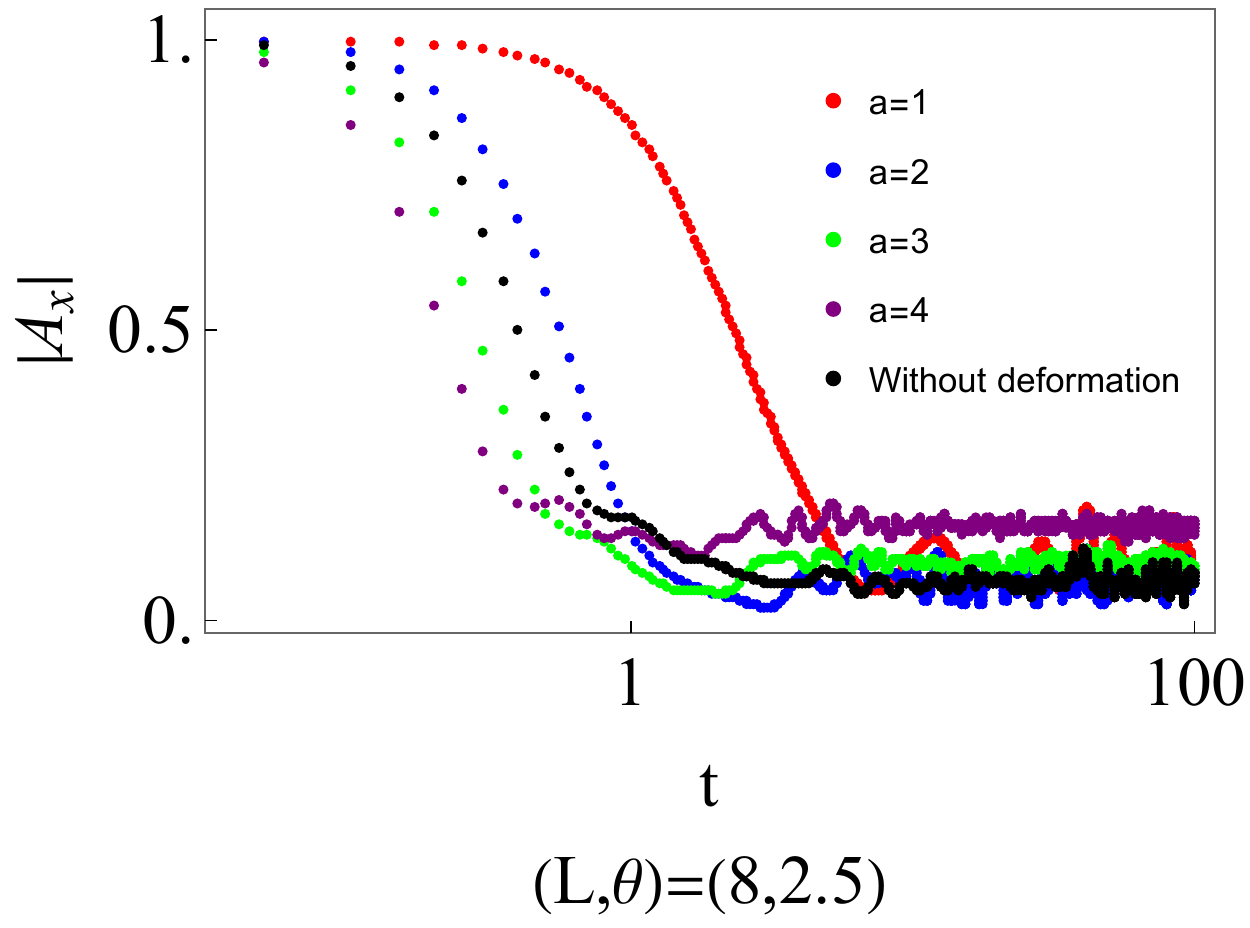}   
\includegraphics[width=0.32\textwidth]{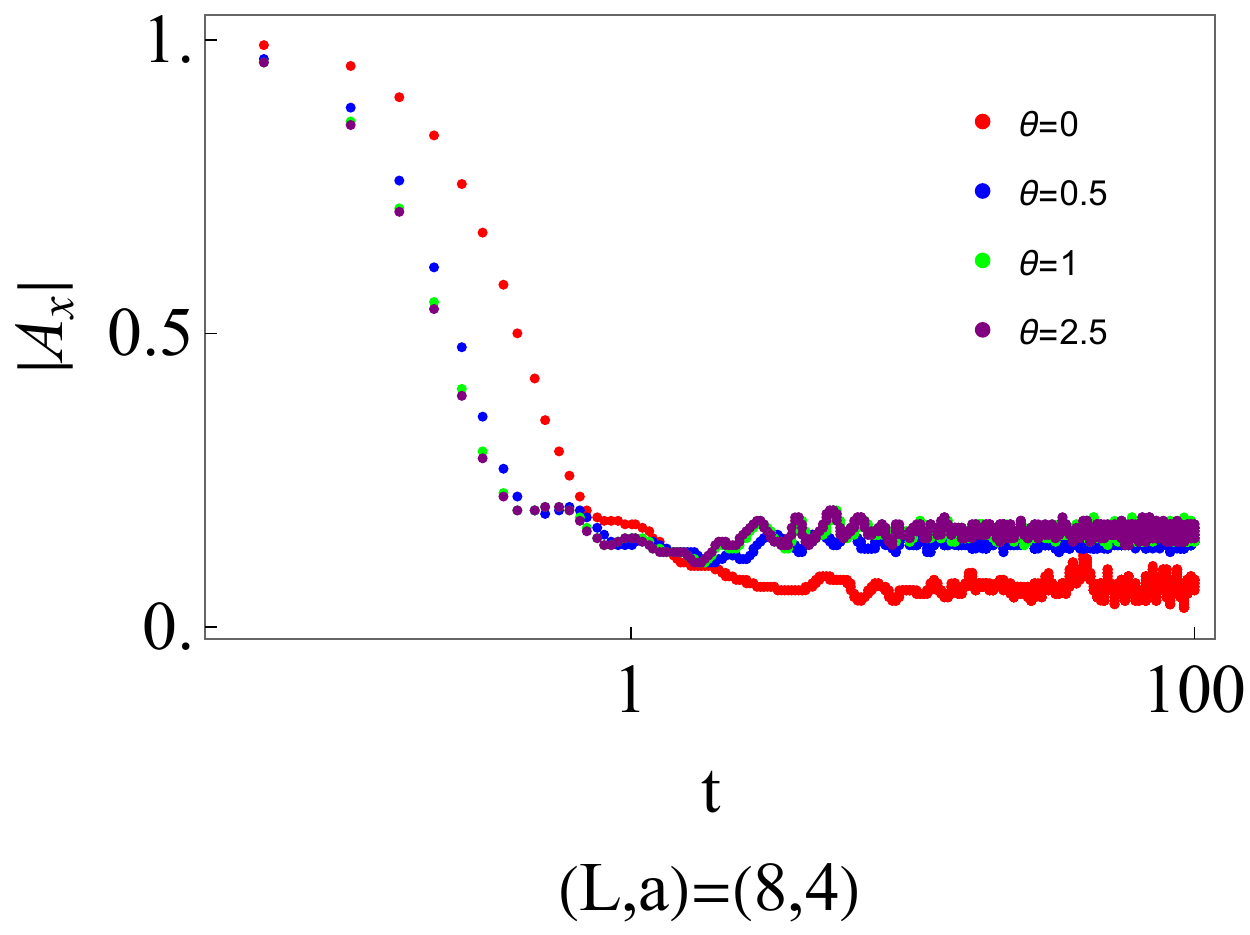} 
%
%
\includegraphics[width=0.32\textwidth]{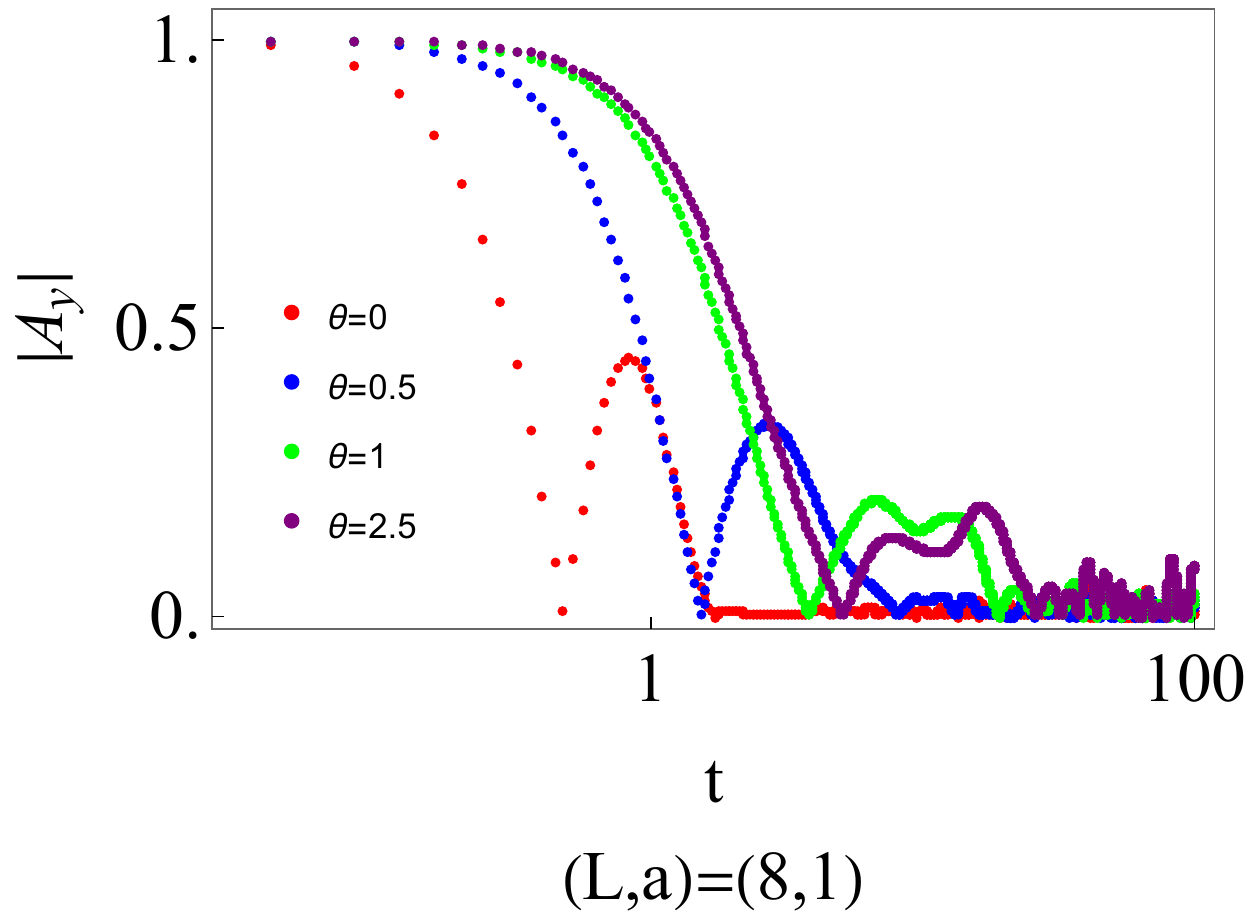}
 \includegraphics[width=0.32\textwidth]{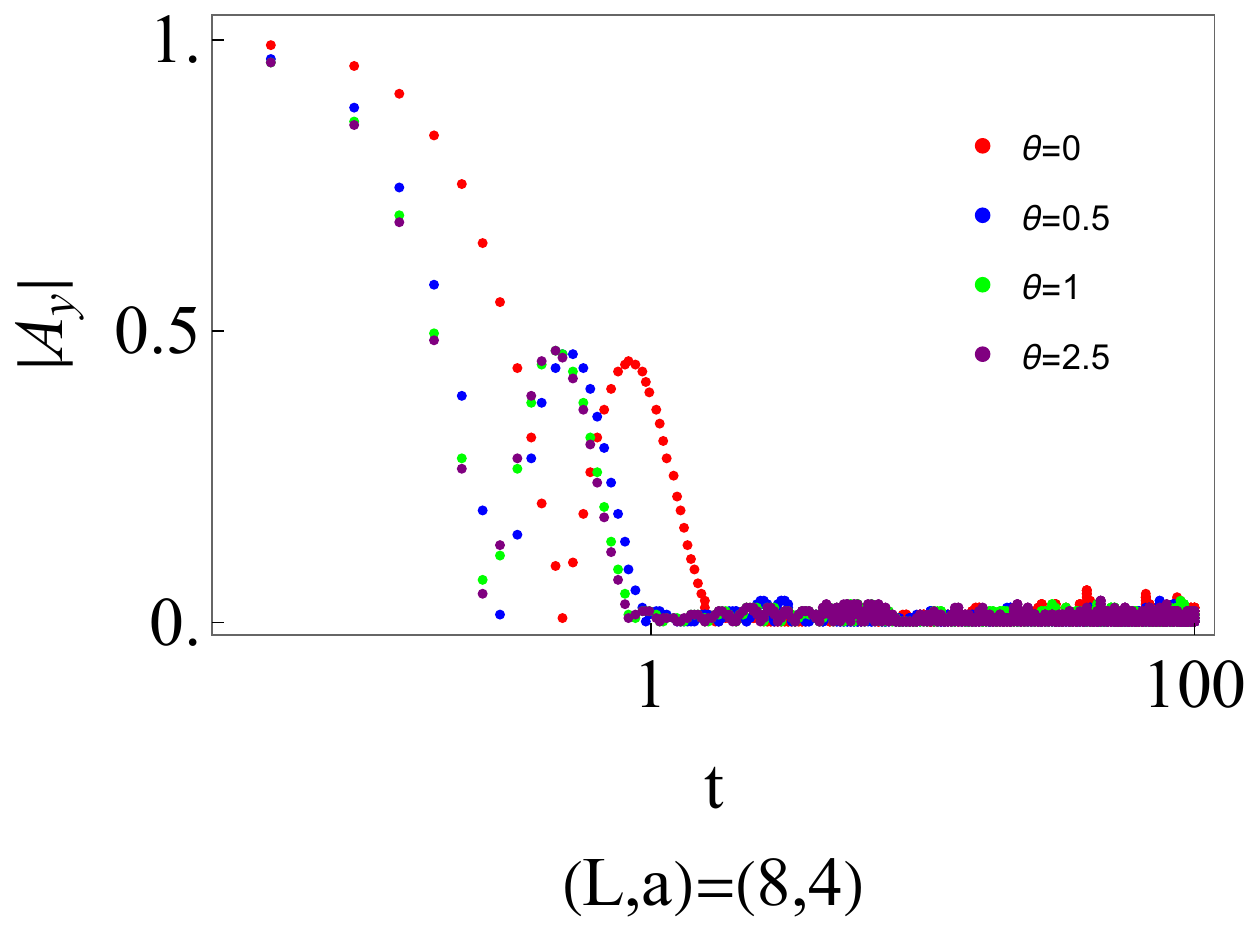}
\includegraphics[width=0.32\textwidth]{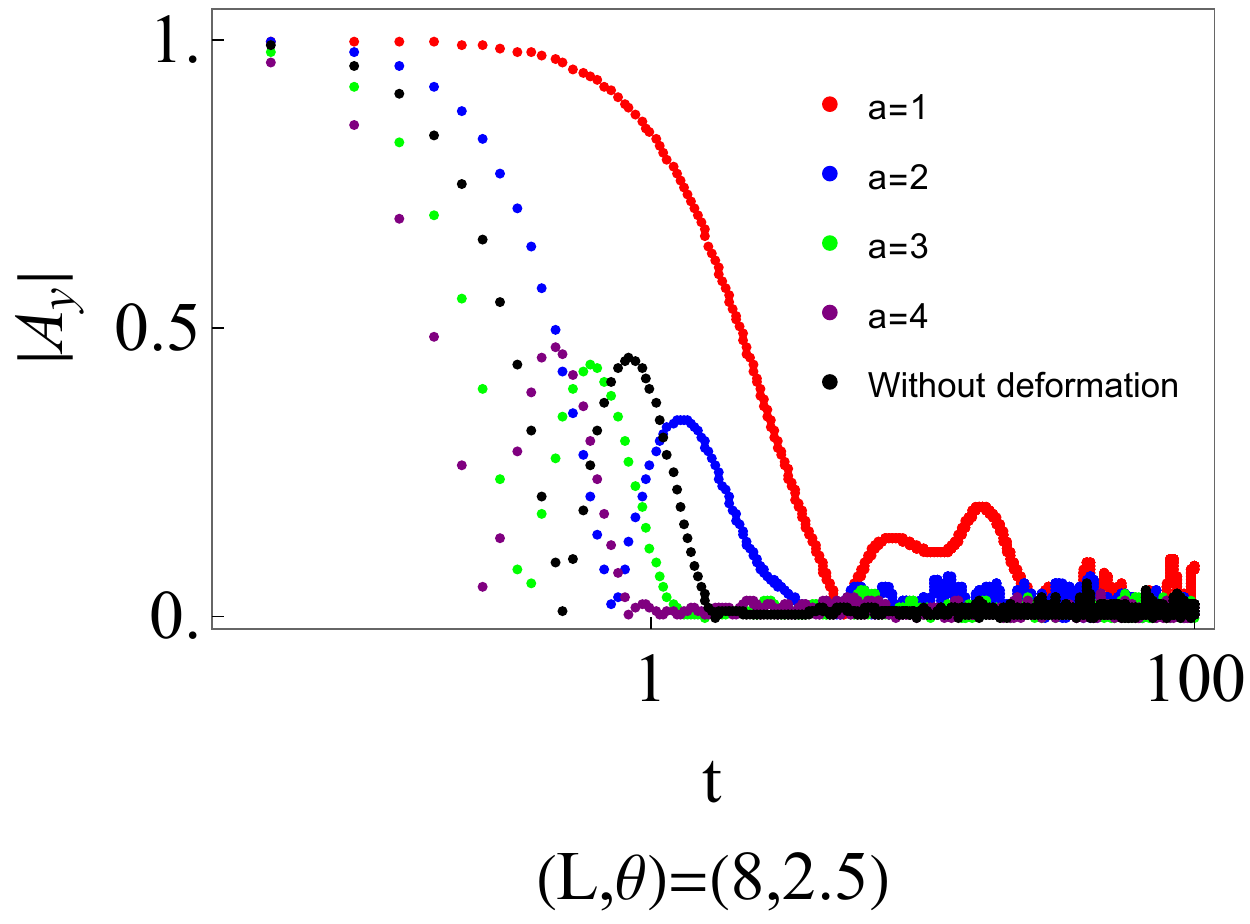}
    \caption{The inhomogeneous affect on $|A_{\alpha=x,y,a}|$ in the chaotic region. In the top panels, we show the $\theta$-dependence of $\left|A_{\alpha=x,y,a=1, 2.5, 4}\right|$.
In the bottom panels, we show the position-dependence of $|A_{\alpha=x,y,a=1,2.5,4}|$. The black points illustrate the time evolution of $|A_{\alpha=x,y,z,a}|$ in the chaotic region of the uniform Ising spin chain.}
    \label{theta_dependence_of_computational_speed_all}
  \end{figure*}

  In Fig.\ \ref{theta_dependence_of_computational_speed_all}, we depict $|A_{\alpha,a}|$ for various $a$ and $\theta$ as a function of $t$. 
As in the case of $|A_{z,a}|$, the time-dependence of $|A_{\alpha=x,y,a}|$ depends on $a$ and $\theta$.
When the value of $\theta$ becomes larger, the early-decay of $|A_{\alpha=x,y,a=1}|$ becomes slower, while the early-decay of $|A_{\alpha=x,y,a=4}|$ becomes faster.
When $a$ gets further far away from $a=1$ and closer to $a=4$, for $\theta=2.5$, their early-time decays become faster. 

\section{Position-and $\theta$ dependence of two point function in the chaotic region \label{App:twopointfunction}}
\begin{figure*}[tbh!]
\centering
   \includegraphics[width=0.32\textwidth]{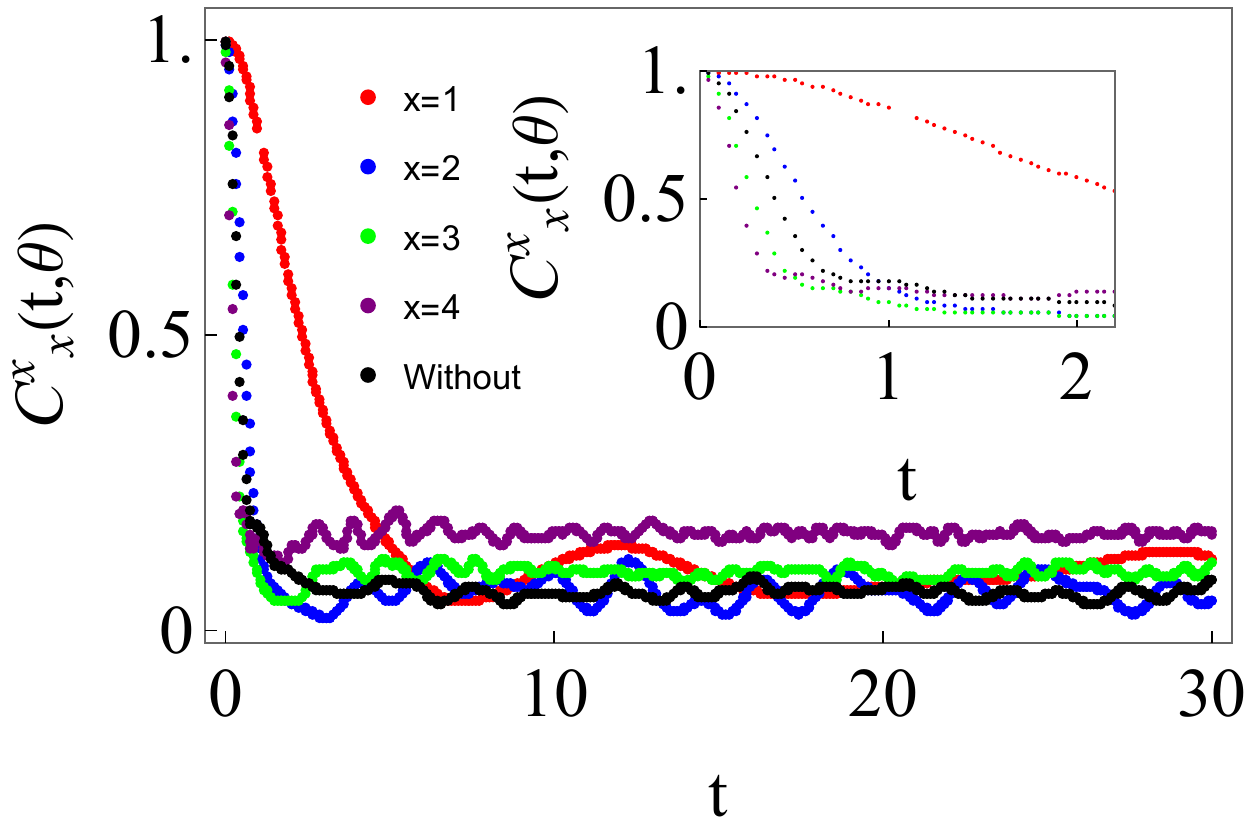}
     \includegraphics[width=0.32\textwidth]{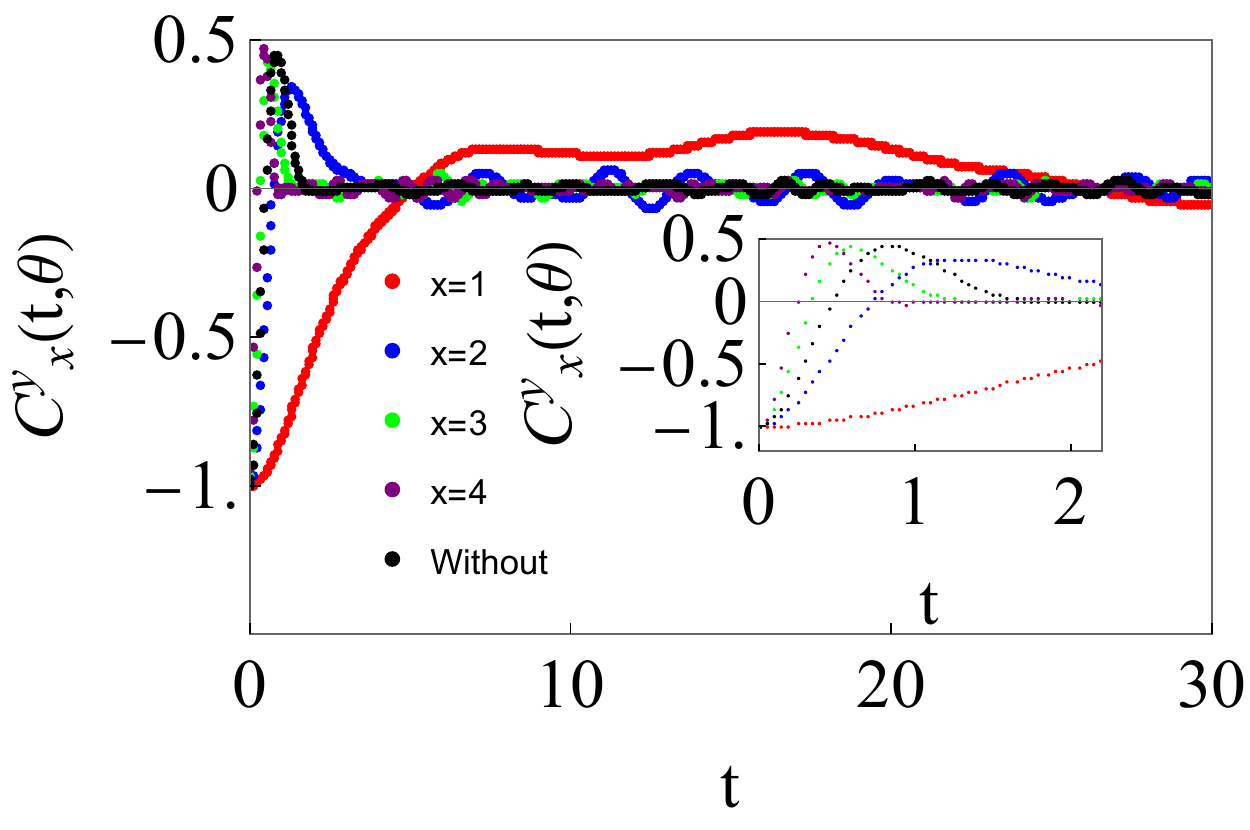}
      \includegraphics[width=0.32\textwidth]{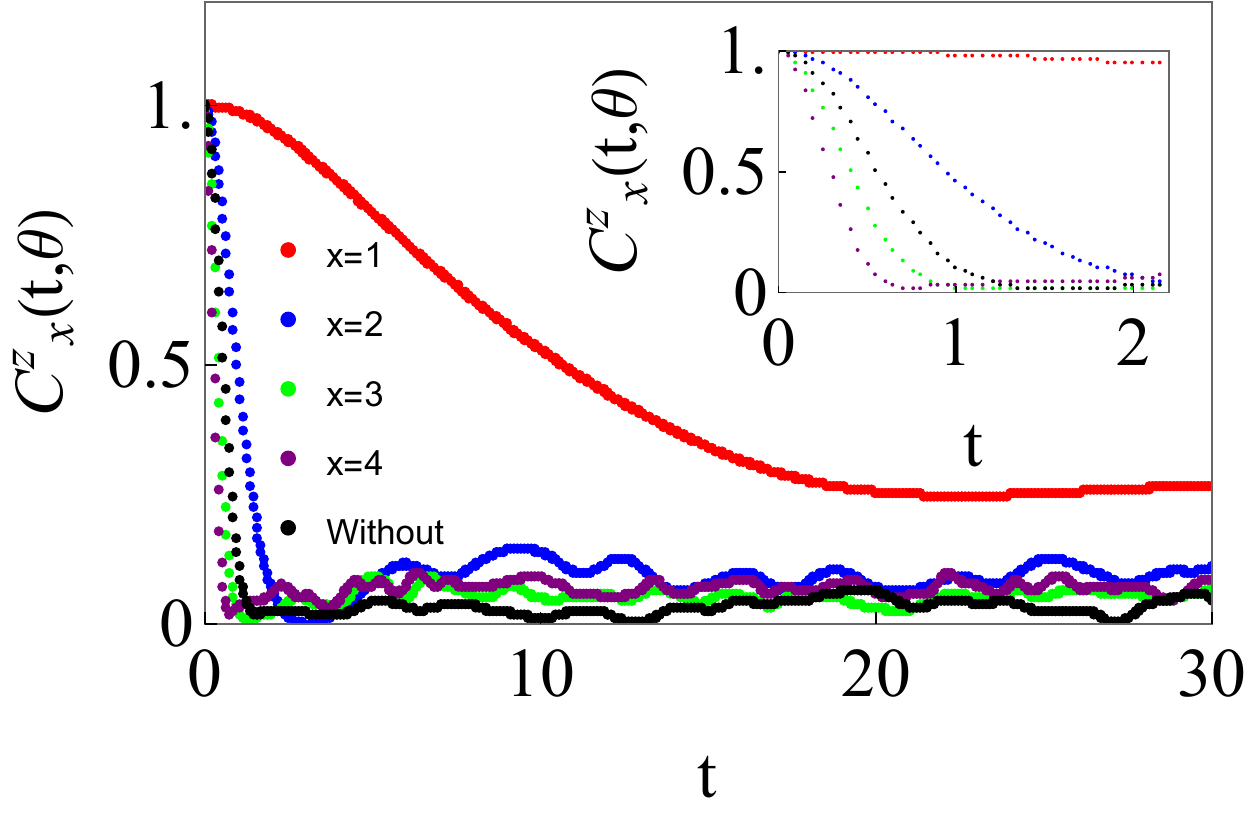}
 
      \includegraphics[width=0.32\textwidth]{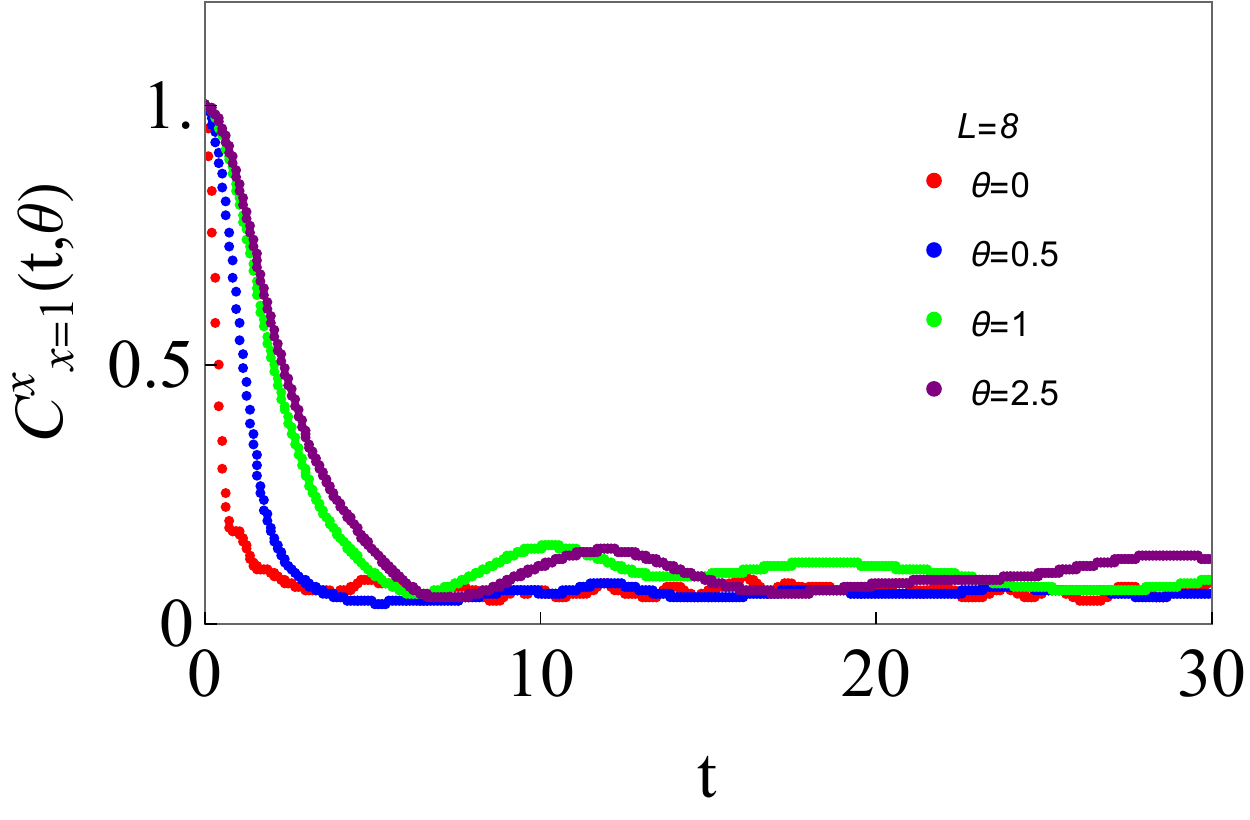}
     \includegraphics[width=0.32\textwidth]{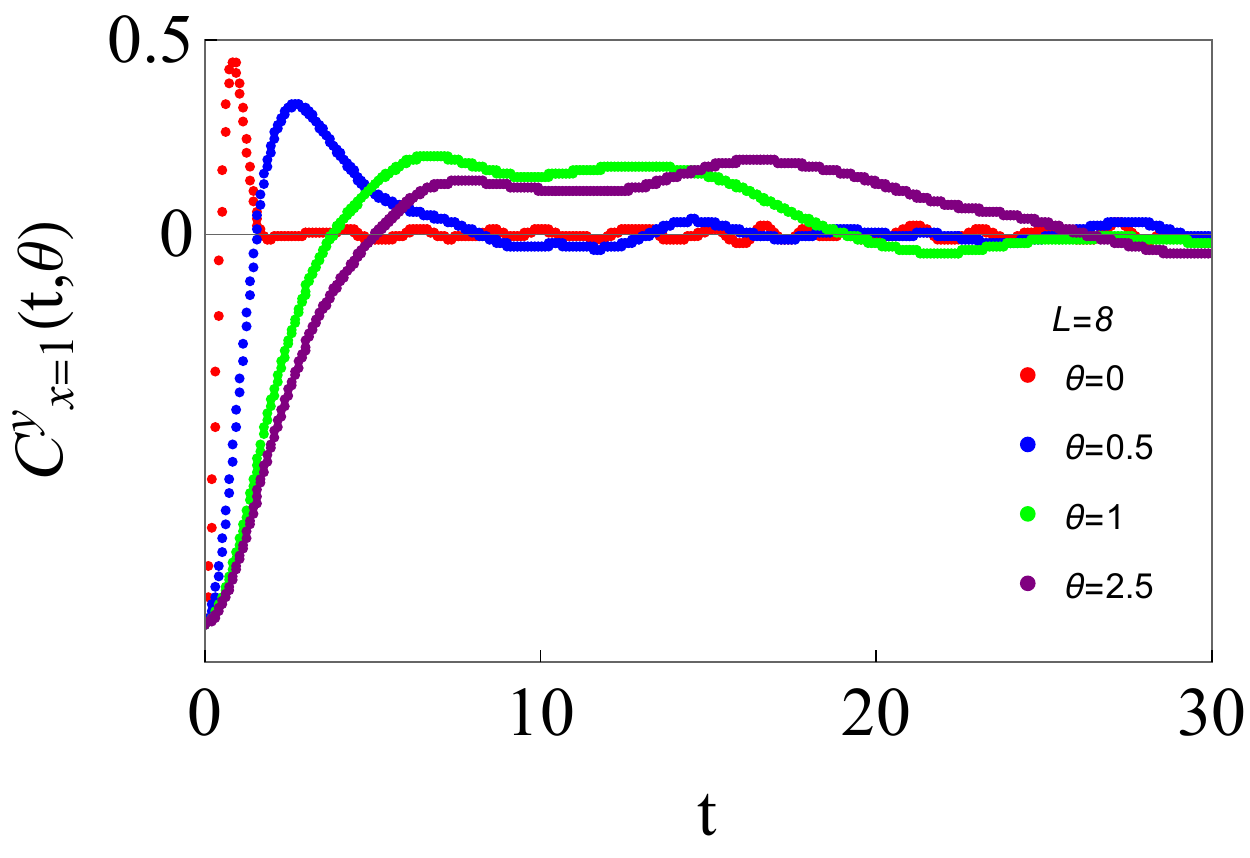}
      \includegraphics[width=0.32\textwidth]{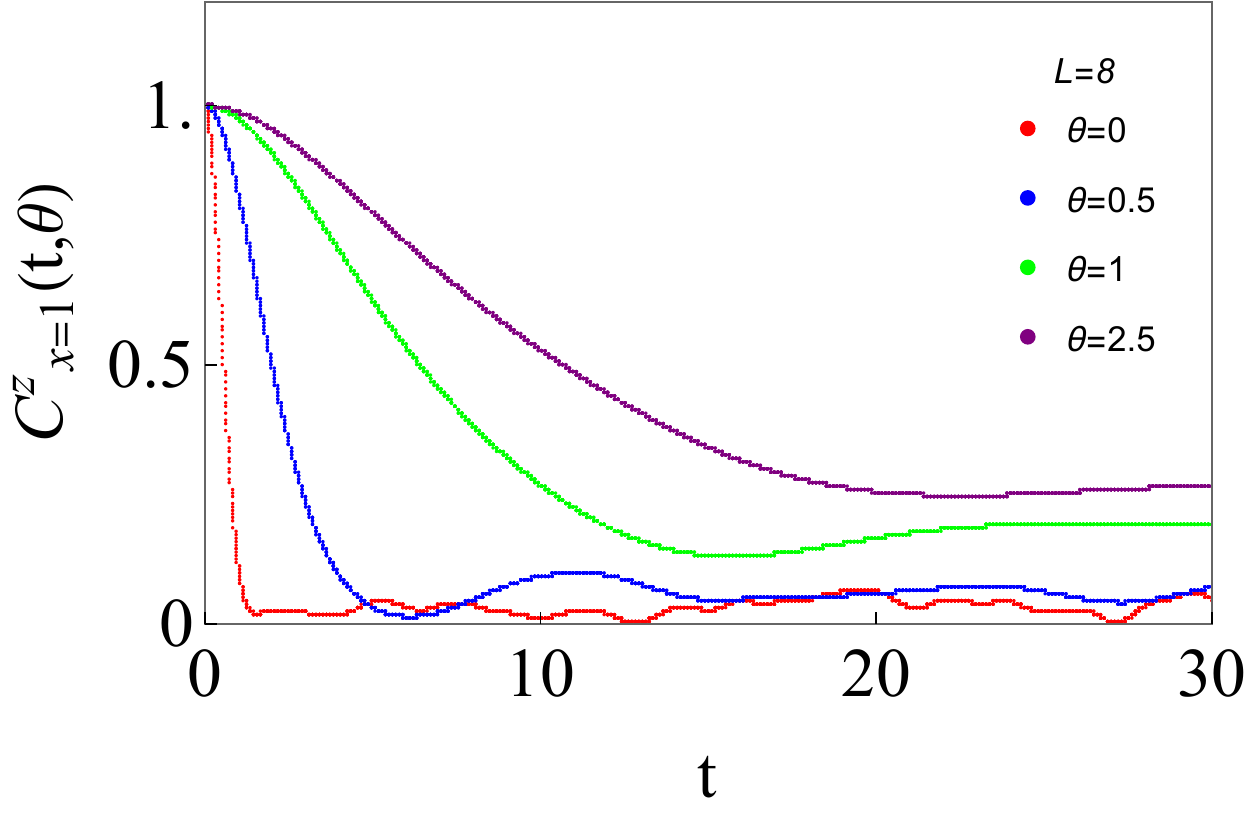}
   
      \includegraphics[width=0.32\textwidth]{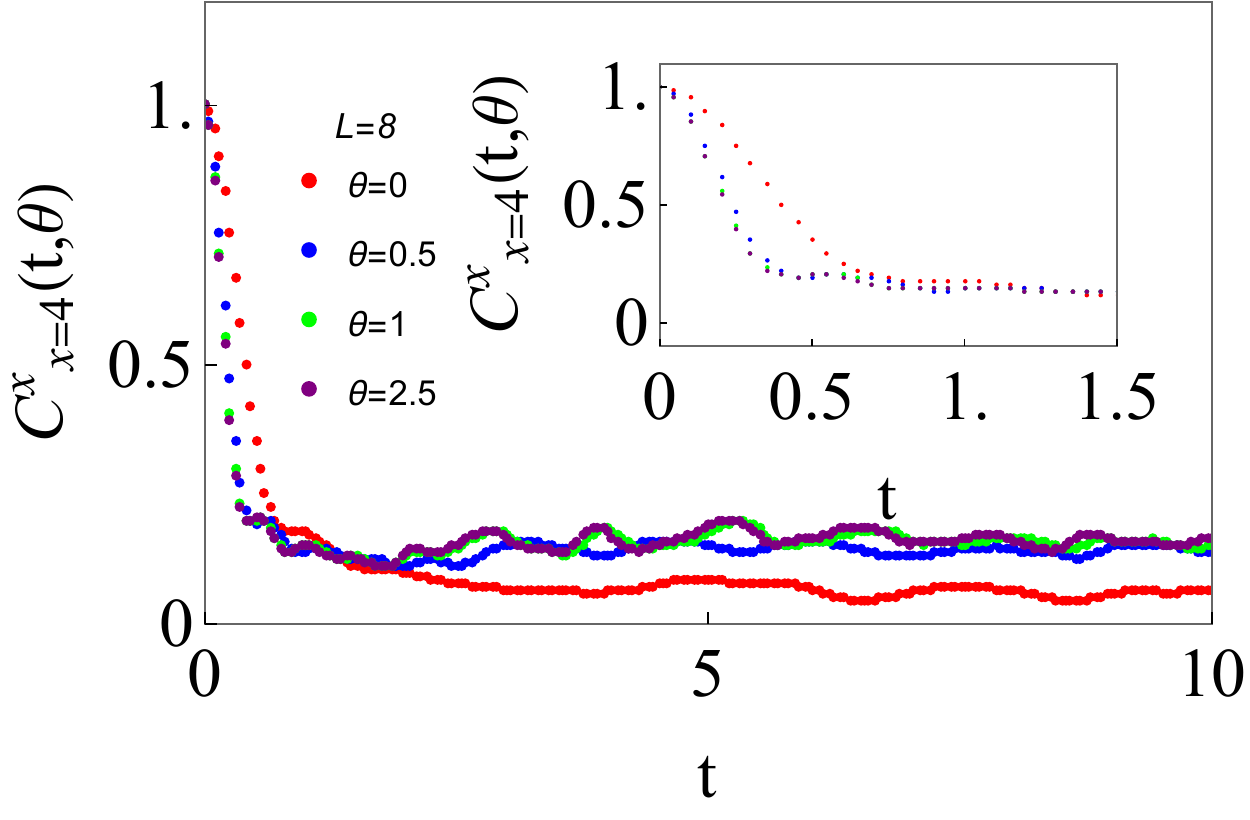}
      \includegraphics[width=0.32\textwidth]{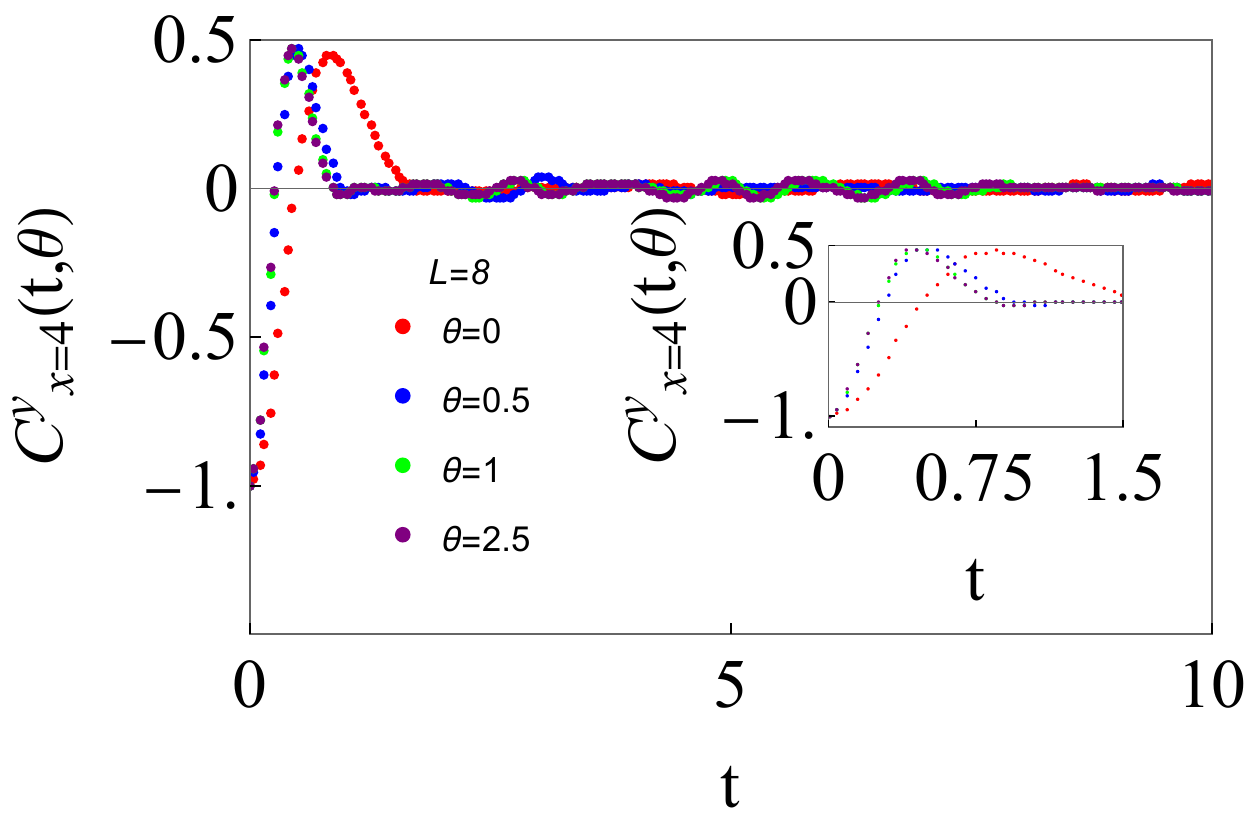}
      \includegraphics[width=0.32\textwidth]{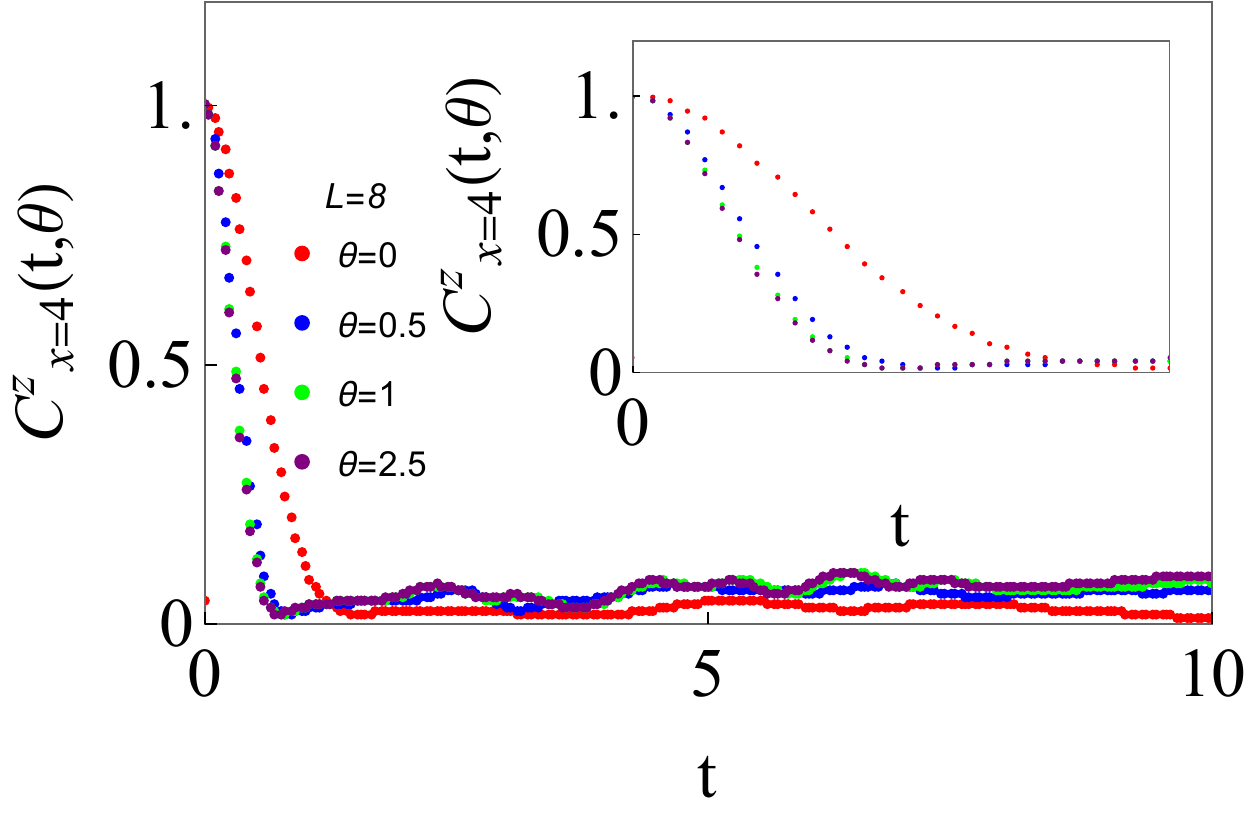}
    \caption{Position- and $\theta$-dependence of the time evolution of $C^{\alpha}_x(\theta,t)$. 
    In the top panels for $C^{\alpha}_x(\theta,t) $, we show how the time evolution of $C^{\alpha}_x(\theta,t) $ for various locations of the operator in the SSD limit.
    In the middle panels for $C^{\alpha}_{x=1}(\theta,t) $, we show how the time evolution of $C^{\alpha}_{x=1}(\theta,t) $ depends on $\theta$. 
    In the bottom panels for $C^{\alpha}_{x=4}(\theta,t) $,, we show how the time evolution of $C^{\alpha}_{x=4}(\theta,t) $ depends on  $\theta$.  In the insets for $C^{\alpha}_{x=4}(\theta,t) $,  we closely look at the early time evolution of two-point functions. The black points correspond to the time evolution of the two-point function in the undeformed chaotic spin chain.
    } 
    \label{theta_and_position_dependence_two_point_function_for_all}
  \end{figure*}

Here, we report how the time dependence of the two-point function depends on the position and $\theta$.
Let us define $C^{\alpha}_{x}$ as 
\be
C^{\alpha}_{x}(t,\theta)=¬\left \langle \sigma^{(1)}_{\alpha,x} \sigma^{(2)}_{\alpha,x} \right \rangle,
\ee
where $\alpha$ denotes the type of Pauli operator and $x$ denotes the location of operators.
In Fig. \ref{theta_and_position_dependence_two_point_function_for_all}, we report the time dependence of $C^{\alpha}_{x}$ for various $\theta$ and $x$. We can see from the time dependence of $C^{\alpha}_{x}$ that during the SSD time-evolution, $C^{\alpha=z}_{x=1}$ does not vanishes even for the large $t$.
\section{Position-and $\theta$- dependence of OTOCs in the chaotic region \label{OTOCs_in_chaotic}}
\begin{figure*}[tbh!]
        \centering
         \includegraphics[width=0.32\textwidth]{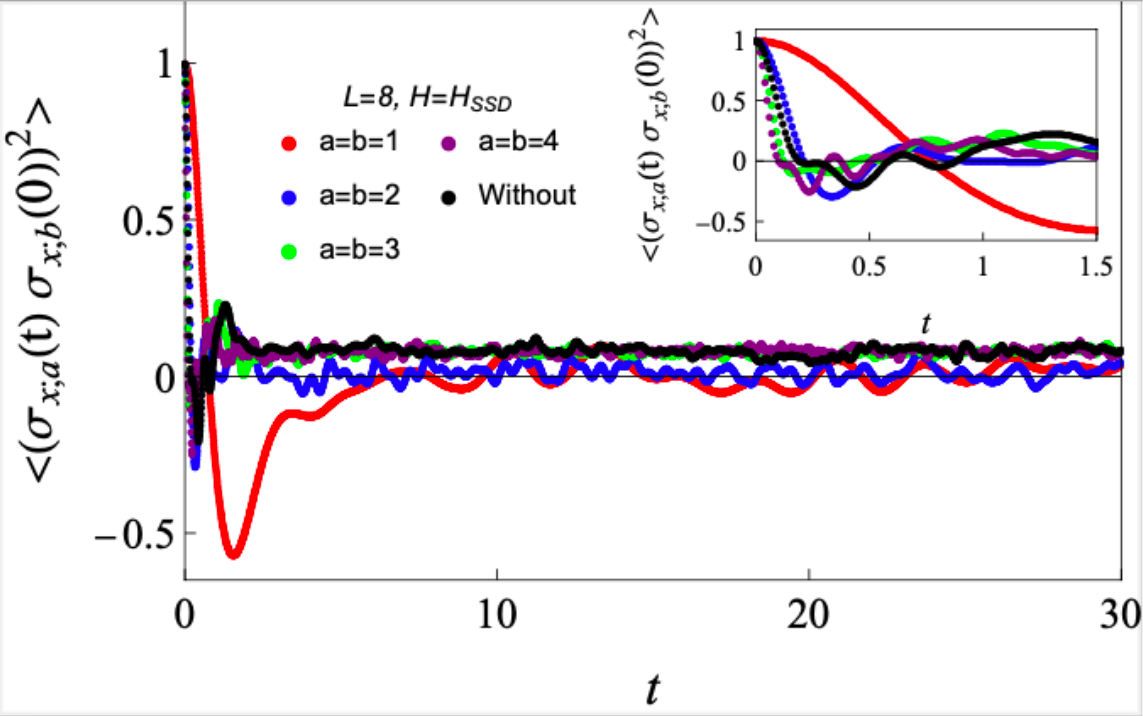}
         \includegraphics[width=0.32\textwidth]{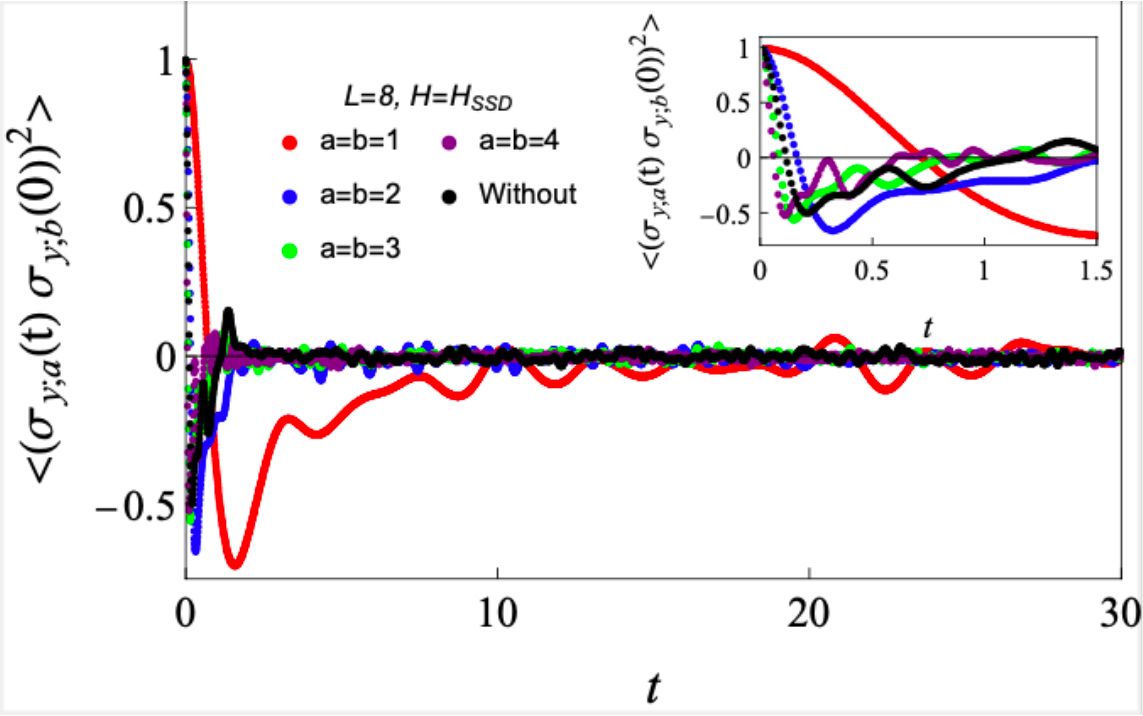}
     \includegraphics[width=0.32\textwidth]{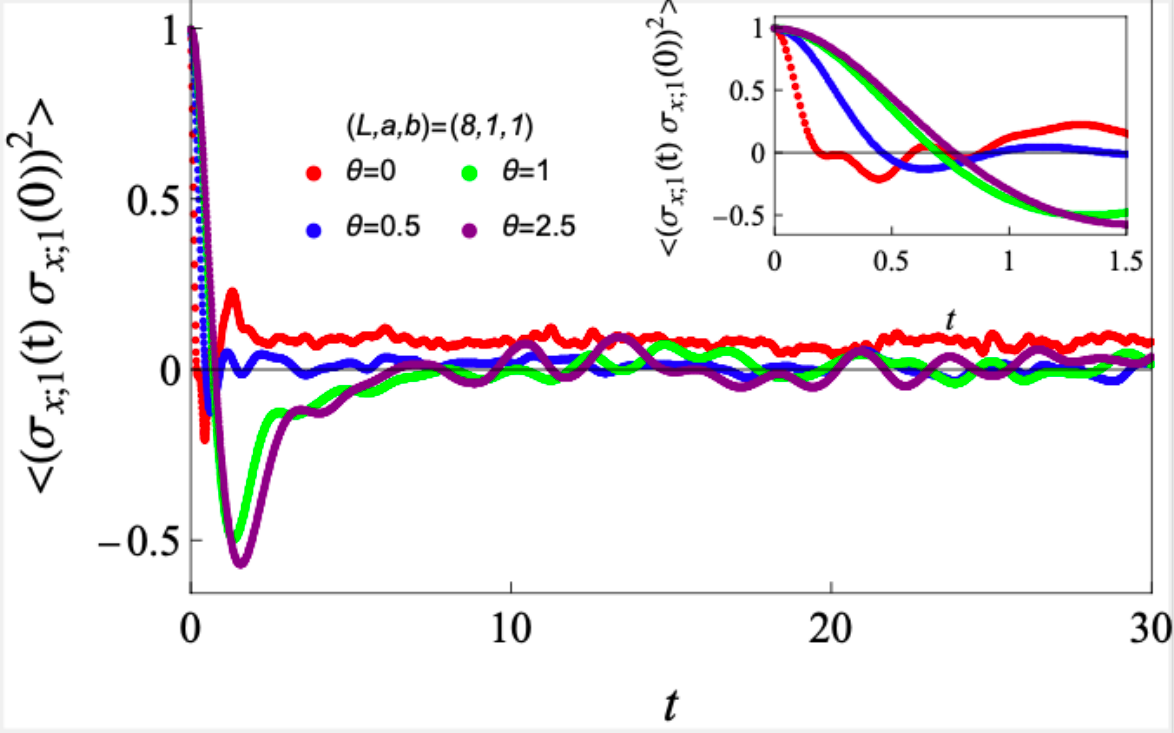}
      
      \includegraphics[width=0.32\textwidth]{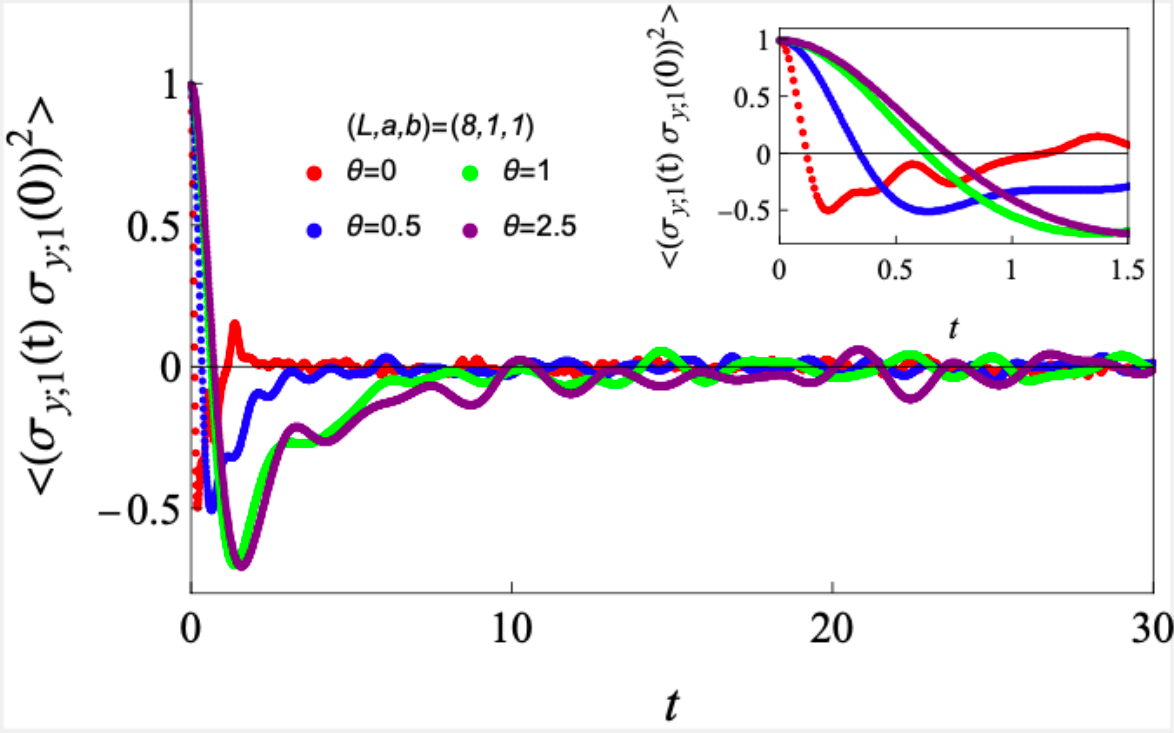}
      \includegraphics[width=0.32\textwidth]{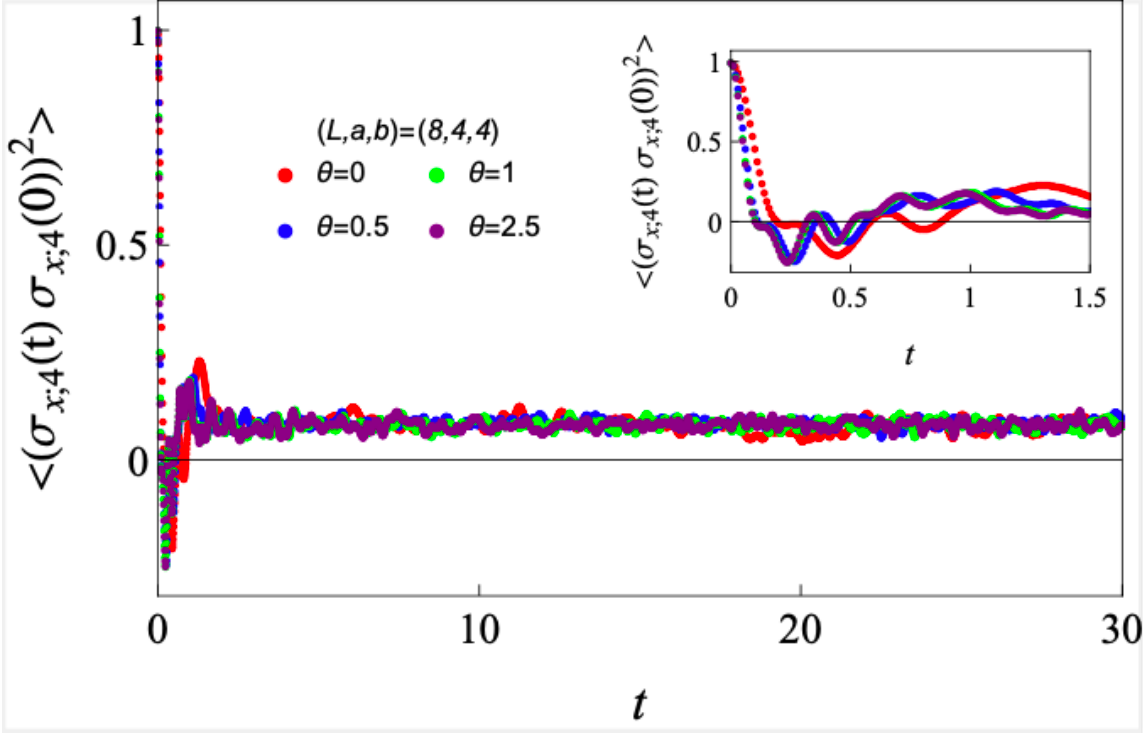}
      \includegraphics[width=0.32\textwidth]{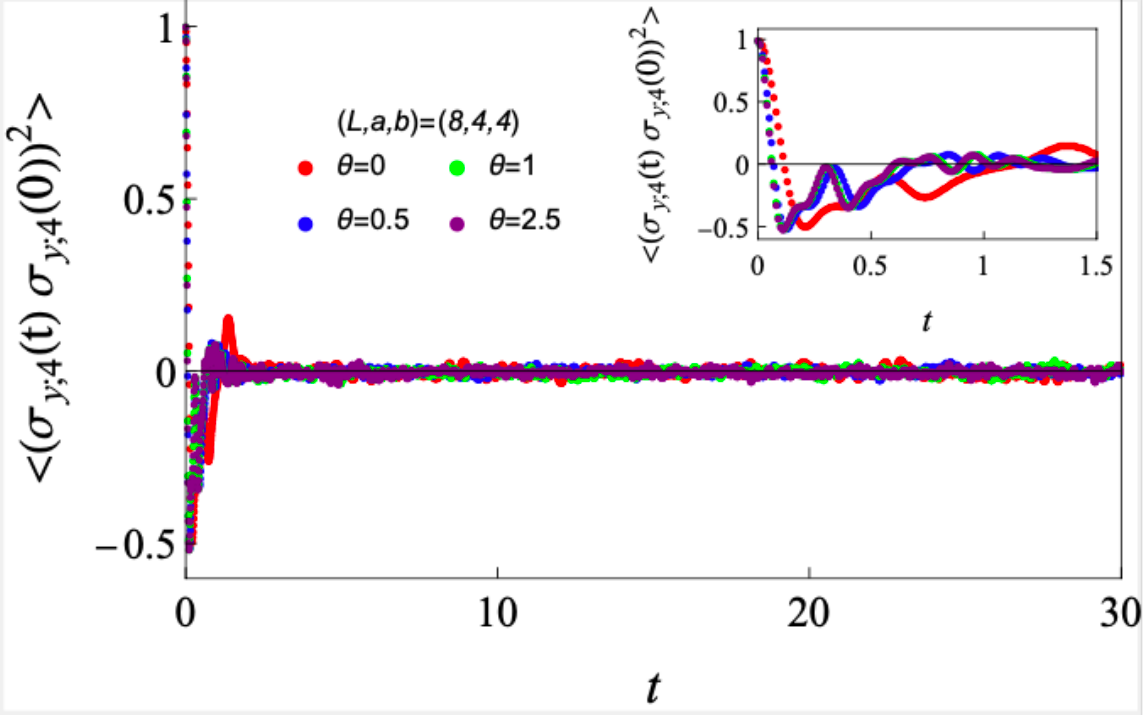}
    \caption{Position- and $\theta$-dependence of the OTOCs $\left\langle (\sigma_{\alpha=x,y,a}(t)\sigma_{\alpha=x,y,a}(0))^2\right\rangle=2^{-L}\text{tr}((\sigma_{\alpha=x,y,a}(t)\sigma_{\alpha=x,y,a}(0))^2)$ under the evolution by the M\"obius/SS deformed Hamiltonian in the chaotic regime, $(h_x,h_z)=(-1.05,0.5)$. 
    In the top two panels , we show how $\left\langle (\sigma_{\alpha=x,y,a}(t)\sigma_{\alpha=x,y,a}(0))^2\right\rangle$ with $H=H_{\text{SSD}}$ differs from that for the undeformed case $H=H_0$ (black points) depending on the location $a$ of the operators.
    In the right plot on the top panel and all the bottom panels, we show how $\left\langle (\sigma_{\alpha=x,y;a=1,4}(t)\sigma_{\alpha=x,y,a=1,4}(0))^2\right\rangle$ with $H=H_{\text{M\"obius}}$ depends on $\theta$.
    Insets show the time dependence of OTOCs in the early-time region, $0\le t\le 1.5$.
}
    \label{theta_and_position_dependence_OTOC_all}
  \end{figure*}

 We present the position- and $\theta$-dependence of OTOCs during the M\"obius and SSD time evolution.
 In Fig.\ref{theta_and_position_dependence_OTOC_all}, we depict the OTOCs as a function of $t$.
 In the top panels, we show how the time-dependence of $\left\langle (\sigma_{\alpha,a}(t)\sigma_{\alpha,a}(0))^2\right\rangle$ depends on $a$ for $\alpha=x,y$ during the SSD evolution.
 The black point illustrates the time-dependence of $\left\langle (\sigma_{\alpha,a}(t)\sigma_{\alpha,a}(0))^2\right\rangle$ during the evolution by the un-deformed Hamiltonian. 
 In the middle panels, we show how that of $\left\langle (\sigma_{\alpha,1}(t)\sigma_{\alpha,1}(0))^2\right\rangle$ depends on $\theta$ for $\alpha=x,y$. 
 In the bottom panels, we show that of $\left\langle (\sigma_{\alpha,4}(t)\sigma_{\alpha,4}(0))^2\right\rangle$ depends on $\theta$ for $\alpha=x,y$. 
 In the middle and bottom panels, the values of $\theta$ are taken to be $\theta=0,0.5,1$ and $2.5$.  
 The insets show the $\theta$-dependence of $\left\langle (\sigma_{\alpha,a=1,4}(t)\sigma_{\alpha,a=1,4}(0))^2\right\rangle$ in the early-time region, $0\le t\le 1.5$.
 
 We can see from the time-dependence of $\left\langle (\sigma_{\alpha,a}(t)\sigma_{\alpha,a}(0))^2\right\rangle$ in the top panels that when $a$ gets further far way from $a=1$ and closer to $a=4$, the early-time decay of $\left\langle (\sigma_{\alpha,a}(t)\sigma_{\alpha,a}(0))^2\right\rangle$ becomes faster.
 For $a=1,2$, the OTOCs under the evolution by the SSD Hamiltonian decay slower than under the un-deformed Hamiltonian, while for $a=3,4$, those under the evolution by the SSD Hamiltonian decay faster than under the un-deformed Hamiltonian.
 The middle and bottom panels indicate that when $\theta$ becomes larger for, the early-time decay of $\left\langle (\sigma_{\alpha,1}(t)\sigma_{\alpha,1}(0))^2\right\rangle$ becomes faster , while that for $a=4$ becomes slower.

\section{$\theta$-dependence of the energy accumulation}
\label{app_energydensity}
\begin{figure*}[tbh!]
    \includegraphics[width=0.32\textwidth]{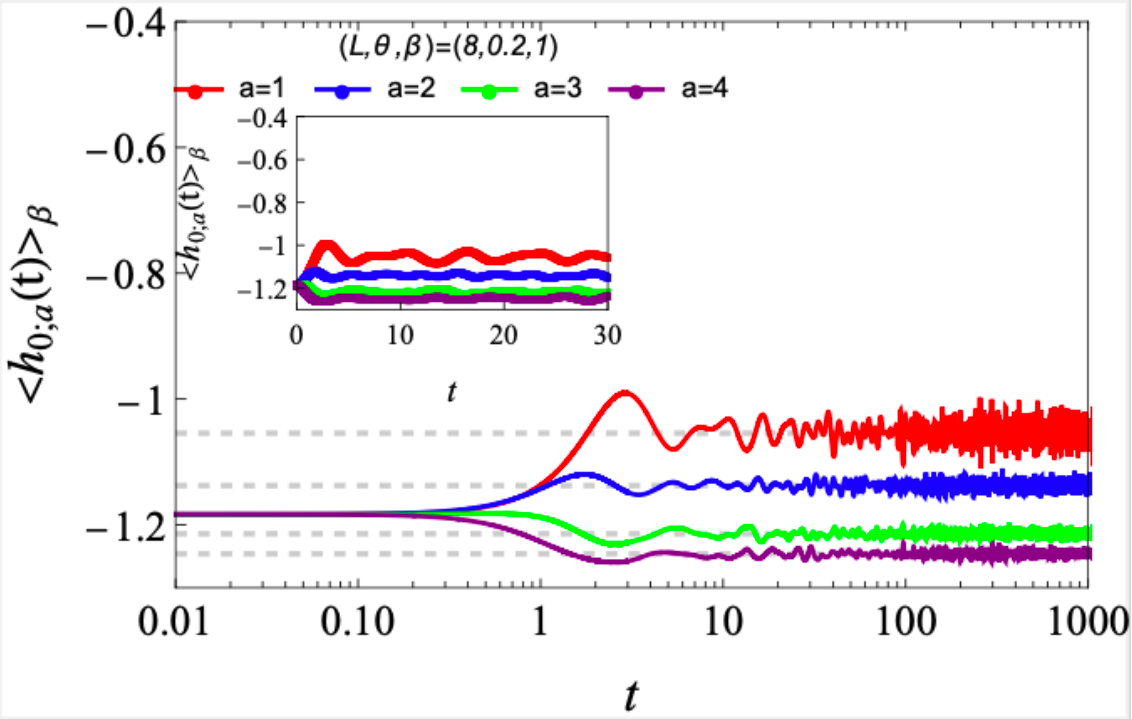}
      \includegraphics[width=0.32\textwidth]{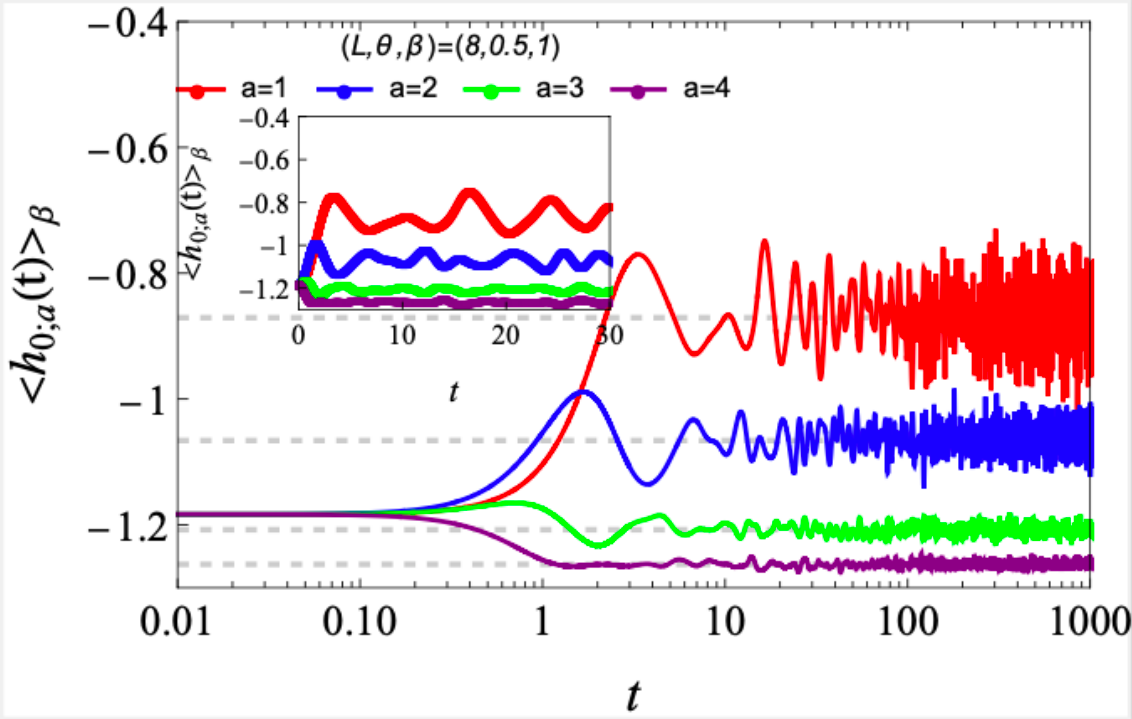}
      \includegraphics[width=0.32\textwidth]{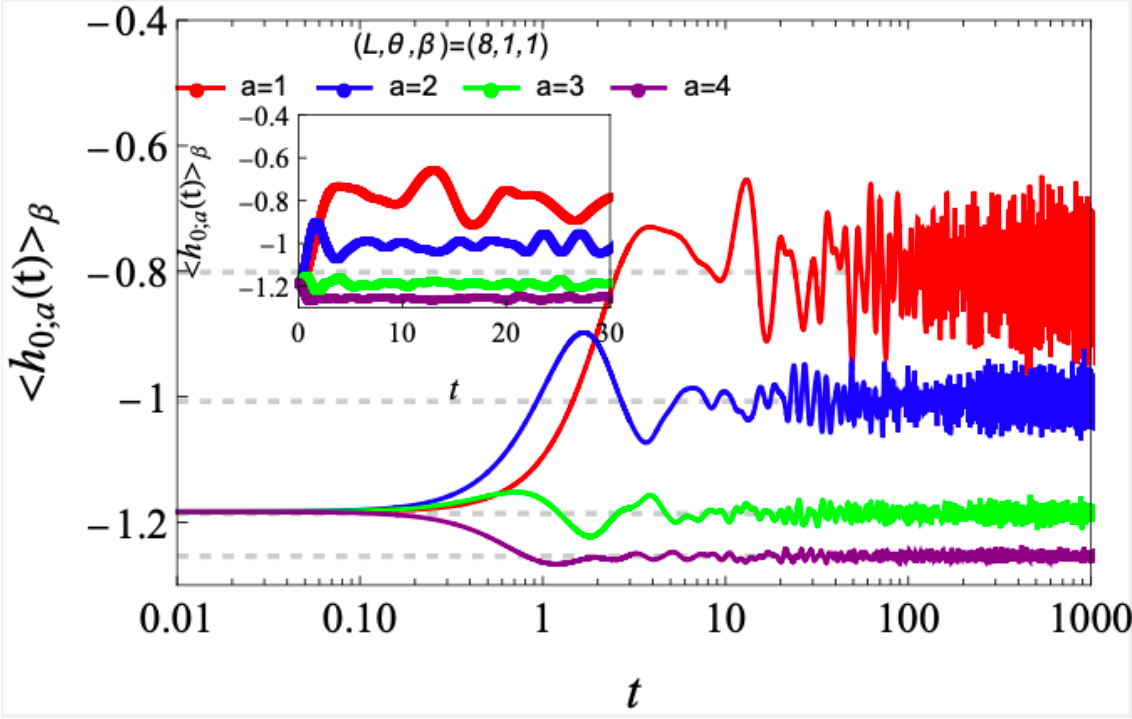}
    \caption{
Position- and $\theta$-dependence of the energy density $\left\langle h_{0,a}(t)\right\rangle_\beta$ during the M\"obius time evolution.
    For visibility, we have connected the discrete data points in $30<t<1000$ with $\Delta \log t=1.62\times 10^{-3}$ with lines.
    The dashed horizontal lines indicate the averaged values of $\left\langle h_{0,a}(t)\right\rangle_\beta$ over late times ($100<t<100000$ with $\Delta \log t=1.62\times 10^{-3}$).
    An animation of the time dependence of $\langle h_{0,a}(t)\rangle_\beta$ for $0<t<30$ is also available as a GIF file (\url{220919_energydensityanimation_L8_100hxm105_100hz50_10000beta10000_10goto.gif}) attached to the source of this paper at arXiv.org.
}
\label{theta_and_position_dependence_of_h0a}
  \end{figure*}

In Fig.\ \ref{theta_and_position_dependence_of_h0a} we show the time dependence of the expectation value of the un-deformed energy density $\langle h_{0,a}(t)\rangle_{\beta}$ \eqref{energydensityunderSSDquench} with $(L,h_x,h_z,\beta)=(8,-1.05,0.5,1)$ during the M\"obius time evolution with $\theta=0.2,0.5$ and $1$ (we have also displayed the result for SSD time evolution in Fig.~\ref{fig:ted}.
At $t=0$, $\langle h_{0,a}(t)\rangle_{\beta}$ are independent of $a$ since the un-deformed spin chain is invariant under the translation.
At early time the $\langle h_{0,a}(t)\rangle_{\beta}$ increases for $a\approx 0$ and decreases for $a\approx \frac{L}{2}$, hence the energy is accumulated around $a=0$.
We observe that the early time accumulation takes place faster as $\theta$ is increased.
At late times $\langle h_{0,a}(t)\rangle_{\beta}$ oscillates around some values which, measured from the initial value $\langle h_{0,a}(0)\rangle_\beta$, is larger as $a$ approaches the origin.
We observe that the late time energy distribution becomes sharper as $\theta$ is increased (see Fig.~\ref{theta_and_position_dependence_of_h0a}).
 \clearpage
  \newpage

\end{document}